\documentclass[journal]{IEEEtran}

\usepackage{graphics} 
\usepackage{epsfig} 
\usepackage{mathptmx} 
\usepackage{times} 
\usepackage{amsmath} 
\usepackage{amssymb} 
\usepackage{algorithm} 
\usepackage{verbatim} 
\usepackage{algpseudocode}
\usepackage{url}            
\usepackage{array}
\usepackage{epsfig}
\usepackage{amsmath}
\usepackage{multicol}
\usepackage{duckuments}
\usepackage[]{graphicx}
\usepackage{capt-of}

\usepackage{float}
\usepackage{xcolor}
\usepackage{graphicx}
\usepackage{tabto}
\usepackage{cleveref}
\usepackage{subfigure}
\usepackage{wrapfig}
\usepackage{textcomp}
\usepackage{float}
\usepackage{booktabs}
\usepackage{graphicx}
\usepackage{url} 
\usepackage{multirow}
\usepackage{amsmath}
\usepackage{flexisym}
\usepackage{cite}
\usepackage{filecontents}
\usepackage{caption}
\usepackage{stackengine}

\newcommand{\RN}[1]{%
\textup{\uppercase\expandafter{\romannumeral#1}}%
}

\ifCLASSINFOpdf
\else
\fi
\begin{document}
%
\title{Fast Strain Estimation and Frame Selection in Ultrasound Elastography
using Machine Learning
}
%
%
%

\author{Abdelrahman~Zayed, \textit{Student Member, IEEE} and Hassan~Rivaz, \textit{Senior Member}, \textit{IEEE}
\thanks{Abdelrahman Zayed and Hassan Rivaz are with the Department of Electrical and Computer Engineering and PERFORM Centre, Concordia University,
Montreal, QC, H3G 1M8, Canada. Email: a\_zayed@encs.concordia.ca and hrivaz@ece.concordia.ca.}
\thanks{}
\thanks{}}

\maketitle

\begin{abstract}
Ultrasound Elastography aims to determine the mechanical properties of the tissue by monitoring tissue deformation due to internal or external forces. Tissue deformations are estimated from ultrasound radio frequency (RF) signals and are often referred to as time delay estimation (TDE). Given two RF frames $I_{1}$ and $I_{2}$, we can compute a displacement image which shows the change in the position of each sample in $I_{1}$ to a new position in $I_{2}$. Two important challenges in TDE include high computational complexity and the difficulty in choosing suitable RF frames. Selecting suitable frames is of high importance because many pairs of RF frames either do not have acceptable deformation for extracting informative strain images or are decorrelated and deformation cannot be reliably estimated. \textcolor{black}{Herein, we introduce a method that learns 12 displacement modes in quasi-static elastography by performing Principal Component Analysis (PCA) on displacement fields of a large training database. In the inference stage, we use dynamic programming (DP) to compute an initial displacement estimate of around 1\% of the samples, and then decompose this sparse displacement into a linear combination of the 12 displacement modes}. Our method assumes that the displacement of the whole image could also be described by this linear combination of principal components. We then use the GLobal Ultrasound Elastography (GLUE) method to fine-tune the result yielding the exact displacement image. Our method, which we call PCA-GLUE, is more than 10 times faster than DP in calculating the initial displacement map while giving the same result. This is due to converting the problem of estimating millions of variables in DP into a much simpler problem of only 12 unknown weights of the principal components. Our second contribution in this paper is determining the suitability of the frame pair $I_{1}$ and $I_{2}$ for strain estimation, which we achieve by using the weight vector that we calculated for PCA-GLUE as an input to a multi-layer perceptron (MLP) classifier. We validate PCA-GLUE using simulation, phantom, and \textit{in vivo} data. \textcolor{black}{Our classifier takes only 1.5 $ms$ during the testing phase and has an F1-measure of more than 92\% when tested on 1,430 instances collected from both phantom and \textit{in vivo} datasets.}
\end{abstract}


%
\IEEEpeerreviewmaketitle
\bstctlcite{IEEEexample:BSTcontrol}

\section{Introduction}
%
%
%
%
\IEEEPARstart{U}{ltrasound} elastography has numerous applications in medical diagnosis of diseases and in image-guided interventions ~\cite{gennisson2013ultrasound,app1,app2,app3,app4,app5,app6,app7}. For example, it could be used in imaging cancer tumors by estimating the strain image since tumors are normally more rigid than the surrounding tissue. Ultrasound elastography has two main branches which are dynamic and quasi-static elastography~\cite{j2011recent}. Dynamic elastography refers to the quantitative estimation of the mechanical properties of the tissue. Quasi-static elastography, which is our focus in this paper, is more related to estimating the deformation of the tissue when an external force is applied~\cite{parker2010imaging,external_force}. \textcolor{black}{Recent work has shown success in performing ultrasound elastography using different methods such as spatial angular compounding \cite{he2017performance}, multi-compression strategy \cite{wang2019large}, Lagrangian tracking \cite{pohlman2019physiological} and guided circumferential waves \cite{li2017ultrasound}. In addition, other work has exploited the power of deep learning to achieve the same goal \cite{wu2018direct,hoerig2018data,peng2018convolution,gao2019learning,kibria2018gluenet,peng2020neural,tehrani2020displacement,sun2018pwc}.}

In spite of the various applications that ultrasound elastography has, it also has some challenges. One of these challenges is that time delay estimation (TDE) between frames of radio frequency (RF) data is computationally expensive. The methods used for calculating the TDE are either optimization-based~\cite{hashemi2017global,rivaz2011real,ashikuzzaman2019global} or window-based~\cite{jiang2015coupled,yuan2015analytical,luo2010fast}. In optimization-based techniques, the displacement image is estimated by minimizing a cost function. In window-based techniques, the objective is to find the displacement that maximizes a similarity metric such as normalized cross correlation (NCC) between two windows in the two frames before and after deformation.

Herein, we propose a computationally efficient technique for estimating an approximate TDE between two RF frames. To that end, we first learn the \emph{modes} of TDE by acquiring a large training database of free-hand palpation elastography by intentionally compressing the tissue in different manners. We then perform principal component analysis (PCA) to extract the \emph{modes} of TDE. At the test stage, we first run dynamic programming (DP) on only 1\% of RF data to extract a sparse TDE between two frames $I_1$ and $I_2$. We then estimate the weights of principal components that best approximate this sparse TDE, and subsequently use the weighted principal components as an initial TDE for GLobal Ultrasound Elastography (GLUE)~\cite{hashemi2017global}. We therefore call our method PCA-GLUE. PCA-GLUE was inspired by the success of~\cite{Wulff:CVPR:2015} in natural images. Similar work by Pohlman and Varghese~\cite{pohlman2018dictionary} has shown promising results on displacement estimation using dictionary representations.

Another challenge that ultrasound elastography faces is the suitability of the RF frames to be used for strain estimation. The two RF frames used are collected before and after applying an external force. Depending on the direction of the applied force, different qualities of strain images would be obtained. To be more precise, in-plane displacement results in high-quality strain images, whereas out-of-plane displacement results in low-quality strain images~\cite{hall2003vivo,chandrasekhar2006elastographic}. This means that collecting ultrasound data needs the person to be experienced in applying purely axial force. For imaging some organs, it is hard to hold the probe and apply a purely axial force even for experts. Furthermore, even for pure axial compression, two RF frames can be decorrelated due to internal physiological motions, rendering accurate TDE challenging.

Many solutions have been introduced for solving this problem. Lubinski \textit{et al.}~\cite{lubinski1999adaptive} suggested averaging several displacement images to improve the quality. The weights used are not equal, they rather depend on the step size (i.e. certain images would have higher weights than others).
Hiltawsky \textit{et al.}~\cite{hiltawsky2001freehand} tried to tackle the out-of-plane displacement by developing a mechanical compression applicator to force the motion to be in-plane. Jiang \textit{et al.}~\cite{jiang2006novel} defined a metric that informs the user whether or not to trust the pair of RF frames for strain estimation. This metric is the multiplication of the NCC of the motion compensated RF field and the NCC of the motion compensated strain field.
Other approaches~\cite{rivaz2009tracked,foroughi2013freehand} used an external tracker so as to pick up the RF frames that are collected roughly from the same plane. They used the tracking data to find pairs that have the lowest cost according to a predefined cost function.

Although all of the previously mentioned approaches showed an improvement in the quality of the strain image, they also have some drawbacks. The approaches introduced in~\cite{hiltawsky2001freehand,rivaz2009tracked,foroughi2013freehand} need an external device such as the mechanical applicator or the external tracker. This not only complicates the process of strain estimation, but also makes it more expensive.
The approach introduced in~\cite{jiang2006novel} gives a feedback on the quality of the strain image only \emph{after estimating TDE}, which means that it is not a computationally efficient method for frame selection. The method we propose in this paper selects suitable frames \emph{before estimating TDE} and is also computationally efficient.

Herein, we introduce a new method with three main contributions, which can be summarized as follows:
\begin{enumerate}[]
\item We develop a fast technique to compute the initial displacement image between two RF frames, which is the step prior to the estimation of the exact displacement image. Our method could also be used to speed up different displacement estimation methods by providing initial estimates.
\item We introduce a classifier that gives a binary decision for whether the pair of RF frames is suitable for strain estimation in only 1.5 $ms$ on a desktop CPU.
\item PCA-GLUE, which relies on DP to compute the initial displacement map, is robust to potential DP failures. 
\end{enumerate}
This work is an extension of our recent work~\cite{zayed_embc,zayed_iciar}, with the following major changes. First, we replace the multi-layer perceptron (MLP) classifier with a more robust one that can generalize better to unseen data. Second, we used automatically annotated images for training the classifier, compared to manual annotation that we previously used in~\cite{zayed_iciar}. Third, testing is now substantially more rigorous and is performed on 5 different datasets from simulation, phantom and \textit{in vivo} data. And last, the criteria for measuring the performance of the classifier used in this paper are the accuracy and F1-measure instead of using the signal to Noise Ratio (SNR) and Contrast to Noise Ratio (CNR) in~\cite{zayed_iciar}. Our code is available at \url{https://code.sonography.ai} and at \url{https://github.com/AbdelrahmanZayed}

\section{METHODS}
In this work, we have two main objectives which are fast TDE and automatic frame selection. We first propose a method that computes a superior approximate TDE compared to DP~\cite{rivaz2008ultrasound}, while being more than 10 times faster.

\begin{figure*}[]
\begin{center}
\subfigure[Directions of applied force]{\includegraphics[height=4.25 cm,width=2.5 cm]{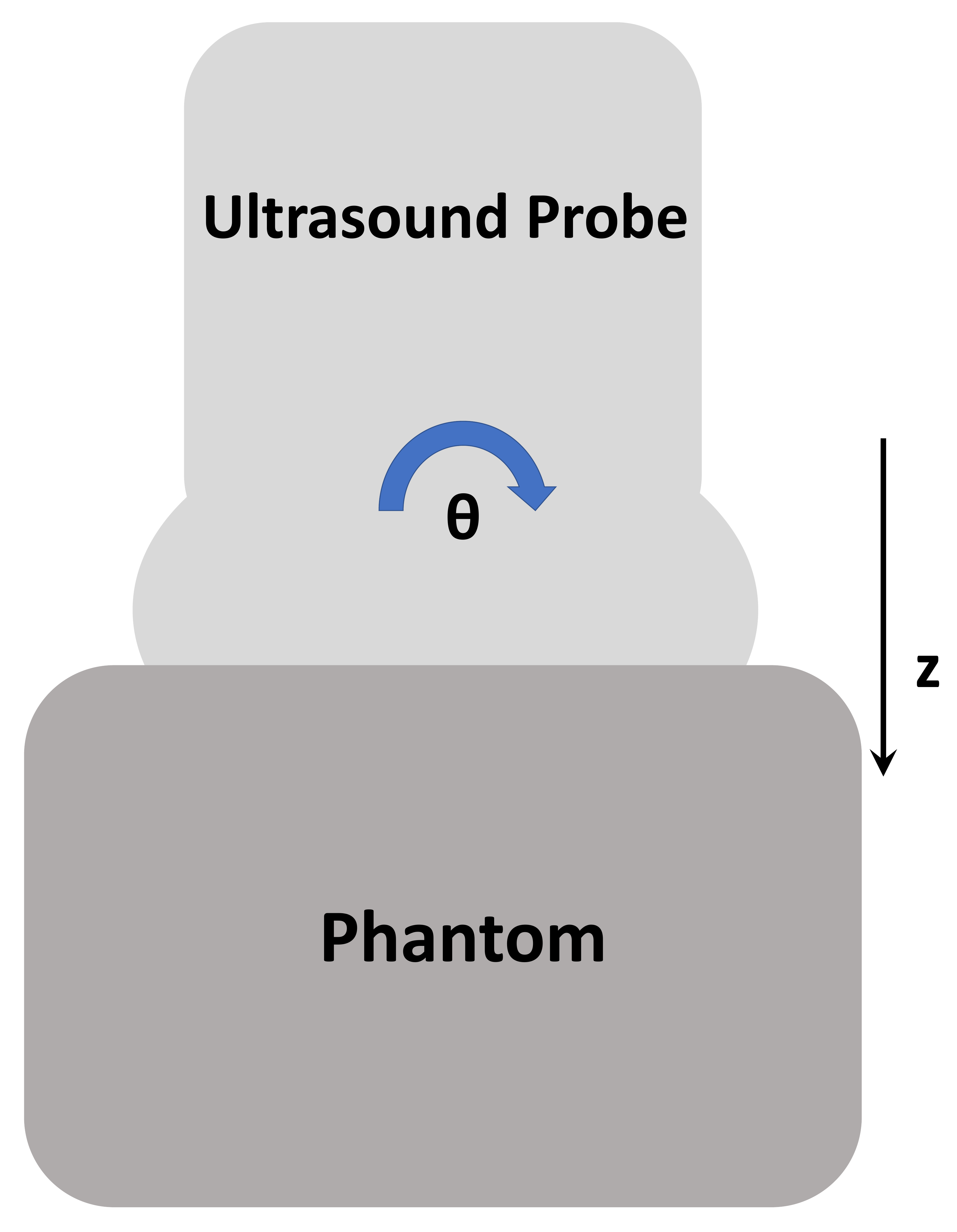}}\hspace*{-0em}
\subfigure[Axial deformation ($z$)]{\includegraphics[height=4.25 cm,width=8 cm]{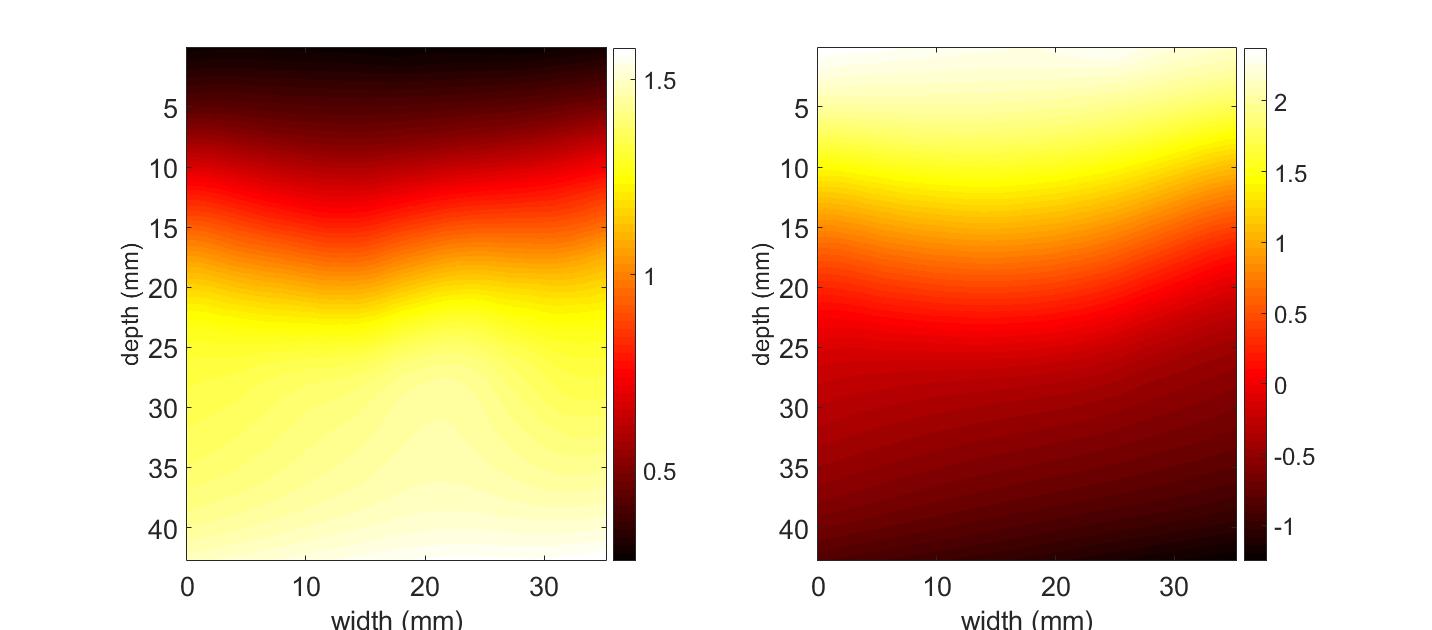}}\hspace*{-2em}
\subfigure[In-plane rotation ($\theta$)]{\includegraphics[height=4.25 cm,width=8 cm]{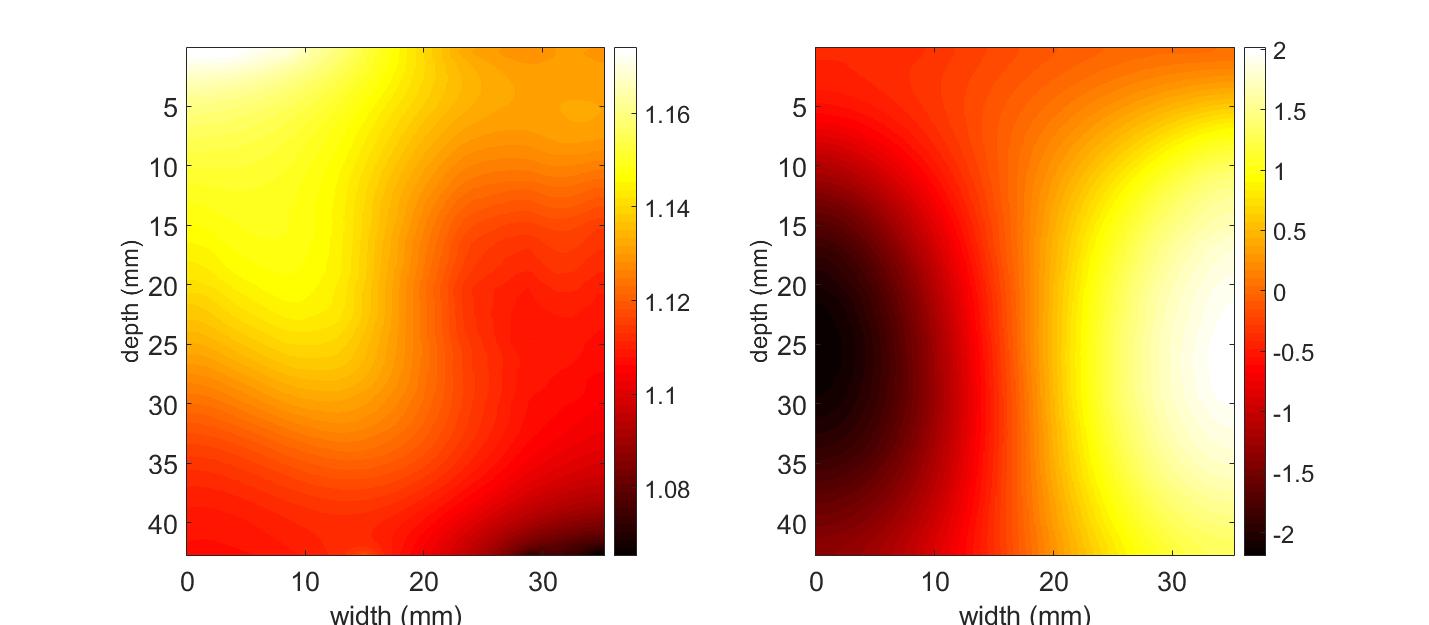}}\hspace*{-0.5em}
\end{center}%
\centering
\caption{Principal components of in-plane axial displacement \textcolor{black}{(in }$\textcolor{black}{mm}$\textcolor{black}{)} learned from both \textit{in vivo} and phantom experiments. In (a), translation of the probe along $z$ and its rotation by $\theta$ generates axial deformation in the phantom. In (b), extension and compression principal components along $z$ are shown. In (c), displacement arising from rotation by $\theta$ is shown.}
\label{fig1}
\end{figure*}

The idea is simple and logical: we compute $N$ principal components denoted by $\textbf{b}_1$ to $\textbf{b}_{N}$ from real experiments that describe TDE under the effect of an external force. In other words, the approximate displacement image is a linear combination of these principal components. During data collection, we applied the force in the 6 degrees of freedom (DOF) to ensure generality and a dataset of displacement images was obtained using GLUE. Using PCA, we were able to compute our principal components. Fig.~\ref{fig1} shows the directions of the applied force as well as some of the principal components learned.

For frame selection, our goal was simply to have a classifier that can classify whether the two RF frames are suitable for strain estimation. One can consider an approach of having a classifier that takes the two RF frames, where the samples are the input features (such as~\cite{peng2018convolution} and~\cite{kibria2018gluenet}), and outputs a binary decision of 1 for suitable frames and 0 otherwise. This approach would need a powerful GPU as the number of samples in each RF frame is approximately 1 million. To simplify the problem, we make use of our representation for the displacement image by the principal components. We can think of this as a dimensionality reduction method for the huge number of features that we had, where the input feature vector can be simply the N-dimensional weight vector $\textbf{w}$, which represents the weight of each principal component in the initial displacement image. Our low-dimensional weight vector $\textbf{w}$ is the input to a multi-layer perceptron (MLP) classifier that would output a binary number 1 or 0 depending on whether the two RF frames are suitable or not for strain estimation.

\subsection{Feature extraction}
\begin{figure}[]

\begin{minipage}[b]{1.0\linewidth}

\subfigure[Before interpolation]{\includegraphics[height=3.5 cm,width=4.35 cm]{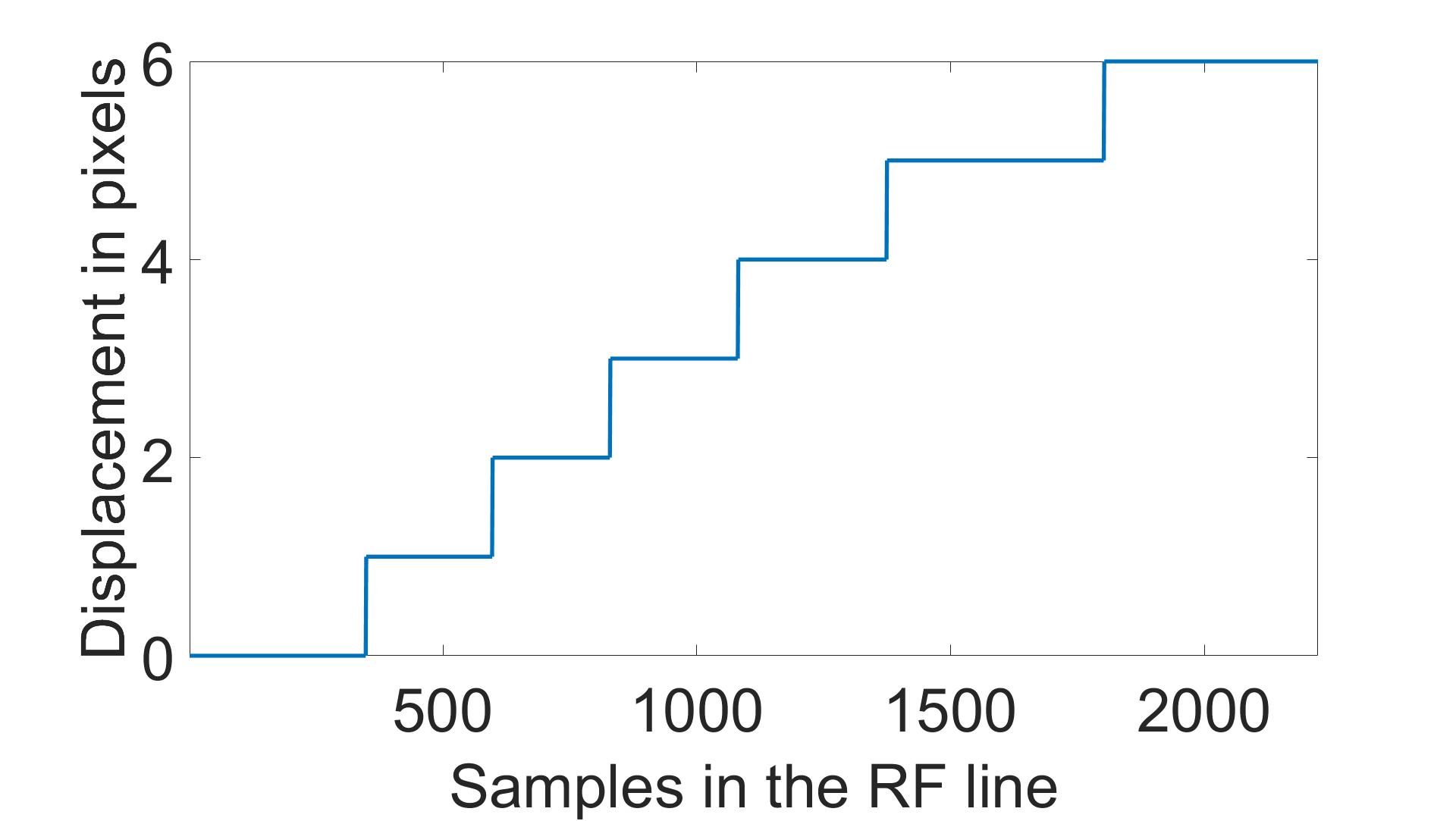}}
\subfigure[After interpolation]{\includegraphics[height=3.5 cm,width=4.35 cm]{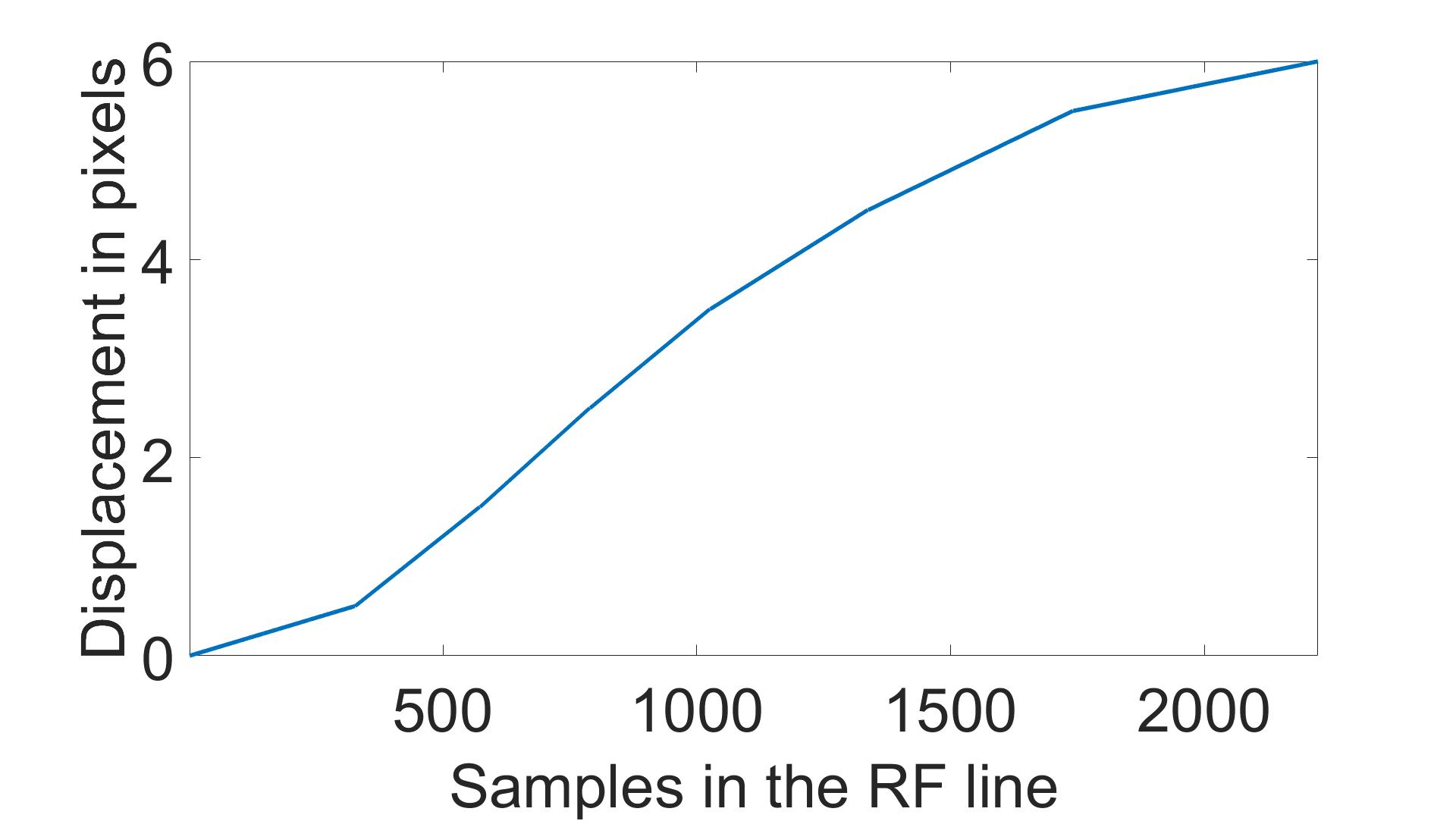}}
\end{minipage}

\caption{The displacement of a certain RF line before and after interpolation.}
\label{staircase}
\end{figure}

Consider having two RF frames $I_1$ and $I_2$ collected before and after some deformation, each of size $m\times l$, where $m$ is the number of samples in an RF-line and $l$ is the number a RF lines. Our goal is to estimate a coarse displacement image that describes the axial motion that each sample has had~\cite{2rf}. We start by running the DP algorithm on only $p$ RF lines out of the total $l$ RF lines (where $p<<l$) to get the integer displacement of $k=m\times p$ pixels. We then form a $k$-dimensional vector named $\textbf c$ after applying a simple linear interpolation to the $k$ estimates to make them smoother, \textcolor{black}{so that the integer estimates become linearly increasing with depth instead of the staircase approximation, as shown in Fig.~\ref{staircase}.}

Next, we construct the matrix $\textbf{A}$ such that 
\begin{equation}
\textbf{A}=
\begin{bmatrix}
\textbf{ b}_1(q_1) &\textbf{ b}_2(q_1) & \textbf{ b}_3(q_1) & \dots &\textbf{ b}_N(q_1) \\
\textbf{ b}_1(q_2) & \textbf{ b}_2(q_2) & \textbf{ b}_3(q_2) & \dots & \textbf{ b}_N(q_2) \\
\hdotsfor{5} \\
\textbf{ b}_1(q_K) & \textbf{ b}_2(q_K) & \textbf{ b}_3(q_K) & \dots & \textbf{ b}_N(q_K)
\end{bmatrix}
\end{equation}

\noindent \textcolor{black}{where the} $\textcolor{black}{N}$ \textcolor{black}{vectors from} $\textcolor{black}{\textbf{b}_1}$ \textcolor{black}{to} $\textcolor{black}{\textbf{b}_{N}}$ \textcolor{black}{represent our $N$ principal components,} $\textcolor{black}{q_1}$ \textcolor{black}{to} $\textcolor{black}{q_K}$ \textcolor{black}{correspond to our 2D coordinates of the sparse features chosen along the} $\textcolor{black}{p}$ \textcolor{black}{RF lines before deformation. For example, for an RF frame of size} $\textcolor{black}{2304 \times 384}$, \textcolor{black}{if we set} $\textcolor{black}{}$\textcolor{black}{$p$ to 1 and choose the sparse features to be along the RF line number 200. Then} $\textcolor{black}{k=2304}$ \textcolor{black}{and} $\textcolor{black}{q_1}$ \textcolor{black}{to} $\textcolor{black}{q_K}$ \textcolor{black}{would be} $\textcolor{black}{\{(1,200),(2,200),…..,(2304,200)\}}$.

Next, we compute the weight vector $\textbf{w}=(w_1,...,w_N)^T$ according to the following equation 
\begin{equation}
\label{lsq}
\hat{\textbf{w}}= \textrm{arg}\min_{\textbf{w}}||\textbf{Aw--c}|| 
\end{equation}

This implies that we choose the weight vector $\textbf{w}$ that decomposes the actual displacement image into a linear combination of the principal components weighted by some coefficients so as to have the minimum sum-of-squared error.
\subsection{Implementation}
\subsubsection{\textcolor{black}{Implementing PCA-GLUE for strain estimation}}
\textcolor{black}{Strain estimation relies on the extracted features to calculate the integer displacement image} $\textcolor{black}{\textbf{\^{d}}}$.

\begin{equation}
\label{eq_sigma}
\textcolor{black}{\textbf{\^{d}} = \sum_{n=1}^{N} \hat{w_n} \textbf{b}_n}
\end{equation}

\textcolor{black}{Eq.~\ref{eq_sigma} shows how to calculate the integer displacement image $\textbf{\^{d}}$ from the weight vector $\textbf{w}$, which is then passed to GLUE to obtain the exact displacement image $\textbf{d}$.} Finally, the resulting image is spatially differentiated to obtain the strain image. Algorithm 1 summarizes the procedure followed by PCA-GLUE.

\begin{algorithm}[]
\caption{PCA-GLUE}
\begin{algorithmic}[1]
\Procedure {PCA-GLUE~}{}
\State \text{Choose $p$ equidistant RF lines.}
\State \text{Run DP to get the integer axial displacement of the}
\text{\hskip 1.5em $p$ RF lines.}
\State \text{Solve Eq.~\ref{lsq} to get the vector $\textbf{w}$.}
\State \text{Compute the initial axial displacement $\textbf{\^{d}}$ of all RF}
\text{\hskip 1.5em lines by Eq.~\ref{eq_sigma}.}
\State \text{Use GLUE to calculate the exact axial displacement.}
\State \text{Strain is obtained by spatial differentiation of the}
\text{\hskip 1.5em displacement.}
\EndProcedure
\end{algorithmic}
\label{algorithm:pca_glue}
\vspace{.12cm}
\vspace{3mm}
\end{algorithm}

\subsubsection{Implementing the MLP classifier for frame selection}
The MLP classifier takes the weight vector $\textbf{w}$ (see Algorithm 1 steps 2 to 4) as the input feature vector. The ground truth (i.e. whether $I_{1}$ and $I_{2}$ are suitable for strain estimation or not) is obtained according to the procedure described in Algorithm 2.

\begin{algorithm}[]
\caption{Labelling the dataset for the MLP classifier}
\begin{algorithmic}[1]
\Procedure {~}{}
\State \text{RF frames $I_{1}$ and $I_{2}$ are passed to GLUE to obtain the}
\text{ \hskip 1em displacement image.}
\State \text{$I_{2}$ is deformed and interpolated according to the}
\text{ \hskip 1em computed displacement image yielding $I_{2}$\textprime.}
\State \text{We calculate the Normalized cross correlation (NCC)}
\text{\hskip 1.5em between $I_{1}$ and $I_{2}$\textprime.}
\State \text{The final decision is 1 if the NCC is higher than 0.9}
\text{\hskip 1.5em and 0 otherwise.}
\EndProcedure
\end{algorithmic}
\label{annotating}
\vspace{.12cm}
\vspace{3mm}
\end{algorithm}

The issue with this algorithm is that it is slow because of three computationally expensive steps of 2, 3 and 4. As such, it cannot be performed on many pairs of RF frames in real-time. Our goal is to train a classifier that predicts the output of step 5 by bypassing steps 2 to 4. The architecture of our classifier is relatively simple, with an input layer, 3 hidden layers, and an output layer. The input layer takes the N-dimensional vector $\textbf{w}$. 
The 3 hidden layers contain 256, 128 and 64 hidden units with a Rectified Linear Unit (ReLU) as the activation function. The output layer contains one unit, where the predicted value corresponds to the Normalized cross correlation (NCC) between $I_{1}$ and $I_{2}\textprime$ such that 

\begin{equation}
\label{eq_NCC}
NCC=\dfrac{\sum_{i} (I_{1}(i)-\bar{I_{1}})(I_{2}(i)\textprime-\bar{I_{2}\textprime}) }{\sqrt{\sum_{i} (I_{1}(i)-\bar{I_{1}})^2\sum_{i}(I_{2}(i)\textprime-\bar{I_{2}\textprime})^2}} \forall i \in I_{1} \cap I_{2}\textprime
\end{equation}

\noindent where $\bar{I_{1}}$ and $\bar{I_{2}\textprime}$ are the mean values of the RF frames $I_{1}$ and $I_{2}\textprime$ respectively.

\begin{figure}[t]
\begin{center}
\includegraphics[width=0.95\linewidth,trim=0 0 0 0,clip]{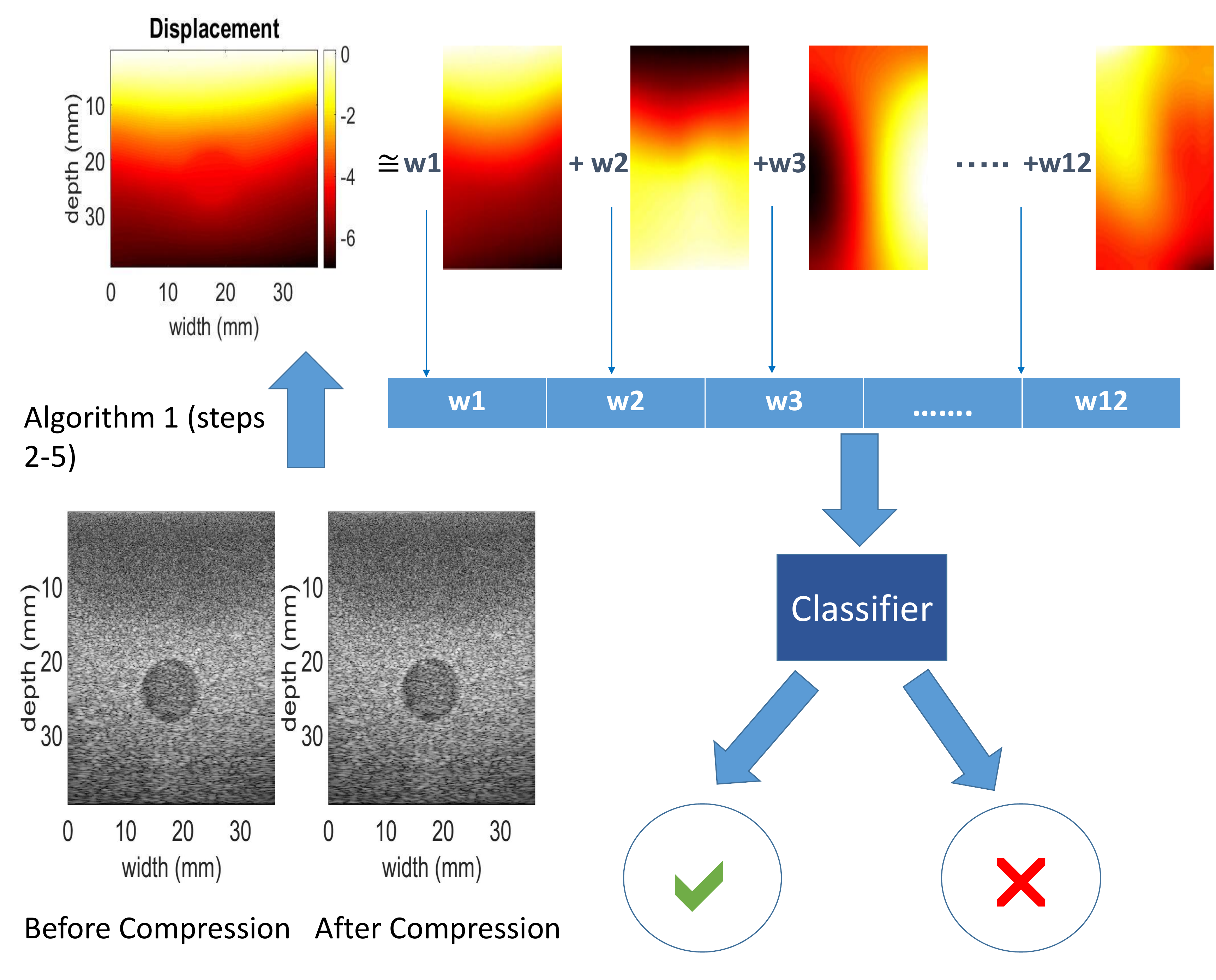}
\end{center}%
\caption{\textcolor{black}{The overall procedure used for frame selection. Given two RF frames (we are showing here the B-mode images for illustration) collected before and after deformation, we first estimate the integer displacement image} $\textcolor{black}{\textbf{\^{d}}}$ \textcolor{black}{(in }$\textcolor{black}{mm}$) \textcolor{black}{using PCA-GLUE, by applying Algorithm 1 (steps 2-5). We then use the weight vector }$\textcolor{black}{\textbf{w}}$ \textcolor{black}{as the input feature vector to the MLP classifier.}}%
\vspace{-.1cm}
\label{algorithm_steps}%
\end{figure}

\begin{figure}[t]
\begin{center}
\includegraphics[width=0.6\linewidth,trim=0 0 0 0,clip]{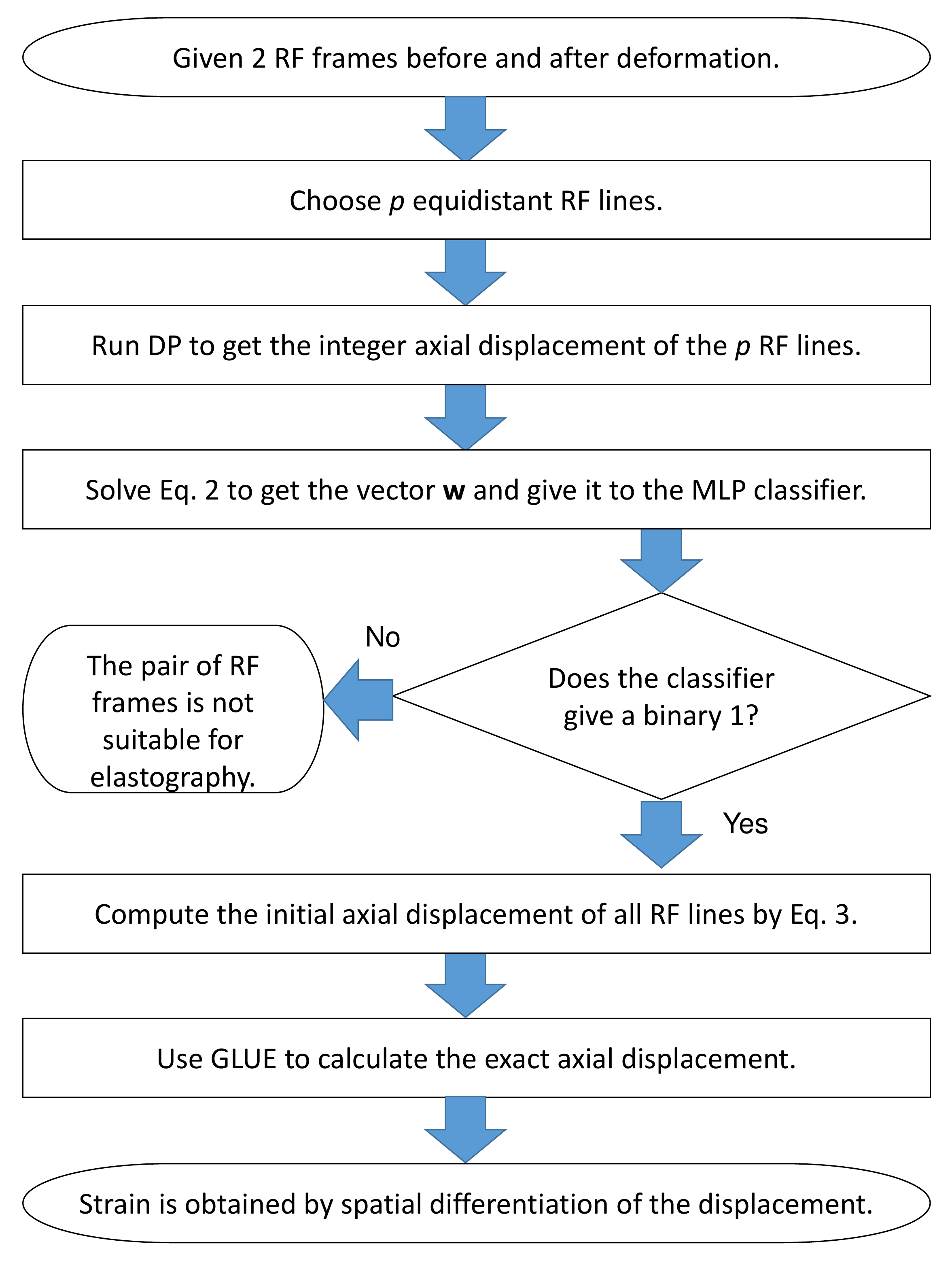}
\end{center}%
\caption{Flowchart of RF frame selection and strain estimation.}%
\vspace{-.1cm}
\label{flowchart}%
\end{figure}

\textcolor{black}{NCC has been widely used as a similarity metric by several image registration methods \cite{de2017end,chicotay2017two,li2018non,subramaniam2018ncc}. In this work, we claim that the NCC between $I_{1}$ and $I_{2}\textprime$ is an indicator for the suitability of $I_{1}$ and $I_{2}$ for elastography. Therefore, we apply a threshold on the value of both the predicted NCC and the ground truth NCC to compute the binary equivalent, which is 1 when the NCC is higher than 0.9 and 0 otherwise}. One possible criticism to our work might be that we do not directly estimate the binary output. This is because better results were obtained when training is done to estimate the NCC, as opposed to training to obtain a binary decision. The reason is that the NCC value provides more information to the network for better training compared to its thresholded binary number. It also makes the derivative of the loss function smoother, resulting in improved backpropagation. Another benefit is to be able to pick up the best possible frame to be paired with a certain frame, where we pair it with the frame with the highest NCC in a specified window of the 16 nearest frames, which has only one solution (assuming that there exist good frames in the window), compared to multiple solutions if the result is just a binary number. Our loss function is the mean square error (MSE) between the estimated NCC and the actual NCC before thresholding. We use Adam optimizer~\cite{kingma2014adam} with a learning rate of $1\mathrm{e}{-3}$. The code is written in Python using Keras \cite{chollet2015keras}. Fig. \ref{algorithm_steps} shows the overall procedure followed by our algorithm for frame selection. \textcolor{black}{Fig. \ref{flowchart} contains a flowchart that shows how strain estimation and frame selection are augmented together.}

\subsection{Data Collection}
\label{data_description}
\subsubsection{PCA-GLUE}

\textcolor{black}{We collected 4,055 RF frames from 3 different CIRS phantoms (Norfolk, VA), namely Models 040GSE, 039 and 059 at different locations at Concordia University’s PERFORM Centre. Model 040GSE has 3 different cylindrical regions with elasticity moduli of 10, 40 and 60 kPa. The 039 and 059 models have spherical inclusions that are distributed throughout the phantoms. The elasticity moduli of the inclusions are 27 kPa for Model 039 and in the range of 10-15 kPa for model 059. The compression was applied in 3 different directions: in-plane axial motion, in-plane rotation and out-of-plane lateral motion. The ultrasound device used is the 12R Alpinion Ultrasound machine (Bothell, WA) with an L3-12H high density linear array probe at a center frequency of 8.5 MHz and sampling frequency of 40 MHz.}

\textcolor{black}{We also have access to 298 RF frames collected at Johns Hopkins Hospital from 3 different patients who were undergoing liver ablation for primary or secondary liver cancers using Antares Siemens system (Issaquah, WA) at a center frequency of 6.67 MHz with a VF10-5 linear array at a sampling rate of 40 MHz. The study has the approval of the institutional review board and an informed consent was obtained from the patients. 3,635 RF frames out of the total 4,055 phantom RF frames, along with 137 \textit{in vivo} RF frames out of the total 298 \textit{in vivo} RF frames were used for obtaining the principal components $\textbf{b}_1$ to $\textbf{b}_{N}$ by following the procedure in Algorithm 3, leaving 420 phantom RF frames and 161 \textit{in vivo} RF frames for validating our method. \textcolor{black}{It is important to note that the training data was excluded from further evaluation.}}

The simulation data was generated using Field \RN{2} software \cite{jensen1996field}. ABAQUS (Providence, RI) software was used to apply some compression, and the ground truth displacement was generated using finite element method (FEM).


\subsubsection{MLP classifier}
We used the data that we collected for PCA-GLUE for training our MLP classifier. It was trained on 4,662 instances from both phantom and \textit{in vivo} data, which were partitioned between training and validation with a ratio 80:20. Testing was done on a different dataset composed of 1,430 frames. The ground truth was obtained by following the procedure in Algorithm~\ref{annotating}.

\subsection{\textcolor{black}{Metrics used for performance assessment}}

\textcolor{black}{In order to be able to quantitatively measure the performance of the strain estimation algorithm PCA-GLUE, we use two quality metrics which are the SNR and CNR~\cite{ophir1999elastography}, such that:}
\begin{equation}
\textcolor{black}{CNR=\frac{C}{N}=\sqrt{\frac{2(\bar{s_{b}}-\bar{s_{t}})^2}{\sigma_{b}^2 + \sigma_{t}^2} }, SNR=\frac{\bar{s}}{\sigma}}
\end{equation}\\
\textcolor{black}{where} $ \textcolor{black}{\bar{s_{t}}}$ \textcolor{black}{and} $ \textcolor{black}{\sigma_{t}^2}$ \textcolor{black}{are the strain average and variance of the target window, }$ \textcolor{black}{\bar{s_{b}}}$ \textcolor{black}{and} $ \textcolor{black}{\sigma_{b}^2}$ \textcolor{black}{are the strain average and variance of the background window respectively. We use the background window for SNR calculation (i.e. }$\textcolor{black}{\bar{s}}$\textcolor{black}{=}$ \textcolor{black}{\textcolor{black}{\bar{s_{b}}}}$ \textcolor{black}{and} $\textcolor{black}{\sigma }$\textcolor{black}{=}$ \textcolor{black}{\sigma_{b})}$. \textcolor{black}{The background window is chosen in a uniform region where the strain values do not vary considerably. It is worth mentioning that the SNR and CNR values are obtained as the average over 10 different experiments.}

\textcolor{black}{Precision and recall are two important metrics for assessing the performance of a classifier. The F1-measure incorporates both metrics as follows:}

\begin{equation}
\textcolor{black}{F1-measure=2\frac{(Precision\times Recall)}{(Precision+Recall} }
\end{equation}

\begin{algorithm}[]
\color{black}
\caption{\textcolor{black}{Obtaining the principal components}}
\begin{algorithmic}[1]
\color{black}
\Procedure {~}{}
\State \text{\textcolor{black}{Run GLUE on the 3,772 RF frame pairs collected}}
\text{ \hskip 1em \textcolor{black}{(3,635 from the phantom dataset and 137 from the}}
\text{ \hskip 1em  \textcolor{black}{\textit{in vivo} dataset), yielding 3,772 displacement images.}}
\State \text{\textcolor{black}{Reshape every displacement image from a $2304 \times 384$ }}
\text{ \hskip 1em \textcolor{black}{matrix into an $884,736 \times 1$ vector.}}
\State \text{\textcolor{black}{Form the data matrix $\textbf{X}$ of size $884,736 \times 3,772$ by }}
\text{ \hskip 1em \textcolor{black}{concatenating the 3,772 vectors.}}
\State \text{\textcolor{black}{Compute the covariance matrix as follows:}}
\text{ \hskip 1em $ \textbf{\textcolor{black}{S}}=\frac{\textcolor{black}{1}}{\textcolor{black}{n}} \textcolor{black}{\times \textbf{X}}$\textcolor{black}{\textprime}$\textcolor{black}{ \times \textbf{X}}$\textcolor{black}{\textprime}$\textcolor{black}{^T}$, \textcolor{black}{where }$ \textcolor{black}{\textbf{X}\textprime}$ \textcolor{black}{is the matrix }$\textcolor{black}{\textbf{X}}$ after \textcolor{black}{ subt-}}
\text{ \hskip 1em \textcolor{black}{racting the mean value of the elements in each row}}
\text{ \hskip 1em \textcolor{black}{(we set $n$ to 3,772).}}
\State \text{\textcolor{black}{Obtain the eigenvalues of the matrix \textbf{S} and sort them}}
\text{ \hskip 1em \textcolor{black}{descendingly.}}
\State \text{\textcolor{black}{Compute the eigenvectors corresponding to the largest}}
\text{ \hskip 1em \textcolor{black}{12 eigenvalues.}}
\State \text{\textcolor{black}{Obtain the 12 principal components for the axial displ-}}
\text{ \hskip 1em \textcolor{black}{acement images (Fig. 1 (b) and (c)) by reshaping each}}
\text{ \hskip 1em  \textcolor{black}{of the 12 eigenvectors from an $884,736 \times 1$ vector into}}
\text{ \hskip 1em  \textcolor{black}{a 2,304 $\times$ 384 matrix.}}
\EndProcedure
\end{algorithmic}
\label{obtaining_pc}
\vspace{.12cm}
\vspace{3mm}
\end{algorithm}

\section{Results}
For our results, we set $N=12$. This means that every displacement image is represented by 12 axial principal components in the form of a 12-dimentional vector \textbf{w}. \textcolor{black}{For results with different number of principal components, please refer to the Supplementary Material of this paper.} We found that this representation captures $95\%$ of the variance in the original data. \textcolor{black}{For the NCC method, we used windows of size (5.42} $\textcolor{black}{\times}$ \textcolor{black}{12.49)} $\textcolor{black}{\lambda}$\textcolor{black}{. For DP estimation, the tunable parameter} $\textcolor{black}{\alpha_{DP}}$  \textcolor{black}{is set to 0.2. For GLUE, the parameters used during phantom experiments are} $\textcolor{black}{\alpha_1=5}$\textcolor{black}{, }$\textcolor{black}{\alpha_2=1}$\textcolor{black}{,} $\textcolor{black}{\beta_1=5}$ \textcolor{black}{and}{ $\textcolor{black}{\beta_2=1}$\textcolor{black}{. During \textit{in vivo} experiments, we change GLUE’s parameters to} $\textcolor{black}{\alpha_1=20}$\textcolor{black}{,} $\textcolor{black}{\alpha_2=1}$\textcolor{black}{,} $\textcolor{black}{\beta_1=20}$ \textcolor{black}{and} $\textcolor{black}{\beta_2=1}$\textcolor{black}{, to account for the increased noise. \textcolor{black}{For NCC, GLUE and PCA-GLUE, the strain image is obtained from the displacement image using least square strain estimation \cite{kallel1997least}.}}

For the running time, we trained PCA-GLUE in 5 hours, but training is done only once. For testing, we estimate the initial displacement in just 258 $ms$ for two very large RF frames of sizes $2304 \times 384$ using an 8th generation 3.2 GHz Intel core i7 compared to 2.6 seconds if we use DP. \textcolor{black}{For the frame selection, feature extraction and labeling the data took 30 hours, which included the procedure in Algorithm~\ref{annotating}}. The actual training of the MLP classifier took $29.16$ seconds, while testing takes only 1.5 $ms$.

\begin{figure*}[]
\begin{center}
\subfigure[B-mode]{\includegraphics[height=3.25 cm,width=4.8 cm]{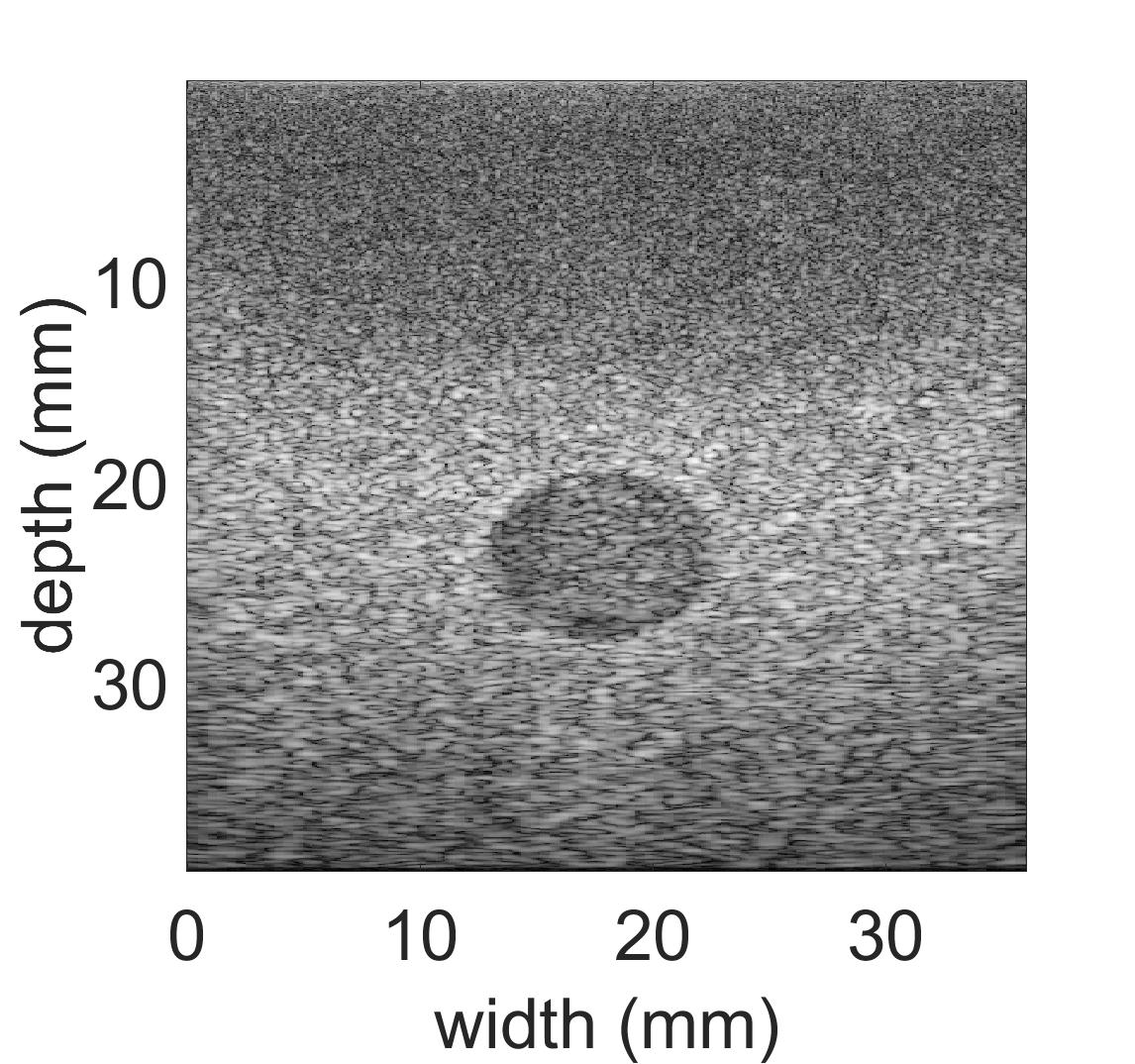}}\hspace*{-0.5em}
\subfigure[NCC]{\includegraphics[height=3.25 cm,width=4.8 cm]{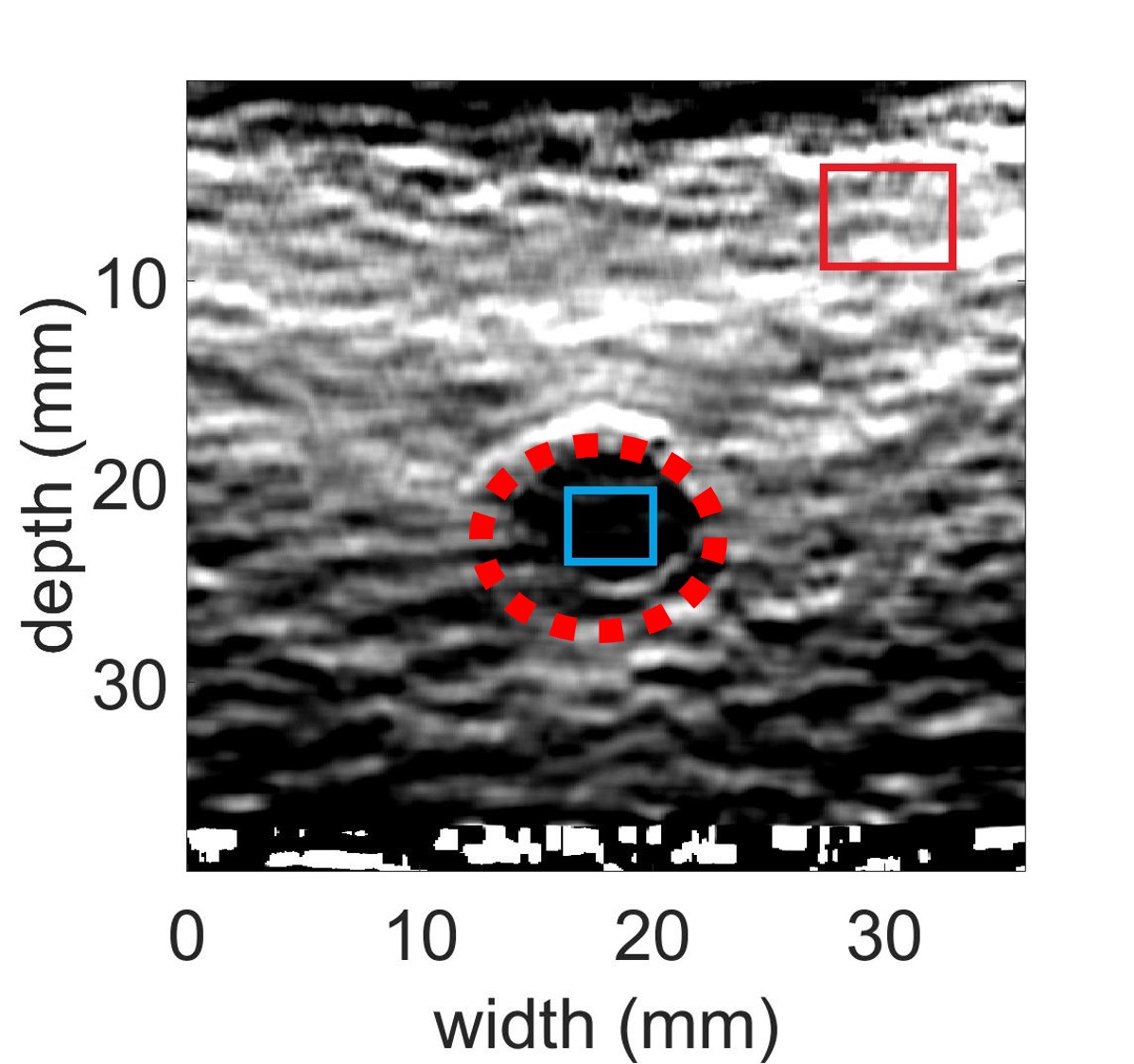}}\hspace*{-0.5em}
\subfigure[GLUE]{\includegraphics[height=3.25 cm,width=4.8 cm]{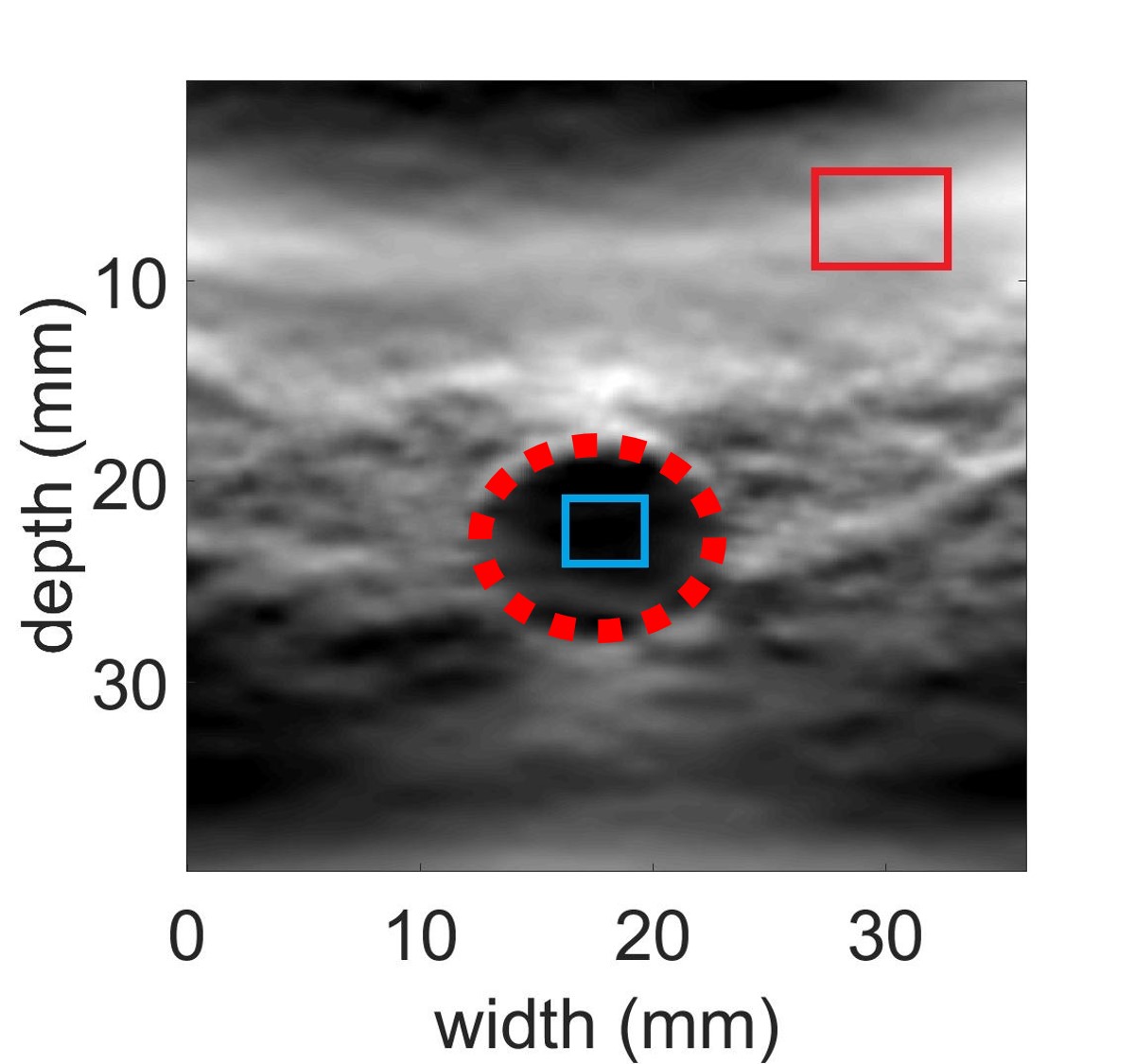}}\hspace*{-0.5em}
\subfigure[PCA-GLUE]{\includegraphics[height=3.25 cm,width=4.8 cm]{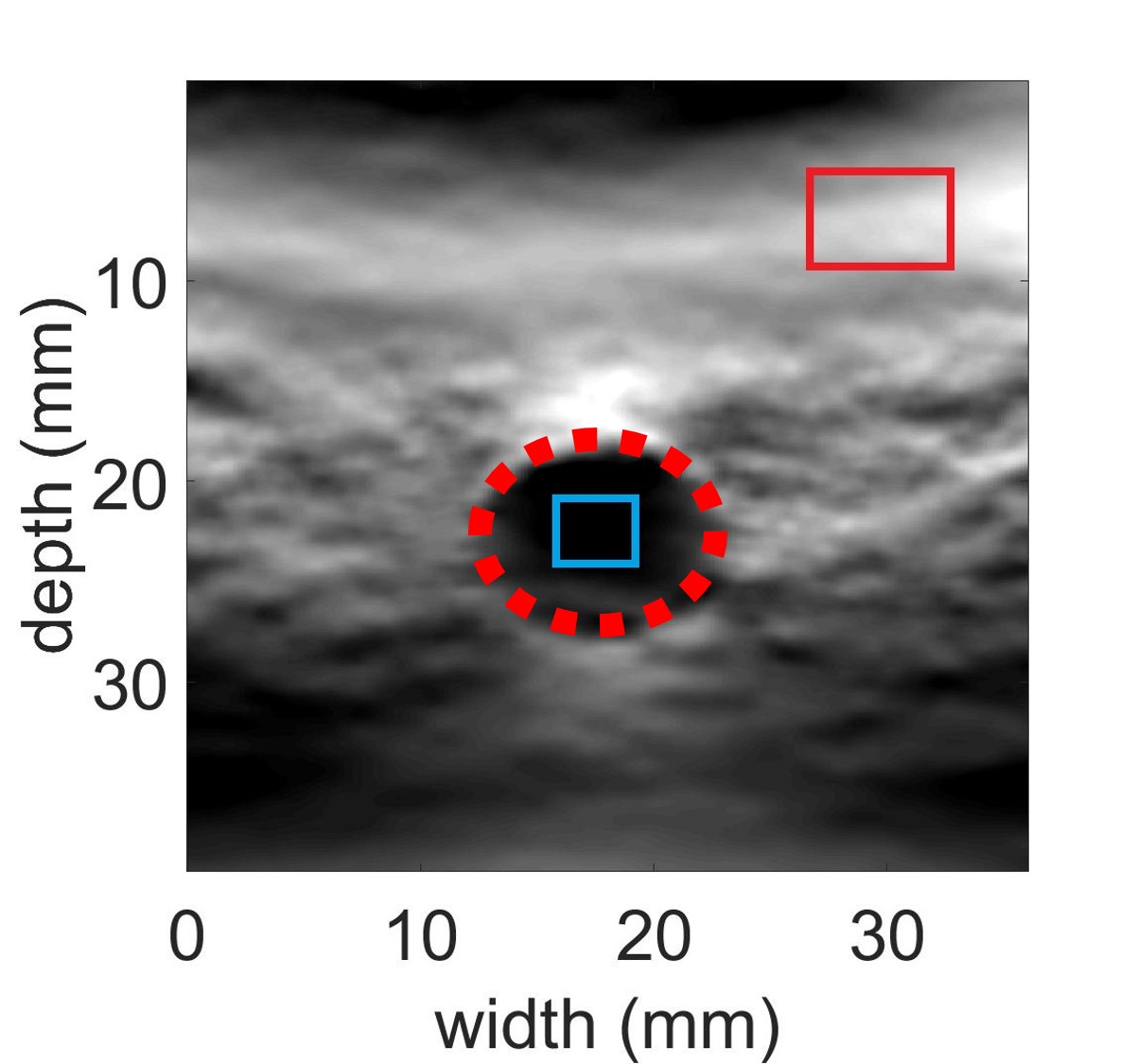}}\vspace*{-0em}
\stackunder[5pt]{\includegraphics[height=0.75 cm,width=4 cm]{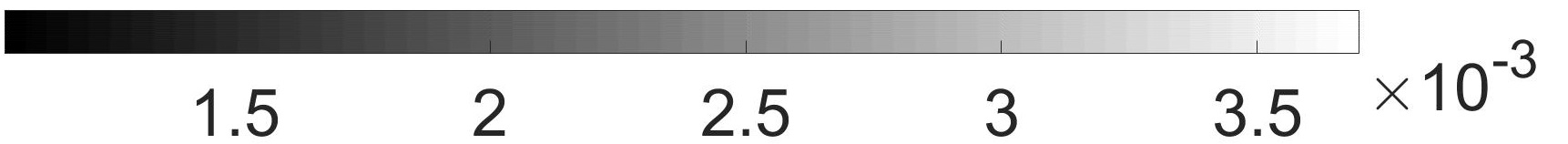}}{\textcolor{black}{Strain color bar}}\vspace*{-0.5em}
\end{center}%
\centering
\caption{The B-mode ultrasound and axial strain image using NCC, GLUE and PCA-GLUE for the real phantom experiment. The target and background windows are used for calculating SNR and CNR. \textcolor{black}{Dashed contour outlines the inclusion.}}
\label{fig_phantom}
\end{figure*}

\begin{table}[h]
\centering
\caption{The SNR and CNR values of the axial strain images for the phantom experiment. Target windows and background windows are of size $3$ $mm$ $\times$ $3$ $ mm$ and $5$ $mm$ $\times$ $5$ $ mm$ respectively as shown in Fig.~\ref{fig_phantom}. SNR is calculated for the background window.}
\begin{tabular}{lrr}
\toprule
\textbf{Method used} & \textbf{SNR} & \textbf{CNR} \\
\midrule
NCC & 18.18 & 16.86 \\
GLUE & 22.31 & 20.65 \\
PCA-GLUE & \textbf{23.52} & \textbf{21.46} \\
\bottomrule
\end{tabular}
\label{tab_phantom}
\end{table}


\begin{table}[]
\centering
\caption{The accuracy and F1-measure of our classifier on the phantom and \textit{in vivo} test data.}\label{tab11}
\begin{center}
\begin{tabular}{lrrrrr}
\toprule
\textbf{Dataset} & \hspace*{-2.5 pt} \textbf{Size} & \textbf{Accuracy} & \textbf{F1-measure}\\
\midrule
Phantom & 353 instances & \hskip 1.5em 85.11\% &\hskip 1.5em 93.20\%\\
Patient 1 & 147 instances &\hskip 1.5em 89.74\% & \hskip 1.5em 96.86\% \\
Patient 2 & 707 instances &\hskip 1.5em 70.43\% & \hskip 1.5em 93.2\% \\
Patient 3 & 223 instances & \hskip 1.5em 91.58\% & \hskip 1.5em 92.52\% \\
\bottomrule
\end{tabular}
\end{center}
\label{MLP_results}
\end{table}

\begin{figure*}[]
\begin{center}
\subfigure[B-mode]{\includegraphics[height=3.25 cm,width=4.8 cm]{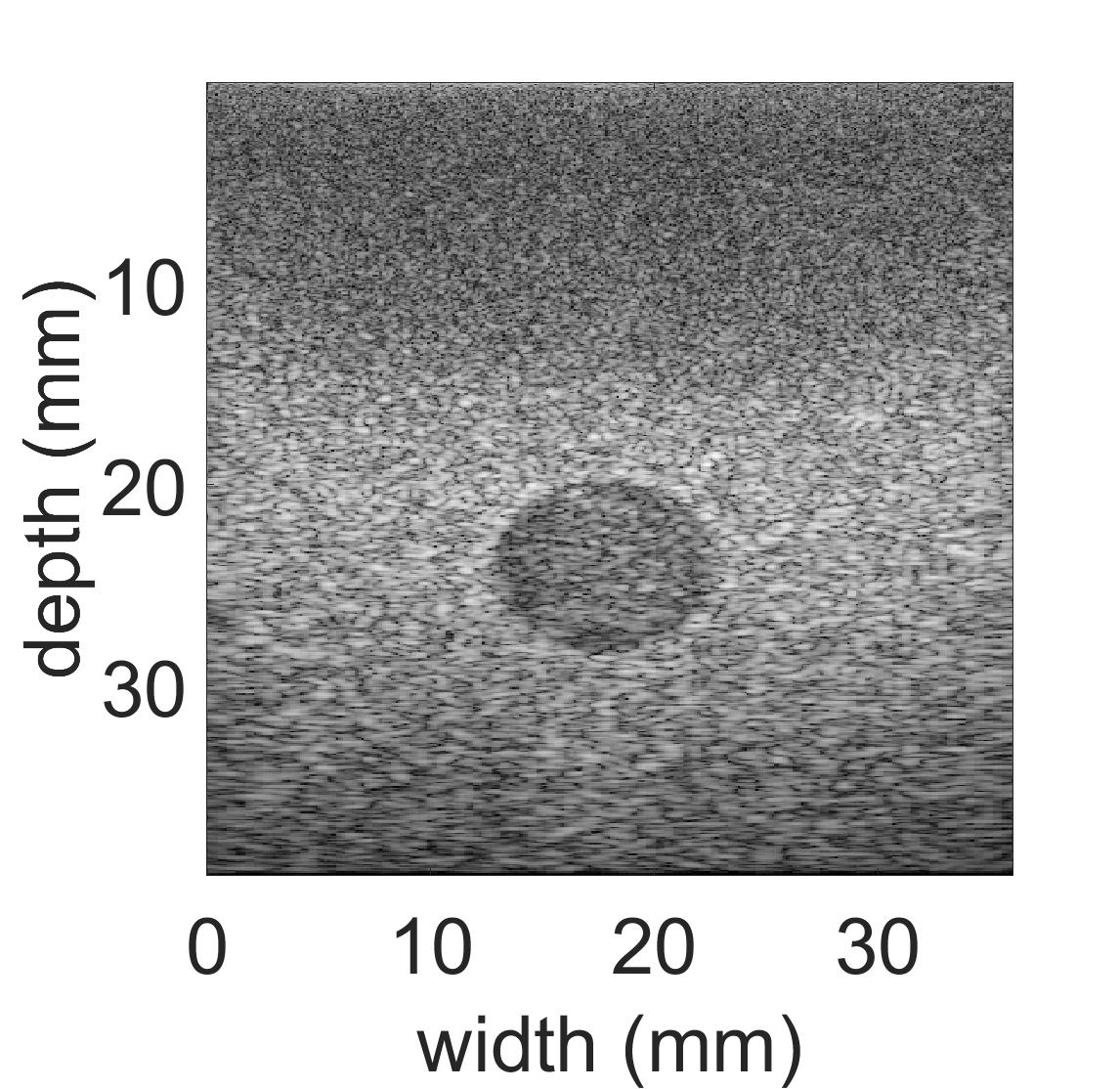}}\hspace*{-0.5em}
\subfigure[Strain from Skip 1 method]{\includegraphics[height=3.25 cm,width=4.8 cm]{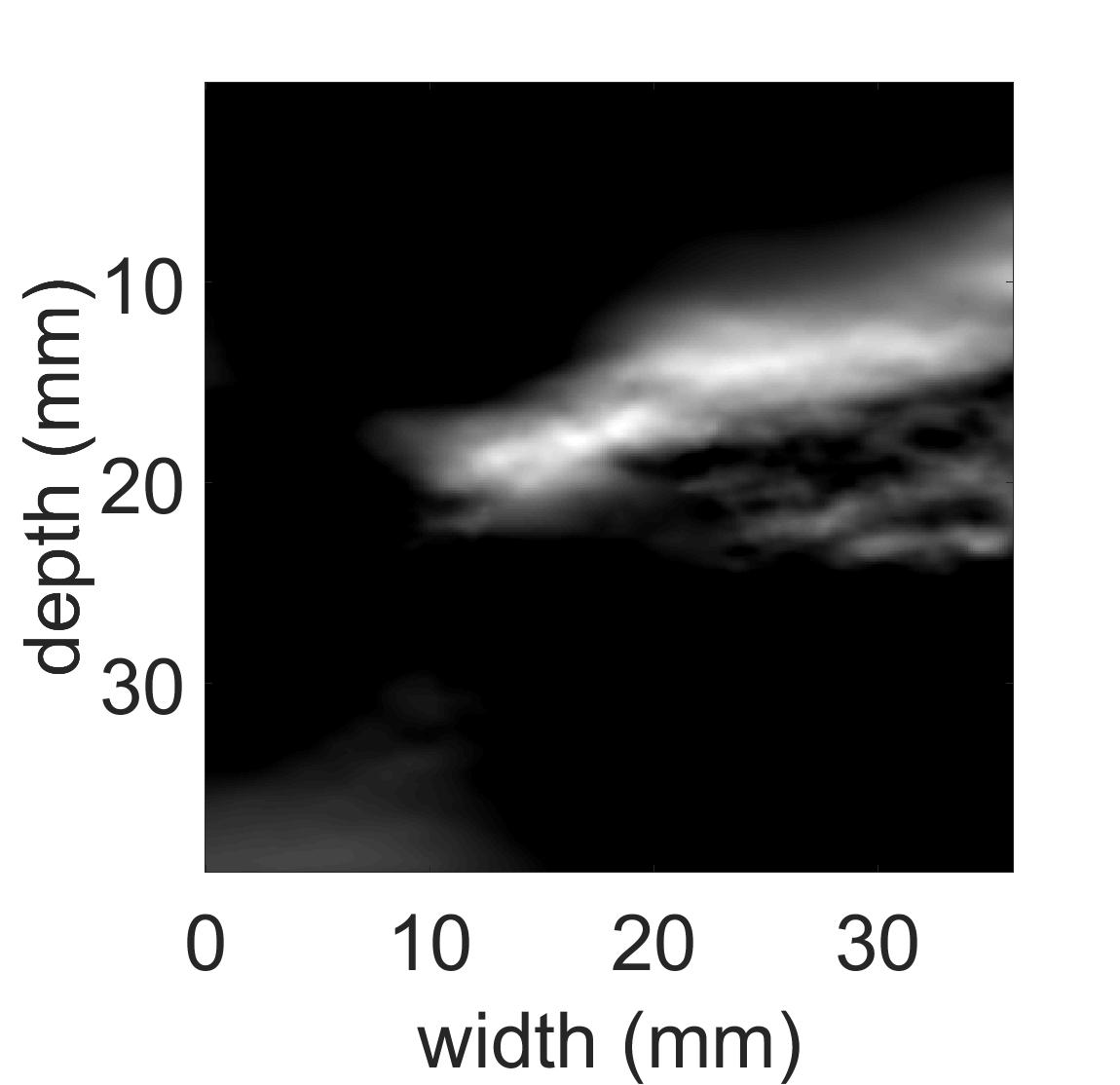}}\hspace*{-0.5em}
\subfigure[Strain from Skip 2 method]{\includegraphics[height=3.25 cm,width=4.8 cm]{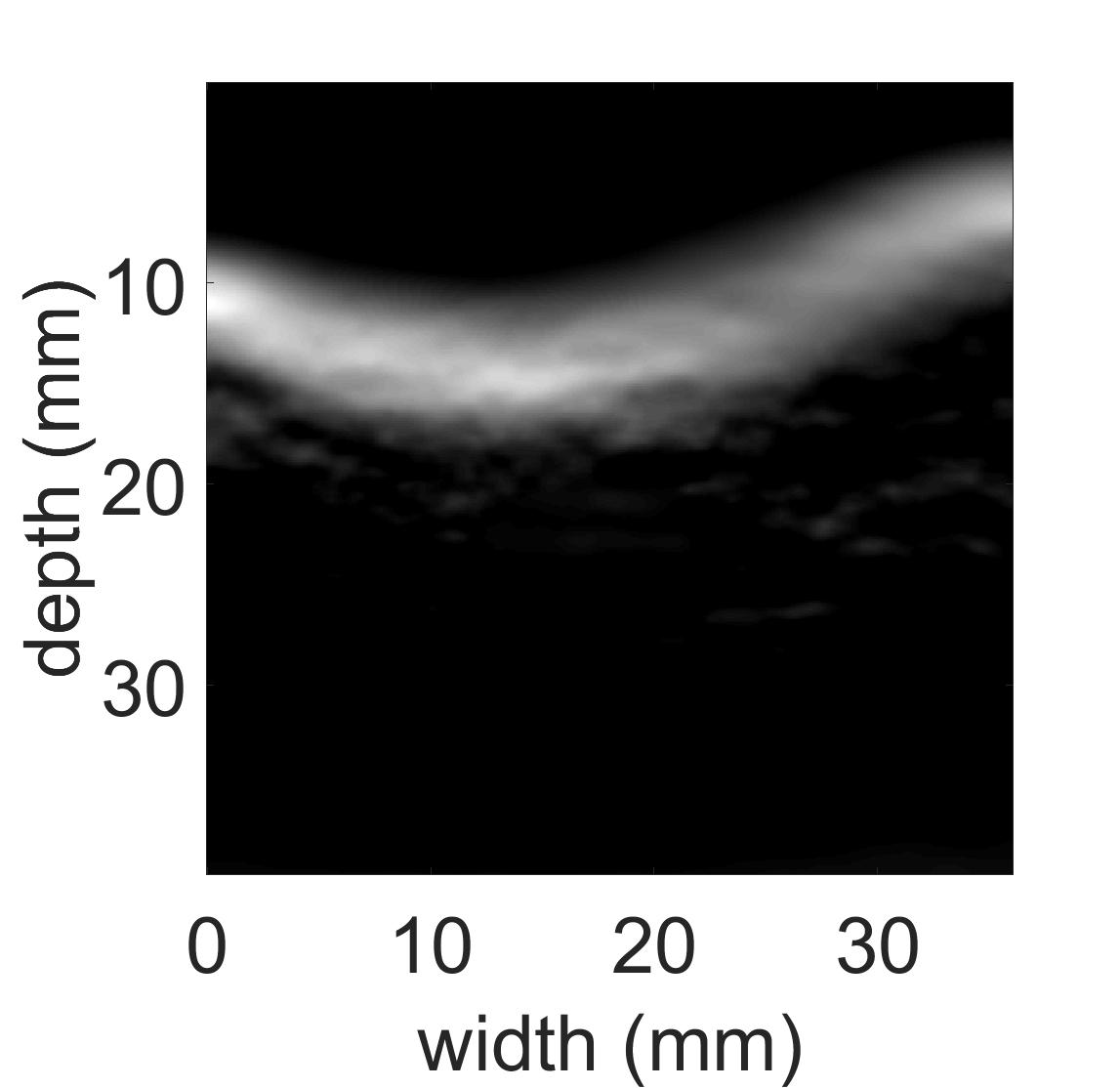}}\hspace*{-0.5em}
\subfigure[Strain from our method]{\includegraphics[height=3.25 cm,width=4.8 cm]{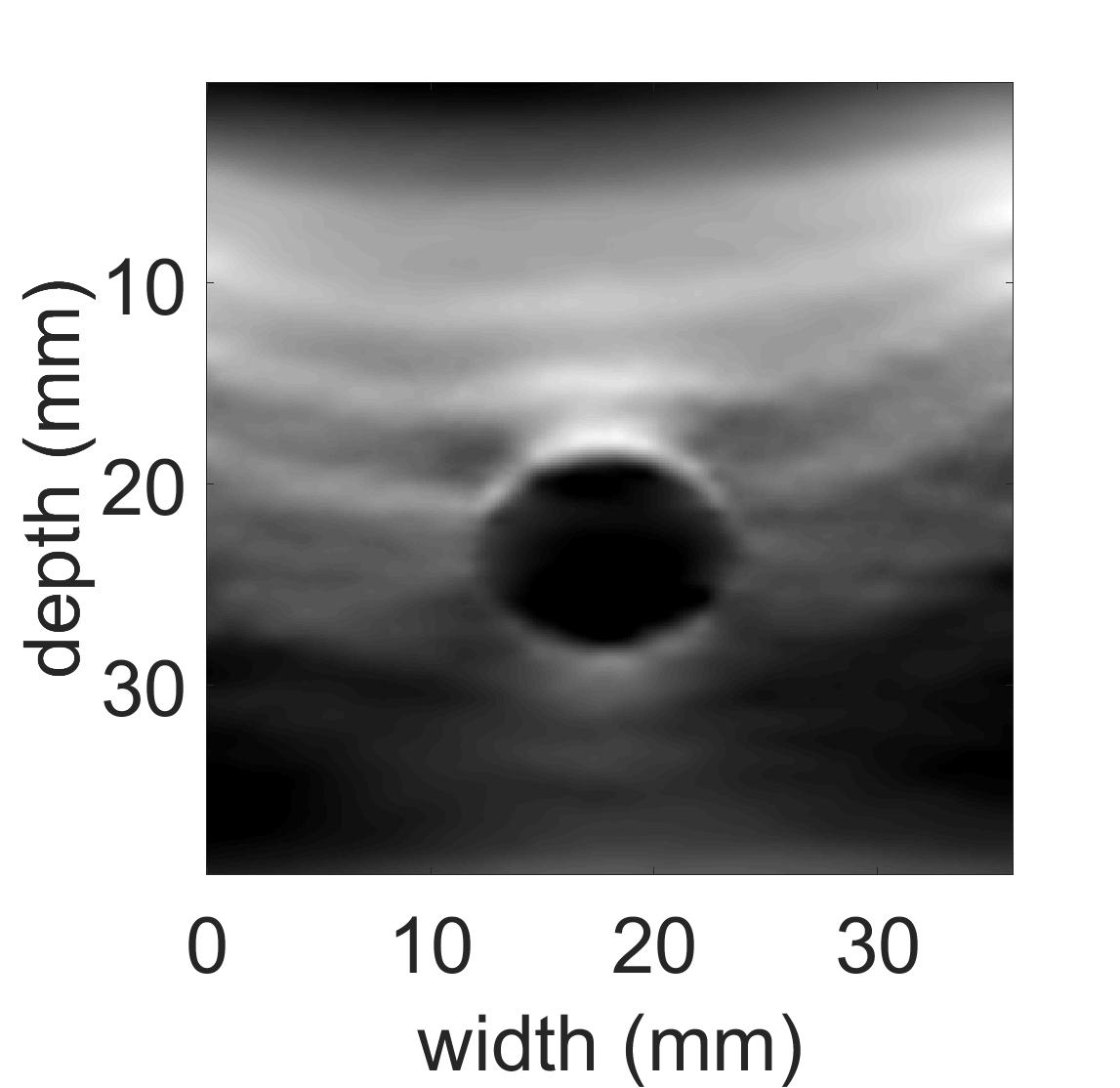}}\vspace*{-0em}
\stackunder[5pt]{\includegraphics[height=0.7 cm,width=4 cm]{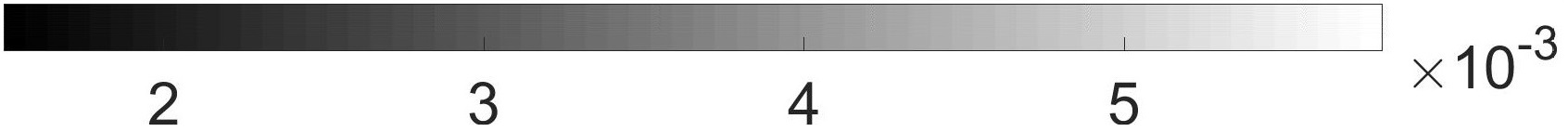}}{\textcolor{black}{Strain color bar}}\vspace*{-0.5em}
\end{center}%
\centering
\caption{The B-mode ultrasound and PCA-GLUE axial strain image for the phantom experiment using different frame selection methods. \textcolor{black}{Note that the pair of RF data used for estimating strain is different from that of Fig.~\ref{fig_phantom}}.}
\label{fig_pahtom_2}
\end{figure*}

\begin{figure*}[h!]
\begin{center}
\centering

\subfigure[B-mode patient 1]{\includegraphics[height=3.25 cm,width=4.8 cm]{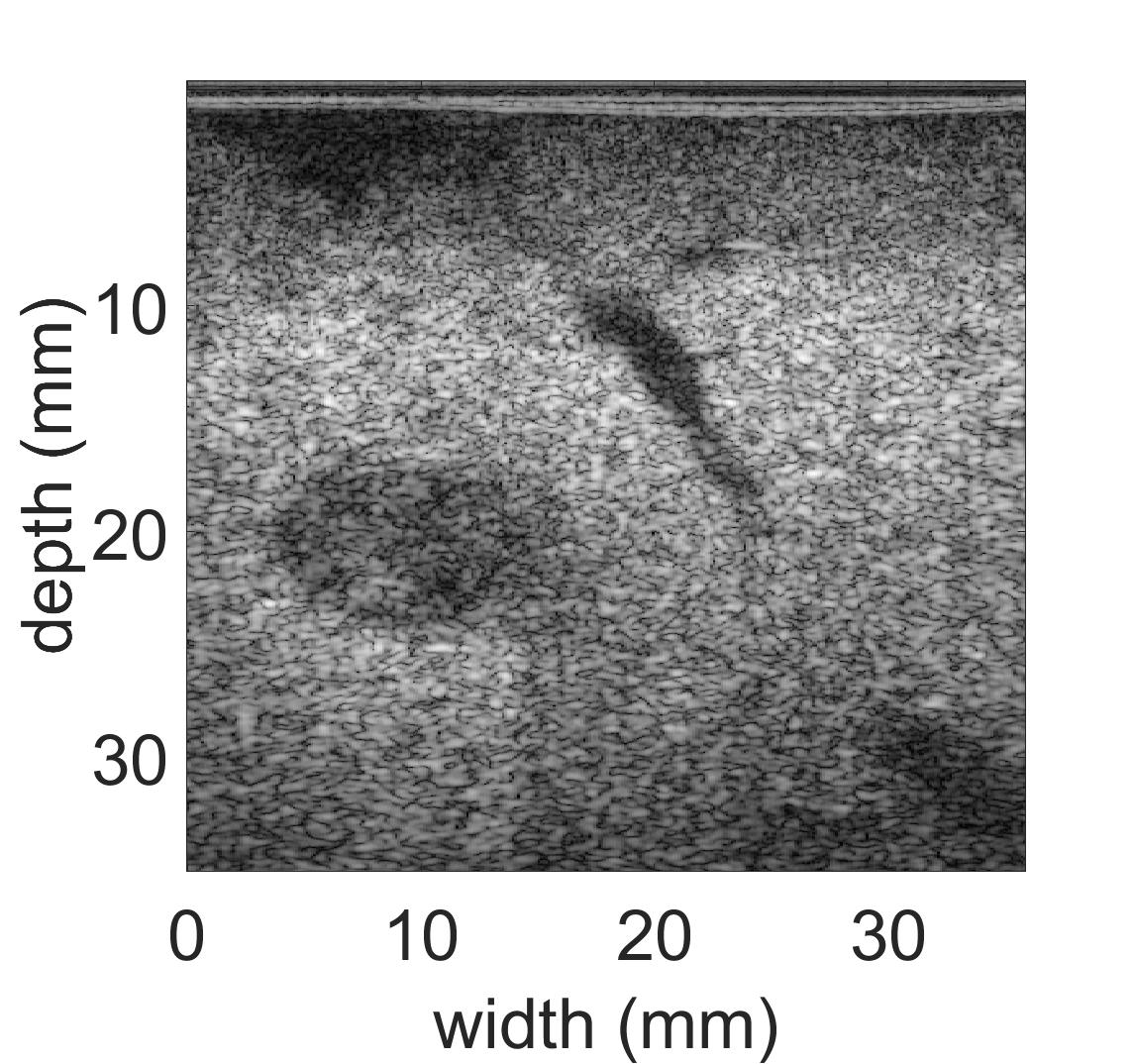}}\hspace*{-0.5em}
\subfigure[NCC]{\includegraphics[height=3.25 cm,width=4.8 cm]{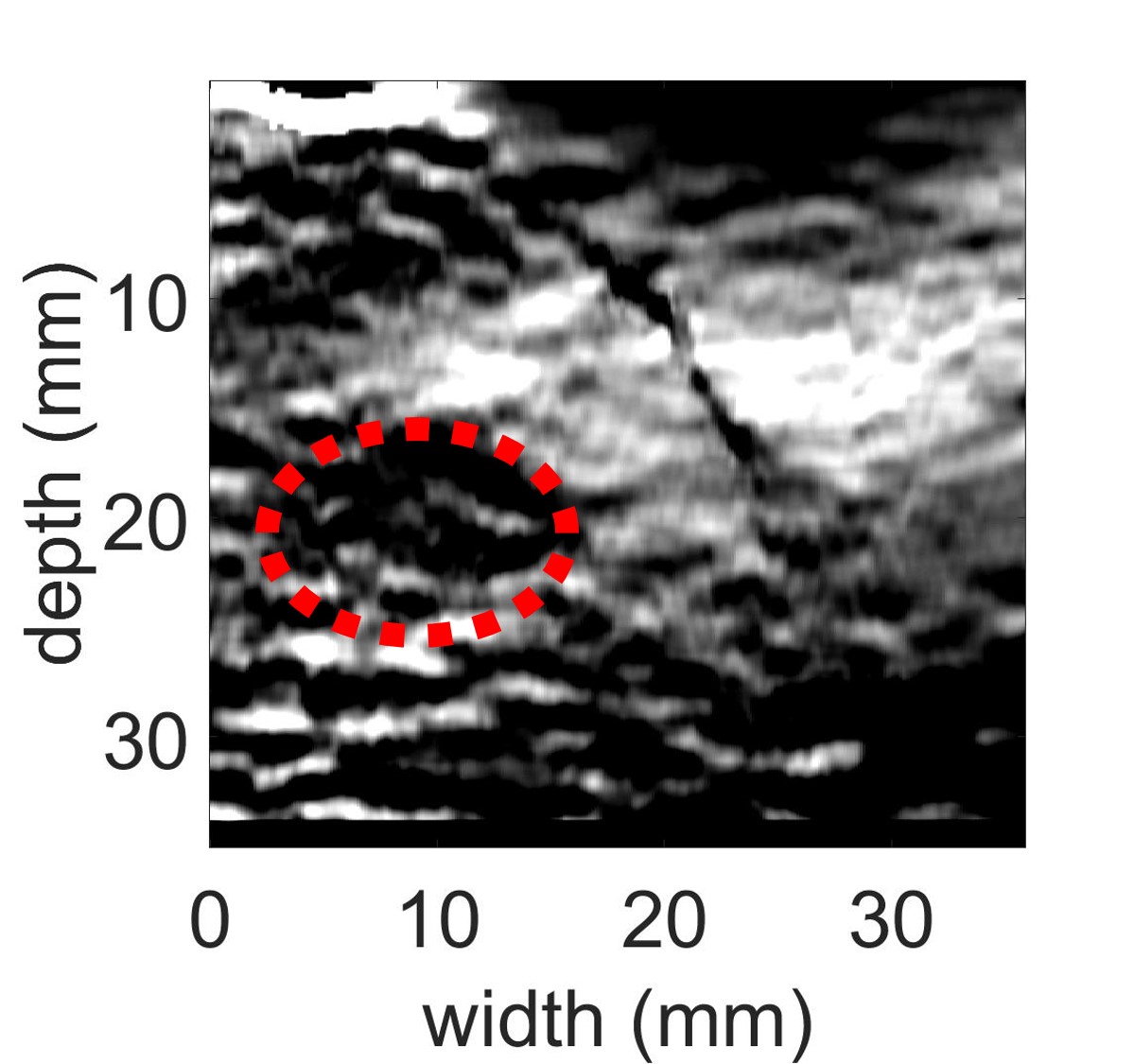}}\hspace*{-0.5em}
\subfigure[GLUE]{\includegraphics[height=3.25 cm,width=4.8 cm]{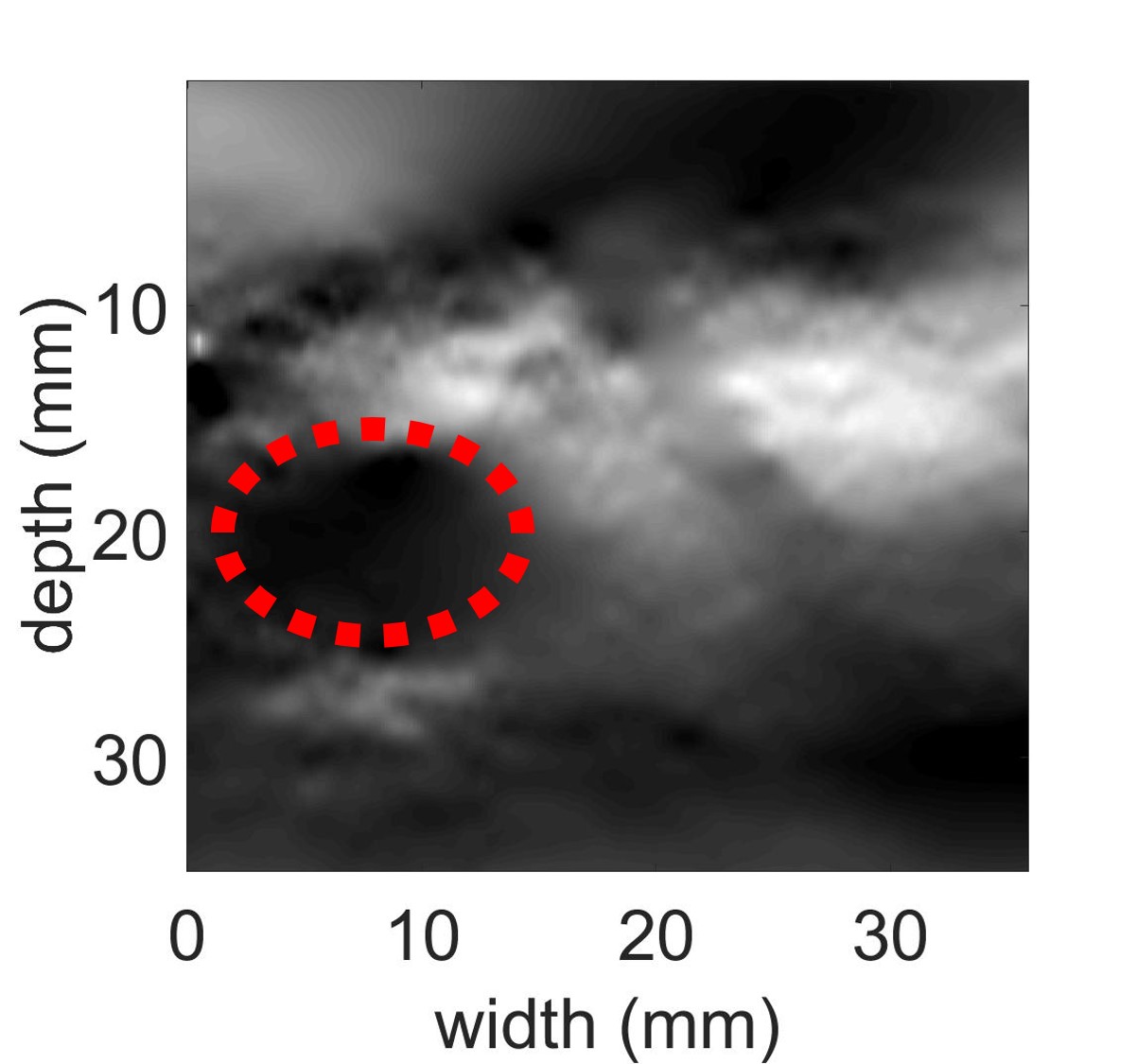}}\hspace*{-0.5em}
\subfigure[PCA-GLUE]{\includegraphics[height=3.25 cm,width=4.8 cm]{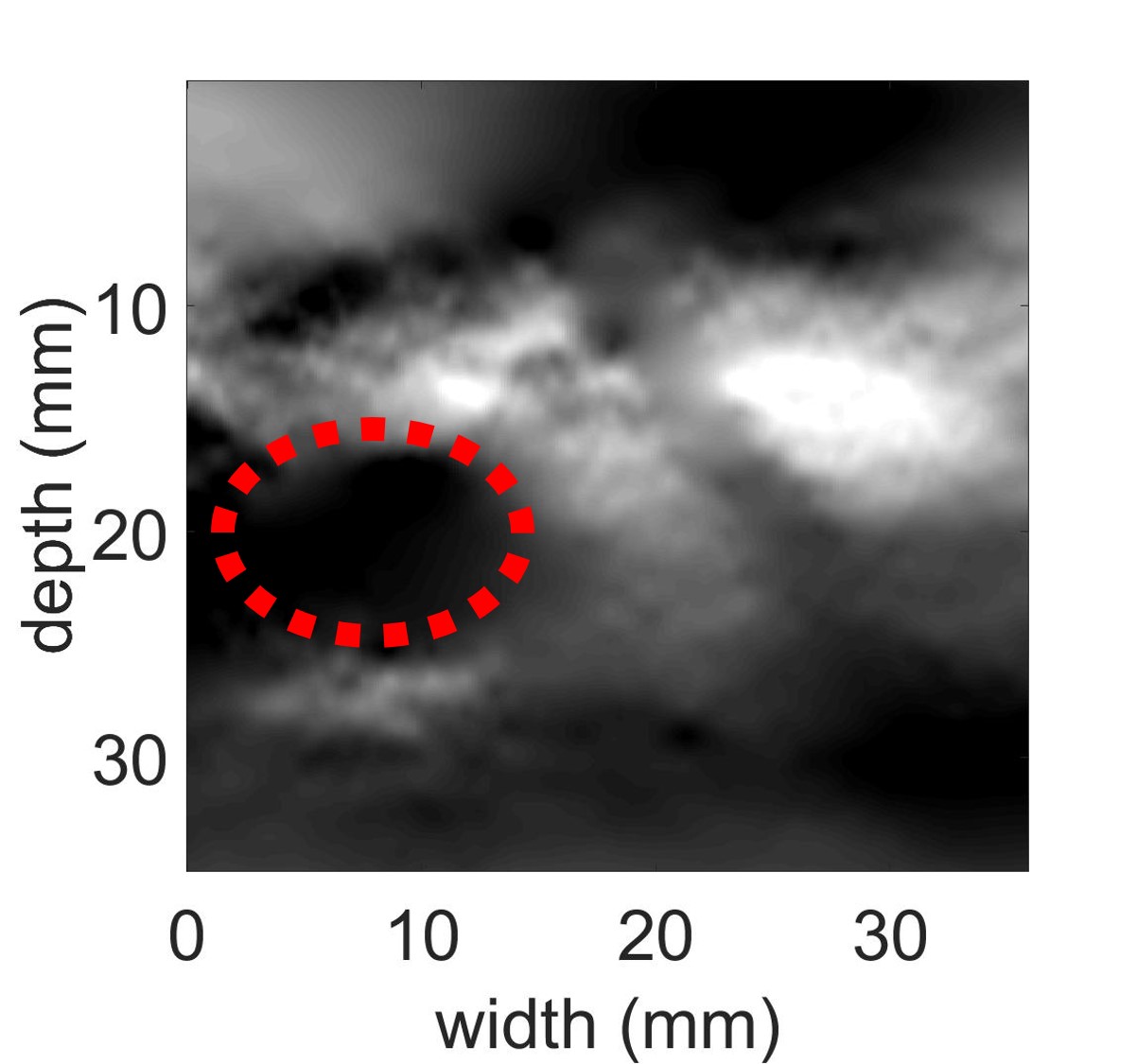}}\vspace*{-0em}
\stackunder[5pt]{\includegraphics[height=0.8 cm,width=4 cm]{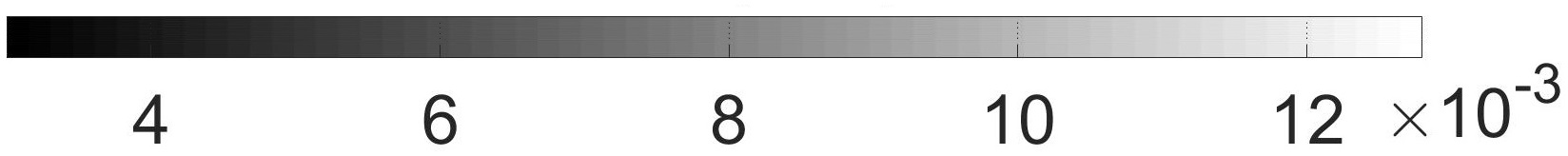}}{\textcolor{black}{Strain color bar}}\vspace*{-0.5em}

\end{center}%
\caption{The B-mode ultrasound and axial strain image using NCC, GLUE and PCA-GLUE for the \textit{in vivo} liver data before ablation. \textcolor{black}{Dashed contour outlines the tumor.}} \label{fig_invivo_pcaglue_before}
\end{figure*}

\begin{figure*}[t]
\begin{center}
\centering

\subfigure[B-mode patient 2 after ablation]{\includegraphics[height=3.25 cm,width=4.8 cm]{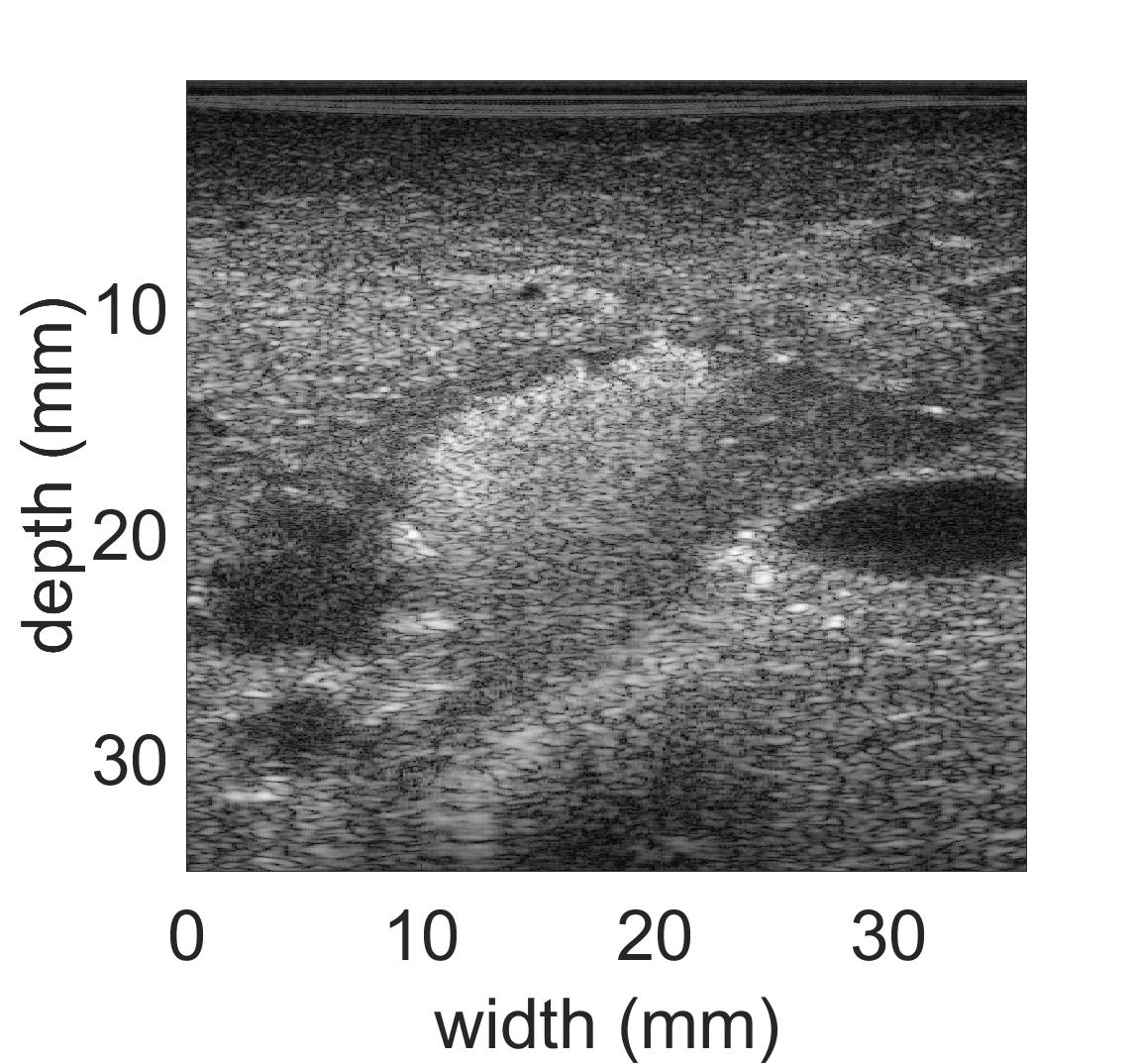}}\hspace*{-0.5em}
\subfigure[NCC]{\includegraphics[height=3.25 cm,width=4.8 cm]{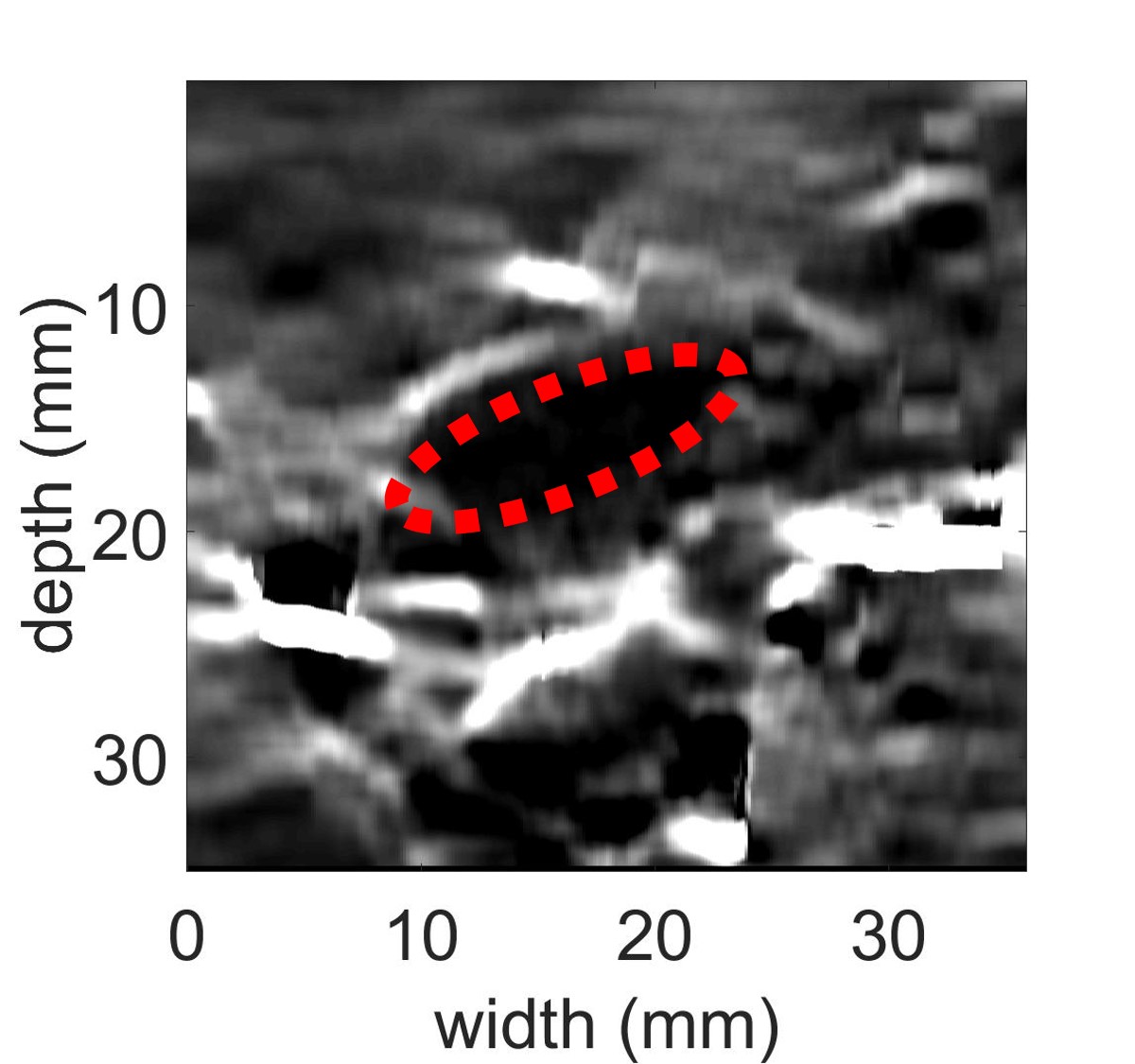}}\hspace*{-0.5em}
\subfigure[GLUE]{\includegraphics[height=3.25 cm,width=4.8 cm]{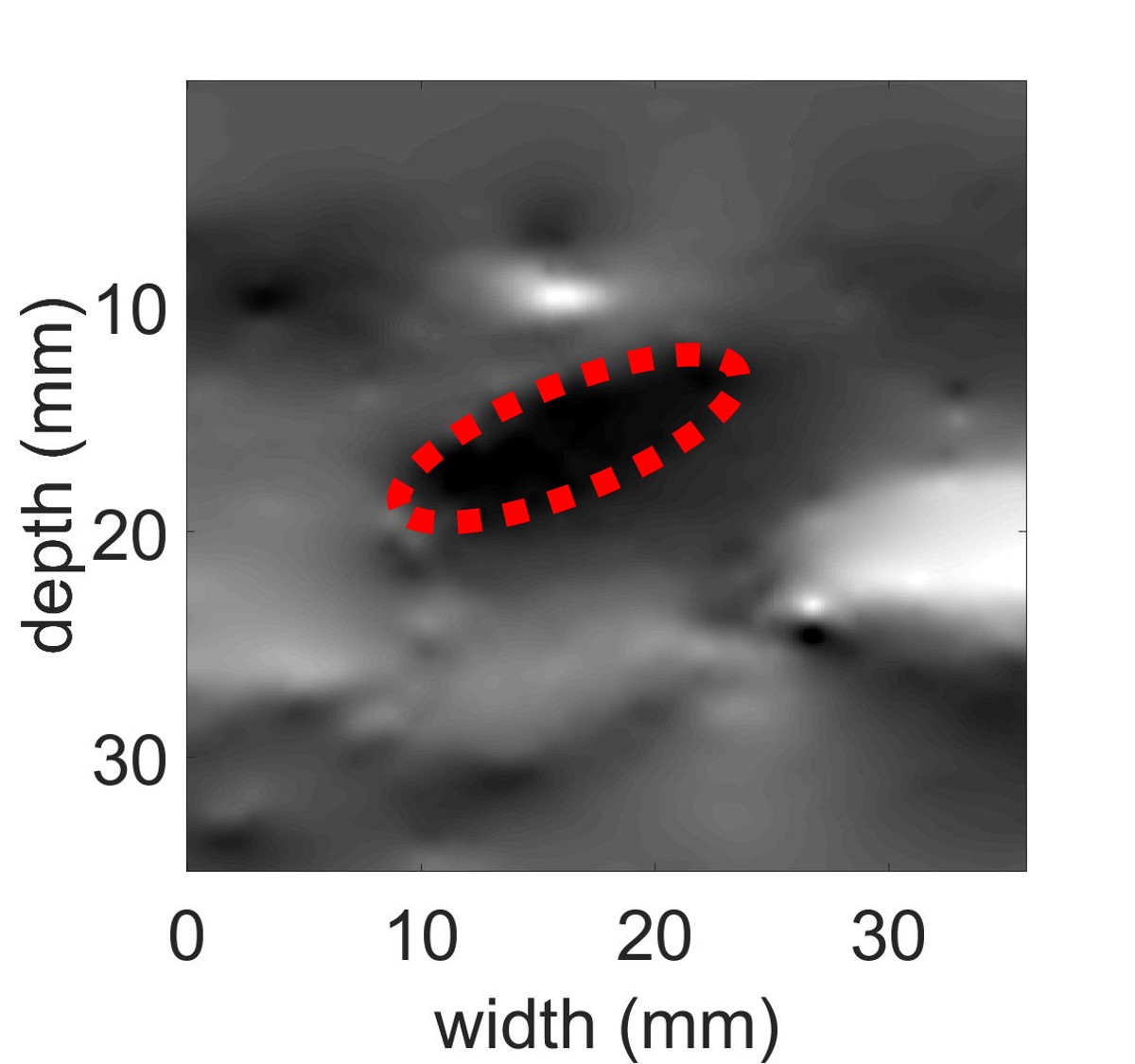}}\hspace*{-0.5em}
\subfigure[PCA-GLUE]{\includegraphics[height=3.25 cm,width=4.8 cm]{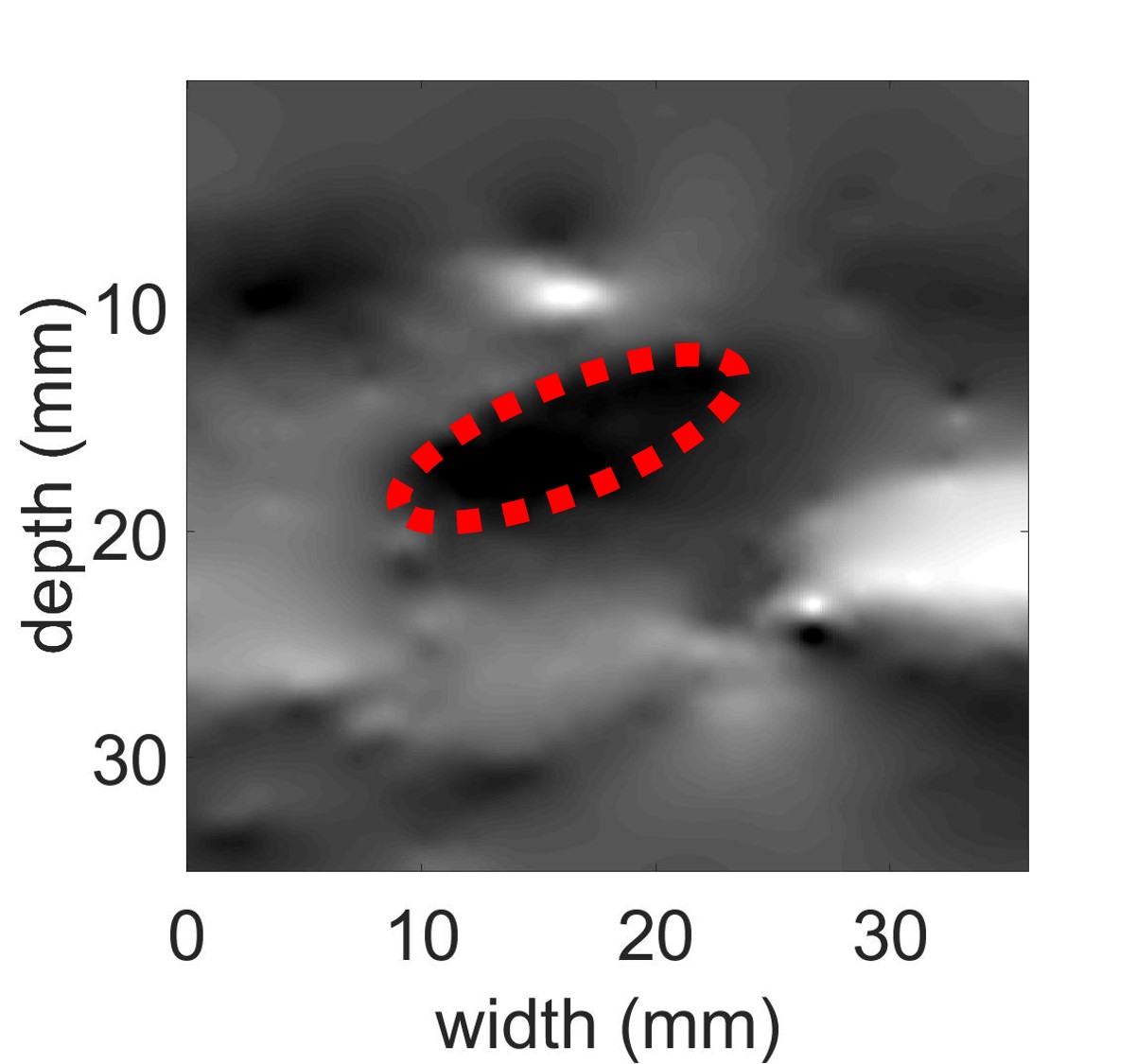}}\vspace*{-0em}
\stackunder[5pt]{\includegraphics[height=0.65 cm,width=4 cm]{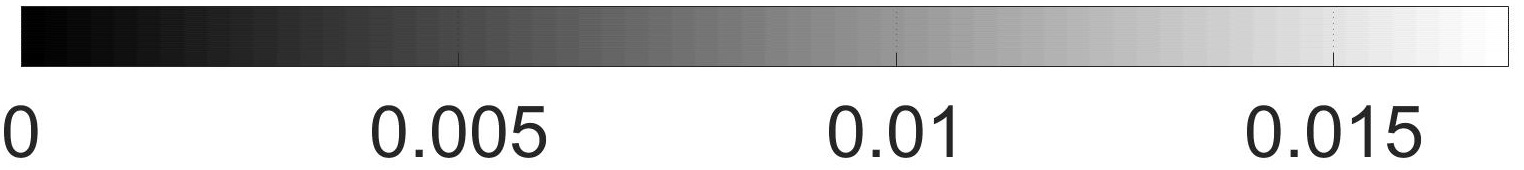}}{\textcolor{black}{Strain color bar}}\vspace*{-0.5em}

\end{center}%
\caption{The B-mode ultrasound and axial strain image using NCC, GLUE and PCA-GLUE for the \textit{in vivo} liver data after ablation. \textcolor{black}{Dashed contour outlines the tumor.}} \label{fig_invivo_pcaglue_after}
\end{figure*}

\begin{table}[]
\centering
\caption{The SNR values of the axial strain images for the \textit{in vivo} data.}
\begin{tabular}{lrrrr}
\toprule
\textbf{Dataset} & \textbf{NCC} & \textbf{GLUE} & \textbf{PCA-GLUE} \\
\midrule
Patient 1 & 13.23 & 21.11 & \textbf{21.19} \\
Patient 2 & 2.09 & \textbf{21.33} & 21.20 \\
Patient 3 & 13.21 & \textbf{25.66} & 23.94 \\
\bottomrule
\end{tabular}
\label{tab_invivo_snr}
\end{table}

\begin{table}[]
\centering
\caption{The CNR values of the axial strain images for the \textit{in vivo} liver data.}
\begin{tabular}{lrrrrr}
\toprule
\textbf{Dataset} & \textbf{NCC} & \textbf{GLUE} & \textbf{PCA-GLUE}\\
\midrule
Patient 1 & 11.01 & 20.34 & \textbf{20.66}\\
Patient 2 & -0.46 & 13.52 & \textbf{17.05}\\
Patient 3 & 9.87 & \textbf{16.66} & 15.95 \\
\bottomrule
\end{tabular}
\label{tab_invivo_cnr}
\end{table}

\begin{figure*}[t]
\begin{center}
\subfigure[B-mode]{\includegraphics[height=3.25 cm,width=4.8 cm]{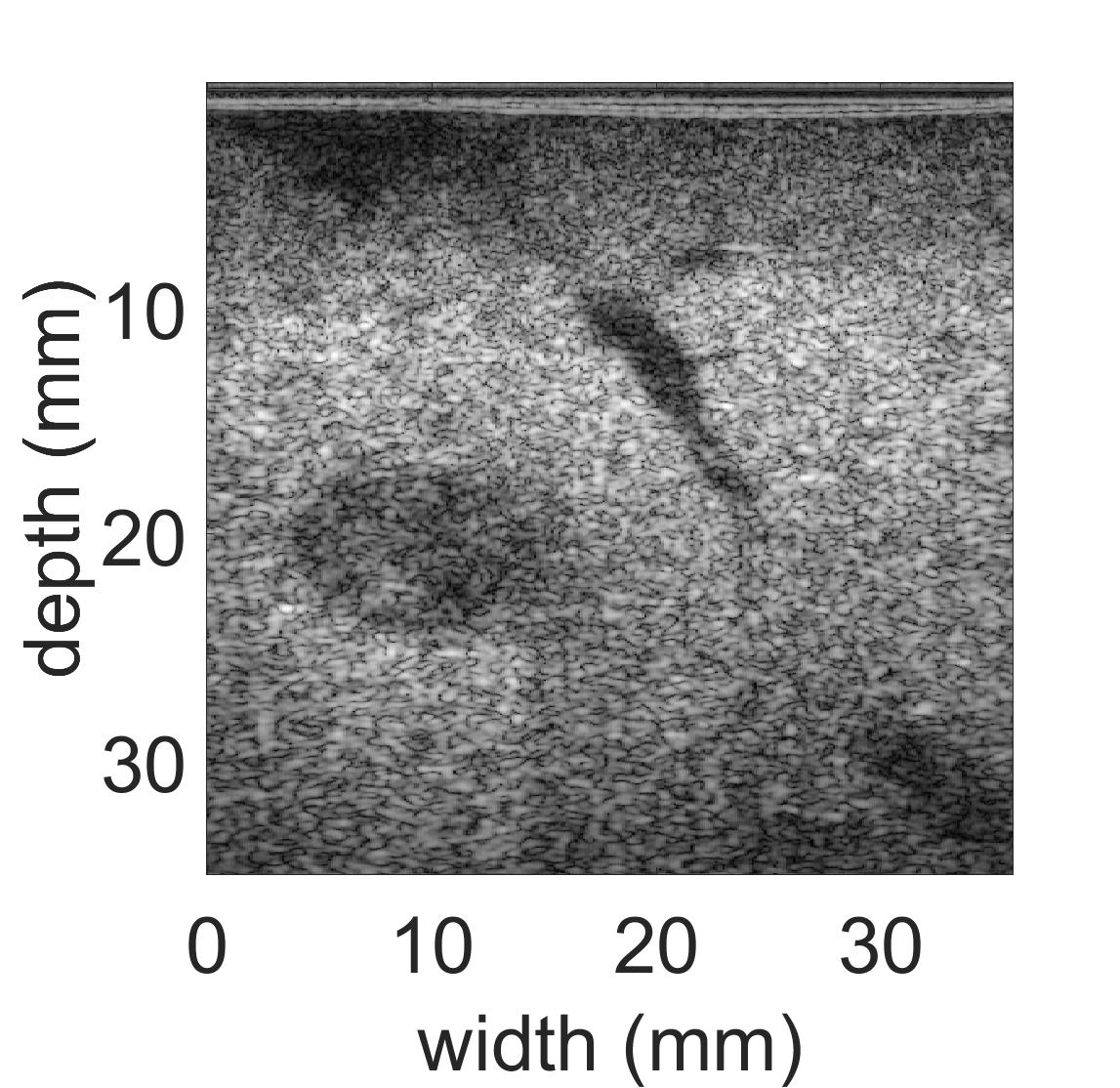}}\hspace*{-0.5em}
\subfigure[Strain from Skip 1 method]{\includegraphics[height=3.25 cm,width=4.8 cm]{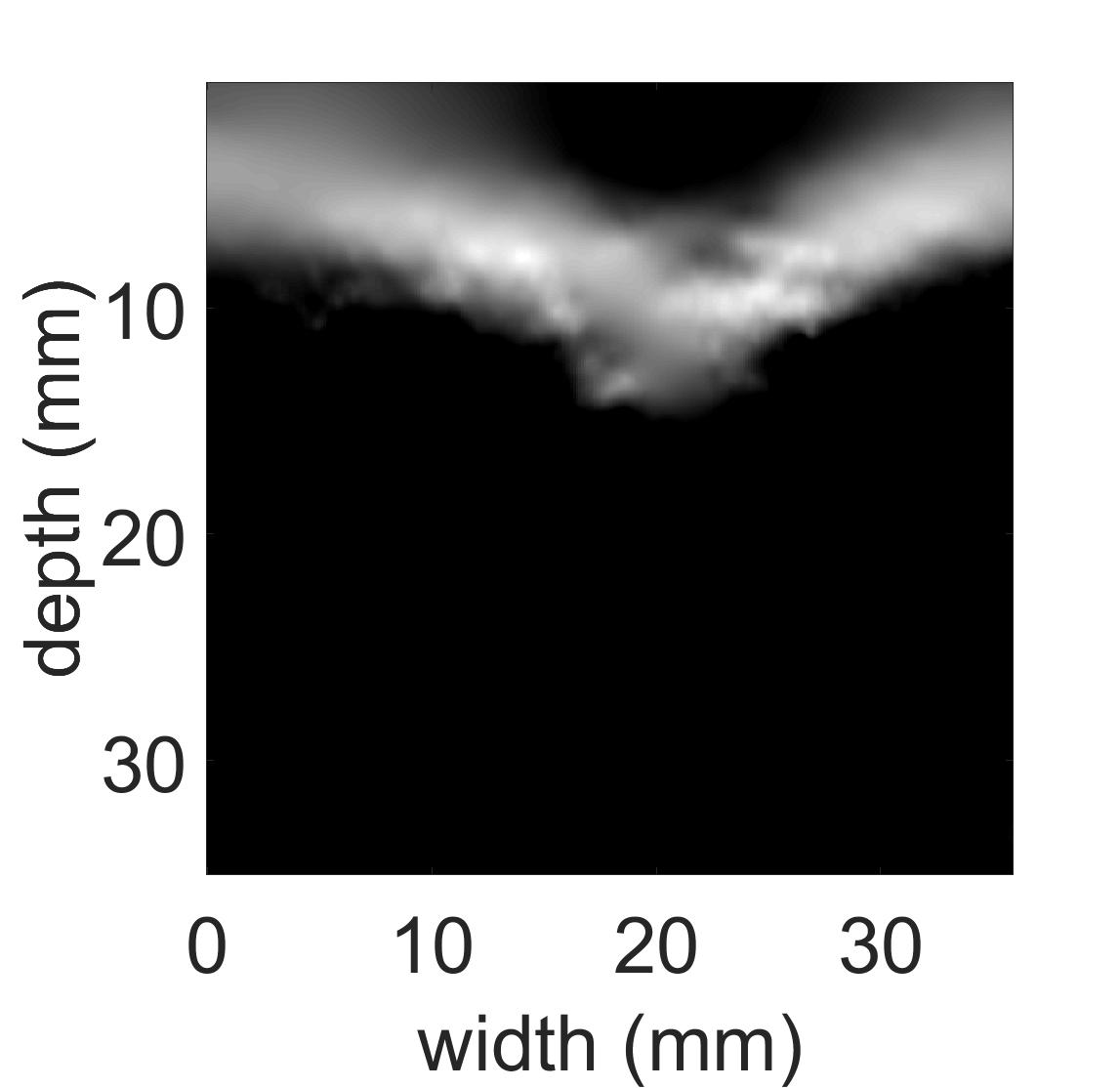}}\hspace*{-0.5em}
\subfigure[Strain from Skip 2 method]{\includegraphics[height=3.25 cm,width=4.8 cm]{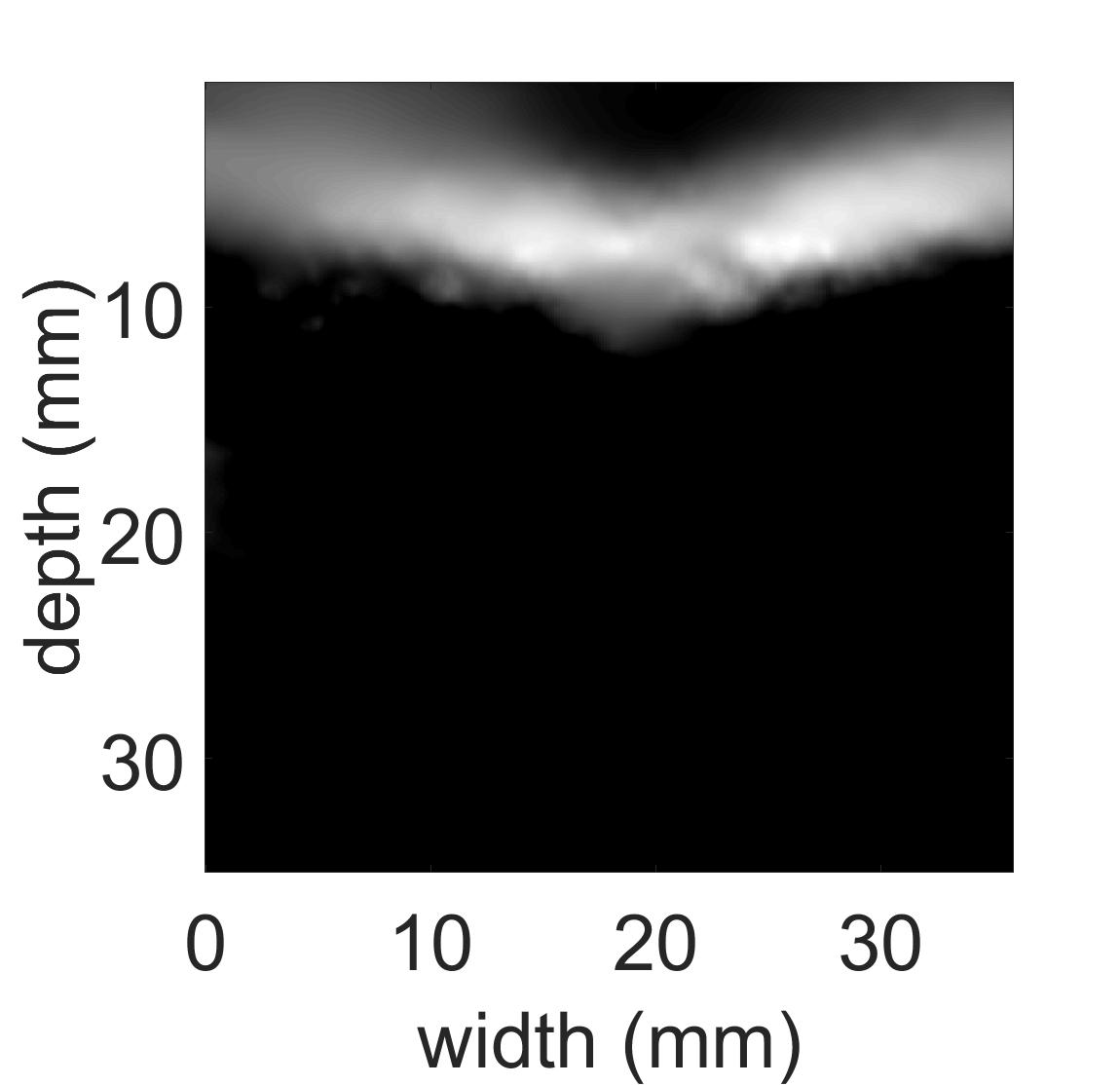}}\hspace*{-0.5em}
\subfigure[Strain from our method]{\includegraphics[height=3.25 cm,width=4.8 cm]{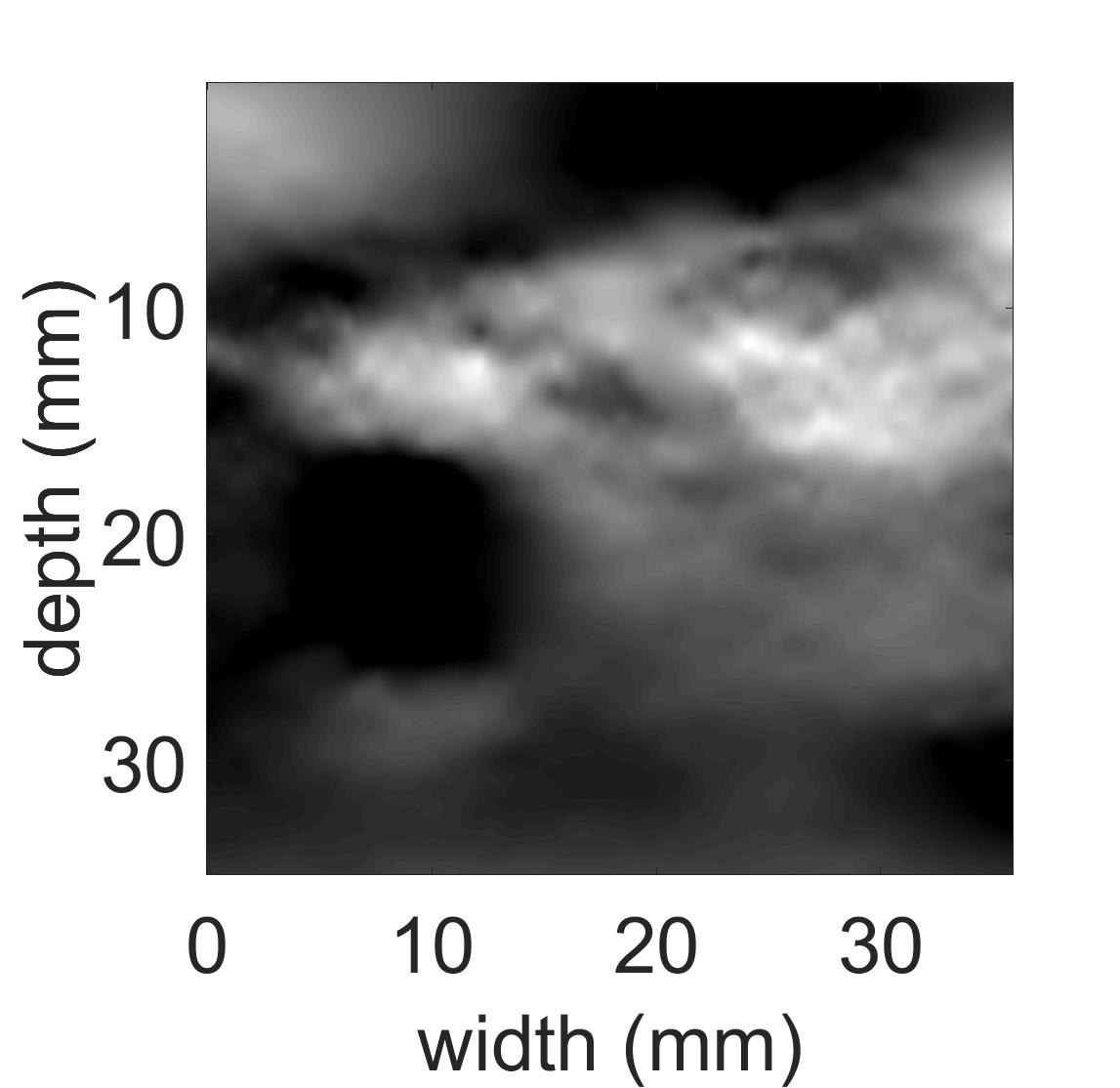}}\vspace*{-0em}
\stackunder[5pt]{\includegraphics[height=0.7 cm,width=4 cm]{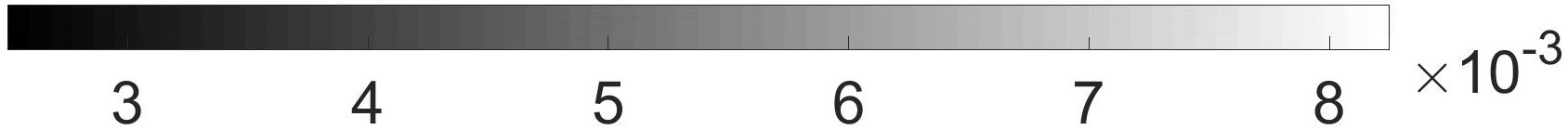}}{\textcolor{black}{Strain color bar}}\vspace*{-0.5em}
\end{center}%
\centering
\caption{The B-mode ultrasound and PCA-GLUE axial strain image for the \textit{in vivo} liver data using different frame selection methods. \textcolor{black}{Note that the pair of RF data used for estimating strain is different from that of Fig.~\ref{fig_invivo_pcaglue_before}}.}
\label{fig_invivo_2}
\end{figure*} 



\begin{figure*}[]
\begin{center}
\subfigure[The correct displacement for a certain RF line]{\includegraphics[height=3.25 cm,width=4.6 cm]{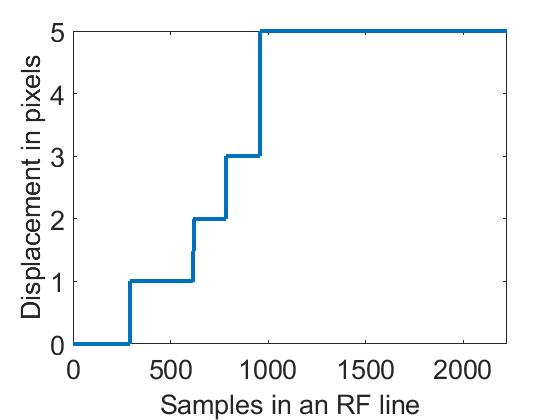}}\hspace*{-0em}
\subfigure[The incorrect displacement for a certain RF line]{\includegraphics[height=3.25 cm,width=4.6 cm]{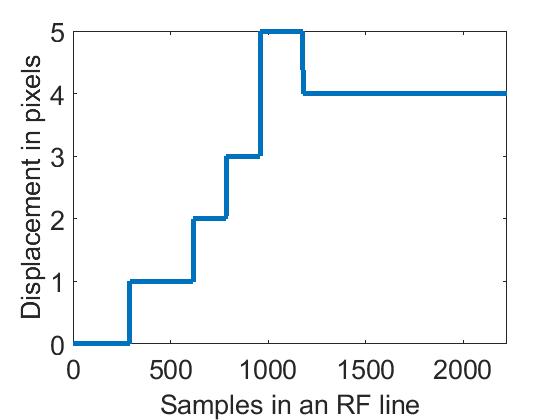}}\hspace*{-0em}
\subfigure[Strain estimated by GLUE ]{\includegraphics[height=3.25 cm,width=5 cm]{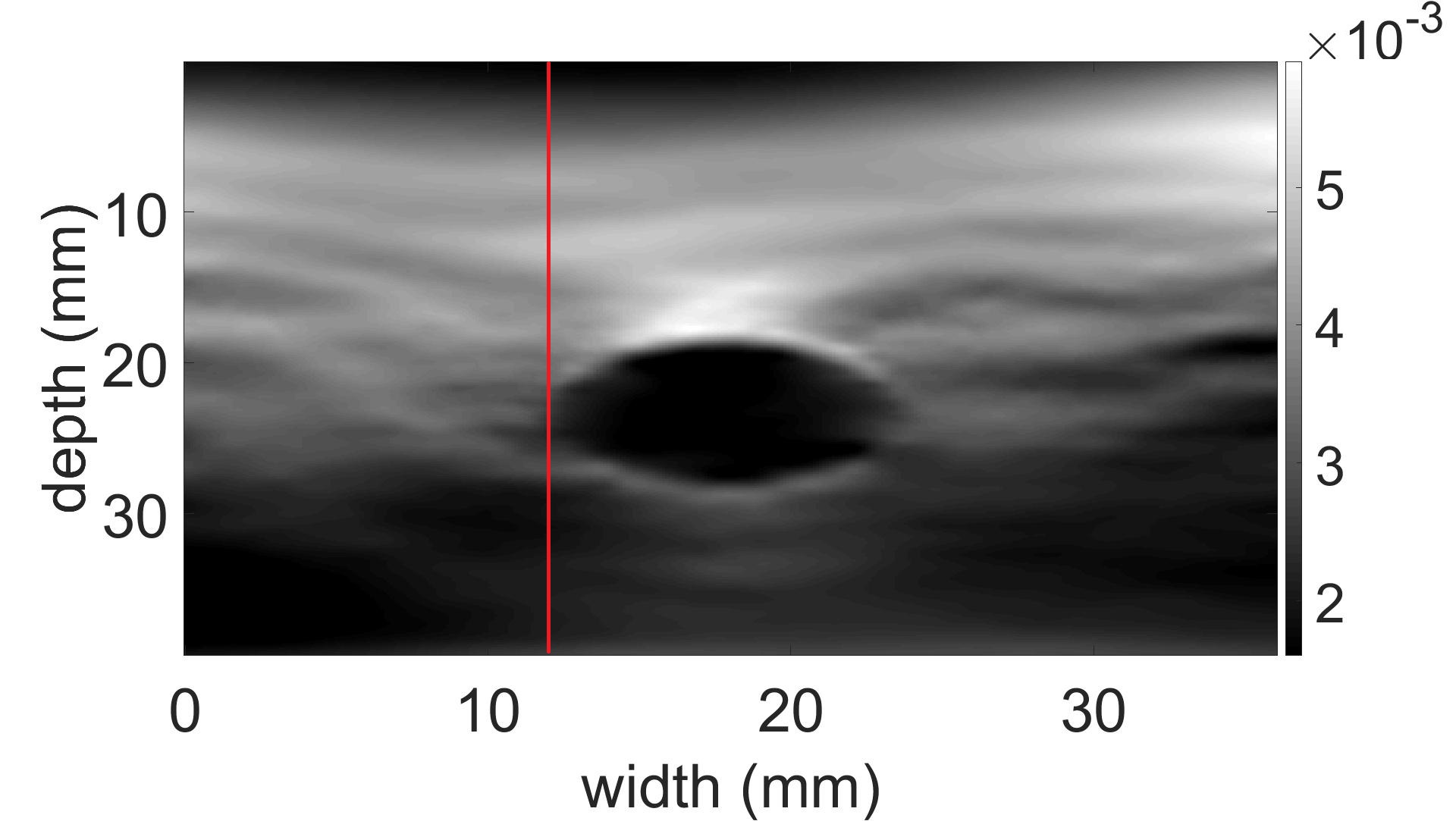}}\hspace*{-0em}
\subfigure[Strain estimated by PCA-GLUE ]{\includegraphics[height=3.25 cm,width=5 cm]{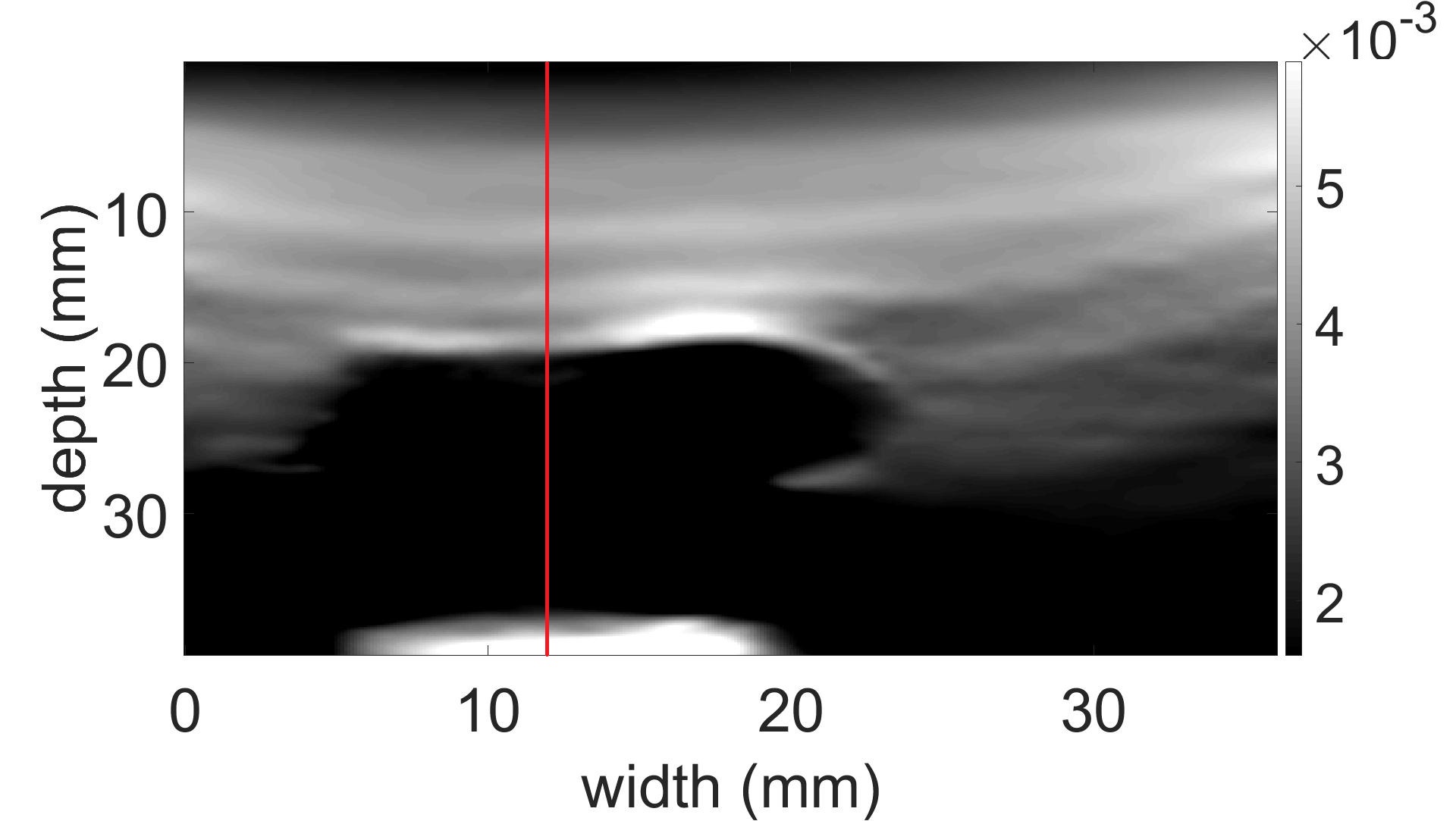}}\hspace*{0em}\vspace*{-0.5em}
\end{center}%
\caption{Strain estimated by both GLUE and PCA-GLUE given that DP failed in computing correct initial estimates. \textcolor{black}{The failure occured in the RF line shown in red.}}%
\label{DP_fail}%
\end{figure*}

\subsection{Phantom Results}
\subsubsection{Strain Estimation}
Fig.~\ref{fig_phantom} shows a comparison between the strain estimated using NCC, GLUE and PCA-GLUE for the phantom experiment, where the dashed circles point to the inclusion. The results of GLUE and PCA-GLUE look almost the same, but the advantage of using PCA-GLUE is that it estimates the initial estimates more than 10 times faster. Table~\ref{tab_phantom} shows the SNR and CNR values obtained using different methods.
\subsubsection{Frame Selection}
Our frame selection algorithm is compared to the simple method that chooses the pair of RF frames such that they are one or two frames apart. Fig.~\ref{fig_pahtom_2} shows the difference between applying our method and the fixed skip frame pairing while using PCA-GLUE for strain estimation. Our method considers a window of 16 frames, 8 of them are before the desired frame and 8 are after it. To choose a good frame to be paired with the desired frame, we run the MLP model on the 16 pairs and choose the pair that has the highest NCC (we don't apply the thresholding here). \textcolor{black}{We can observe that our method selects RF frames that are suitable for strain estimation and it substantially outperforms the fixed skip frame pairing methods such as Skip 1 and Skip 2.}

To make the validation more concrete, we test our classifier on 353 instances to classify them as suitable or not suitable for strain estimation. The ground truth is obtained as previously discussed in Algorithm~\ref{annotating}. Table~\ref{MLP_results} shows the accuracy and F1-measure for our classifier on new data that the model has not seen before. The results show that our classifier is able to generalize well to unseen data, and that it could be used in practice.

\subsection{\textit{In vivo} Results}
\subsubsection{Strain Estimation}
Fig.~\ref{fig_invivo_pcaglue_before} and \ref{fig_invivo_pcaglue_after} show the results obtained when running NCC, GLUE and PCA-GLUE on the liver dataset, where both GLUE and PCA-GLUE yield very similar results. The dashed ellipses point to the tumors. Tables~\ref{tab_invivo_snr} and~\ref{tab_invivo_cnr} show the SNR and CNR calculated.
\subsubsection{Frame Selection}
Fig.~\ref{fig_invivo_2} shows a comparison between the strain estimated using both our frame selection method and the fixed skip frame pairing on two RF frames collected from the \textit{in vivo} liver data. Table~\ref{MLP_results} shows the accuracy and F1-measure obtained for the liver dataset.

\subsection{PCA-GLUE robustness}
Our method is not only capable of estimating strain or selecting suitable RF frames, it is also robust to incorrect initial displacement estimates when DP fails. The main difference between PCA-GLUE and GLUE is in estimating the initial displacement image, where GLUE uses DP to estimate the displacement of every single RF line, whereas PCA-GLUE applies DP for only 5 RF lines, then uses a linear combination of previously computed principal components as an initial displacement image. Therefore, if DP fails in estimating the correct displacement for a certain RF line, that means that GLUE would have an incorrect initial displacement image, which affects the fine-tuned displacement image. 

The reason behind this robustness is that PCA-GLUE relies on the principal components previously computed offline, such that the resulting initial displacement image is represented as a linear combination of them. Therefore, if incorrect results were among the 5 RF lines chosen by PCA-GLUE, it would still be able to estimate the strain correctly due to the additional step of estimating TDE as a sum of principal components.


Fig.~\ref{DP_fail} shows how both GLUE and PCA-GLUE perform when they get incorrect initial estimates from DP. 

Fig.~\ref{simulation_noisy} shows a comparison between the strain estimated by both GLUE and PCA-GLUE on the finite element method (FEM) simulation
data before and after adding a gaussian noise with $\mu=0$ and $\sigma^{2}=0.1225$ to 10\% of the RF lines. The large error on these RF lines could be caused in real life due to air bubbles between the probe and tissue, or large out-of-plane motion in some regions.

\begin{figure*}[t]
\begin{center}
\centering

\subfigure[B-mode before noise addition]{\includegraphics[height=3.25 cm,width=4.8 cm]{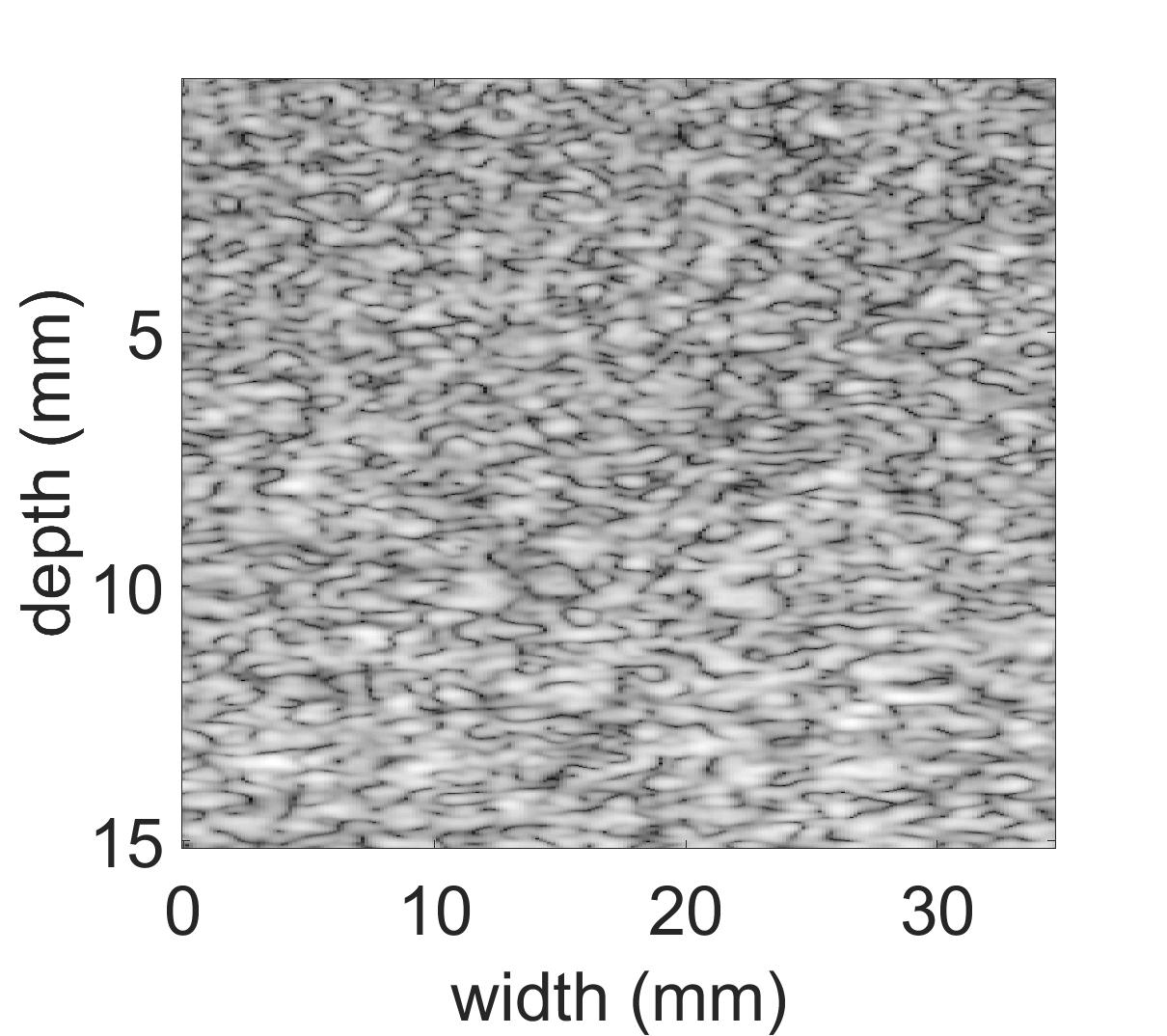}}\hspace*{-0.5em}
\subfigure[ground truth]{\includegraphics[height=3.25 cm,width=4.8 cm]{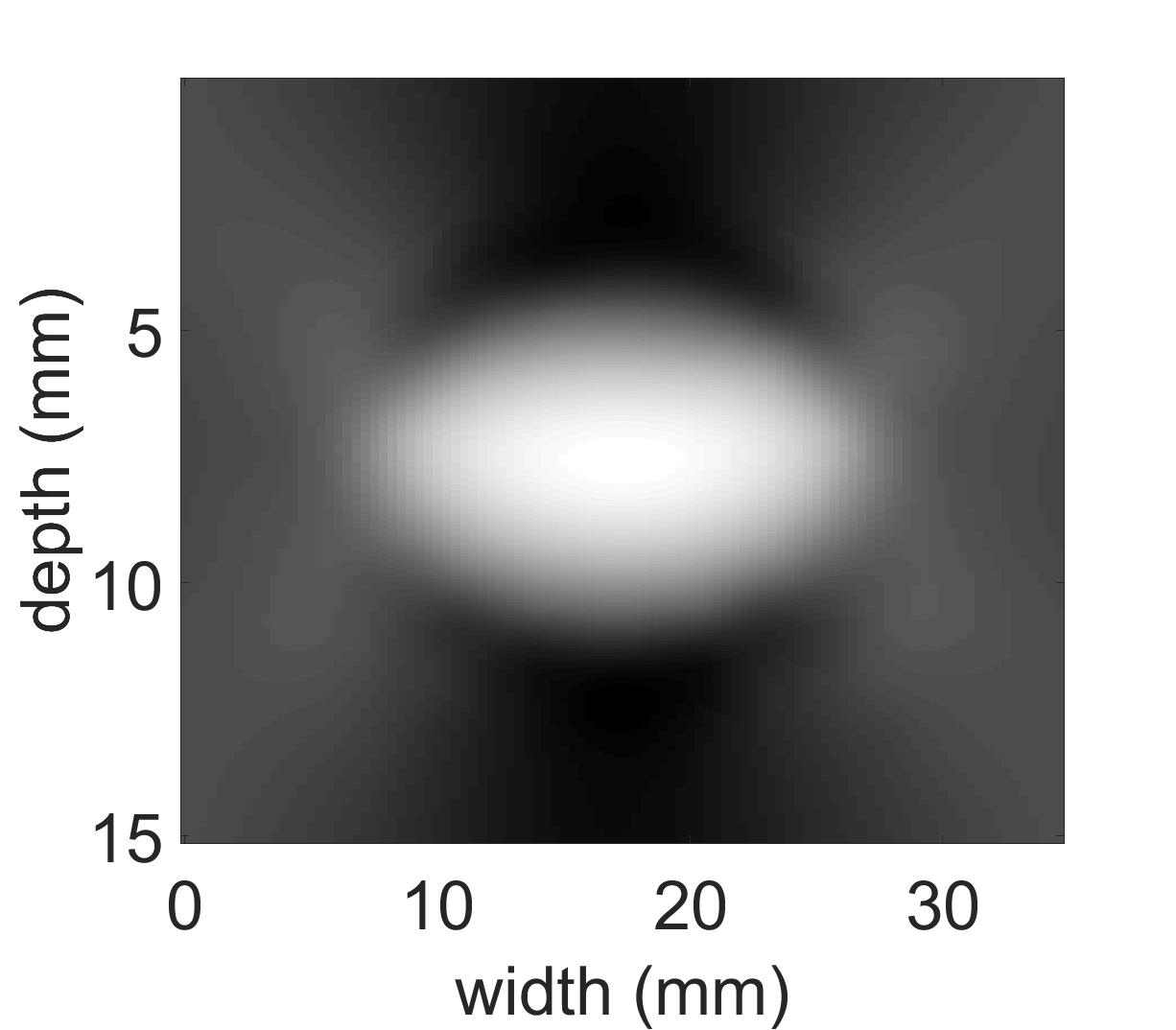}}\hspace*{-0.5em}
\subfigure[GLUE]{\includegraphics[height=3.25 cm,width=4.8 cm]{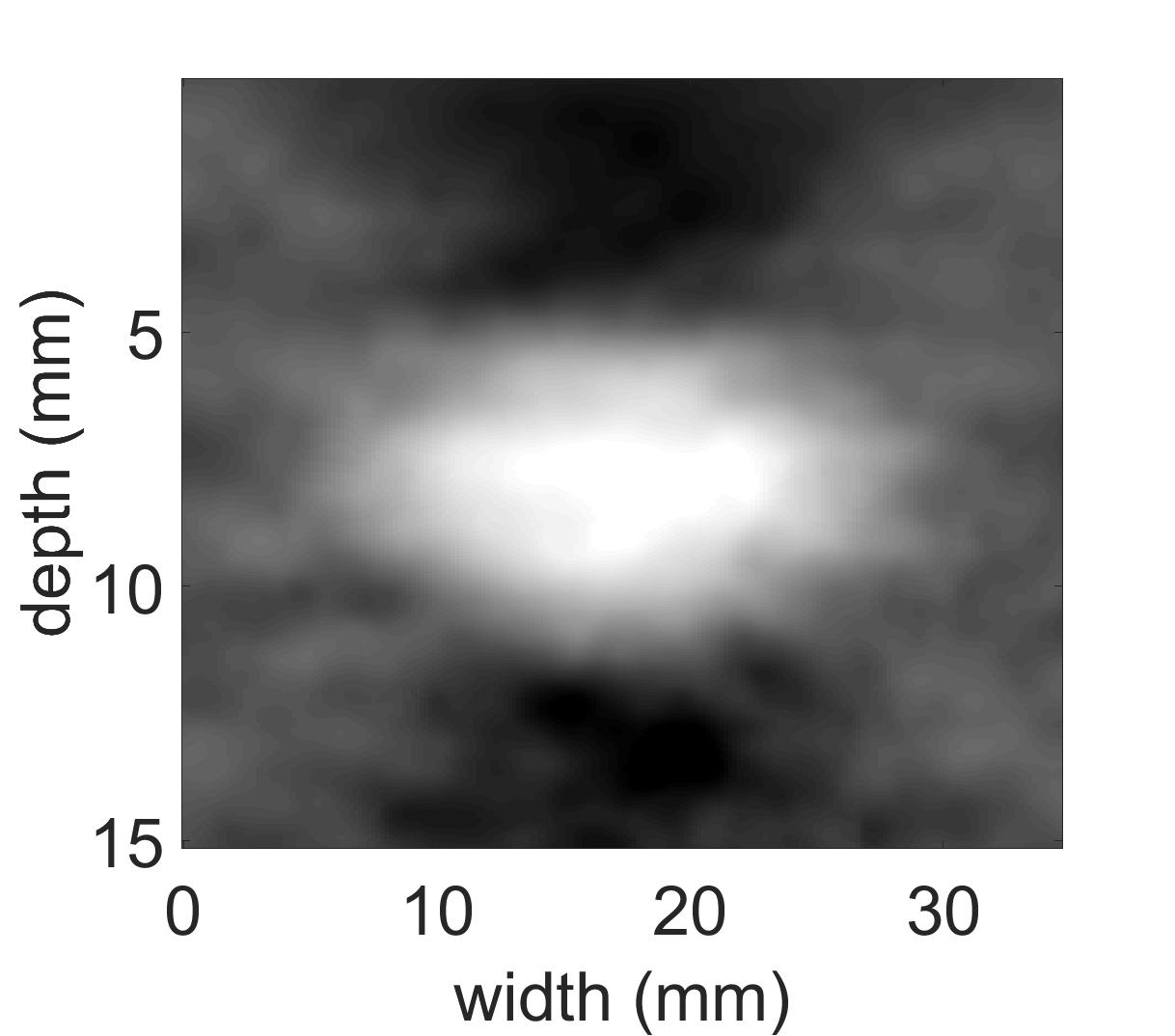}}\hspace*{-0.5em} 
\subfigure[PCA-GLUE]{\includegraphics[height=3.25 cm,width=4.8 cm]{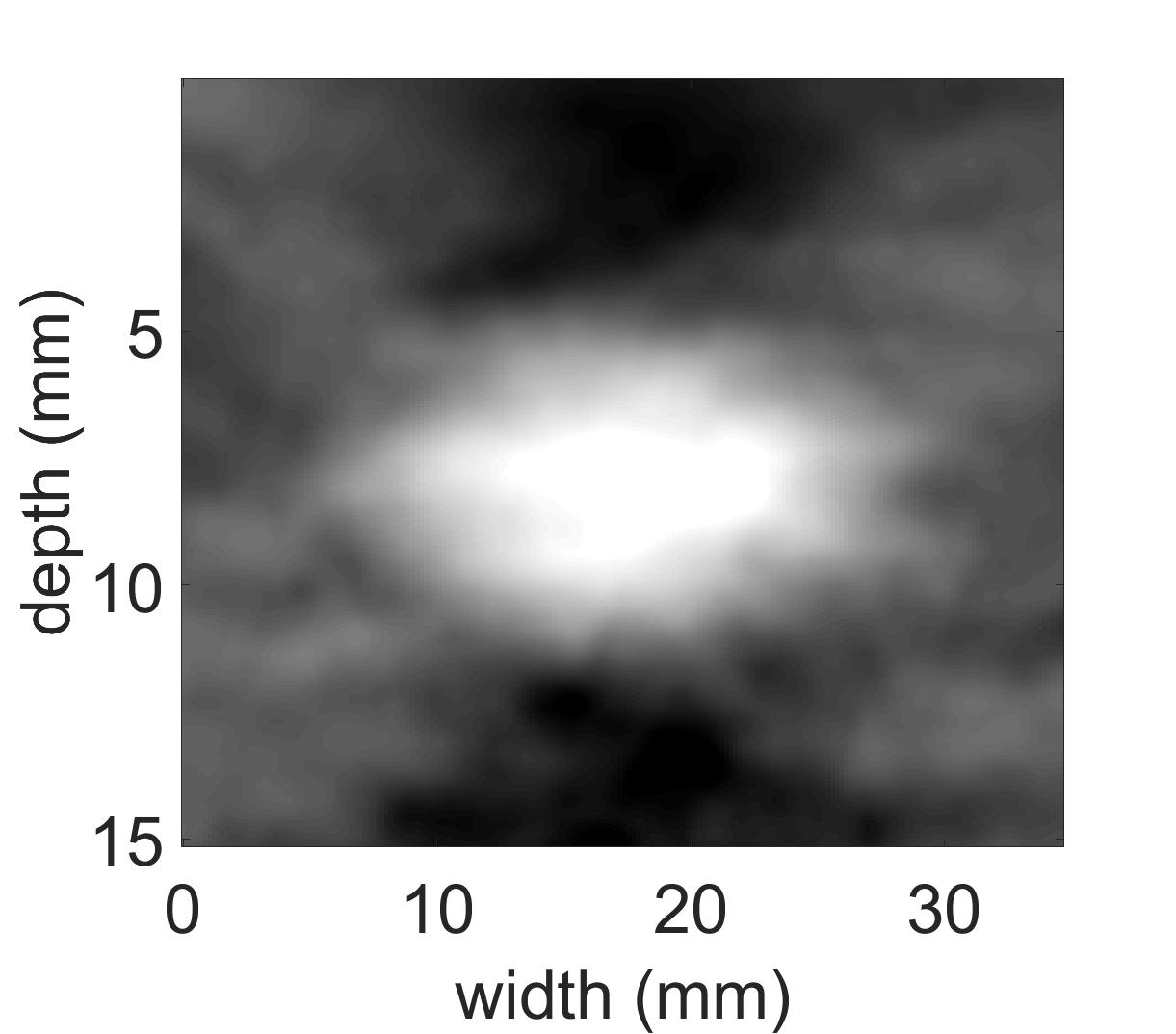}}\vspace*{-0em}
\stackunder[5pt]{\includegraphics[height=0.65 cm,width=4 cm]{p3_after_colorbar_final}}{\textcolor{black}{Strain color bar}}\vspace*{-0.5em}

\subfigure[B-mode after noise addition]{\includegraphics[height=3.25 cm,width=4.8 cm]{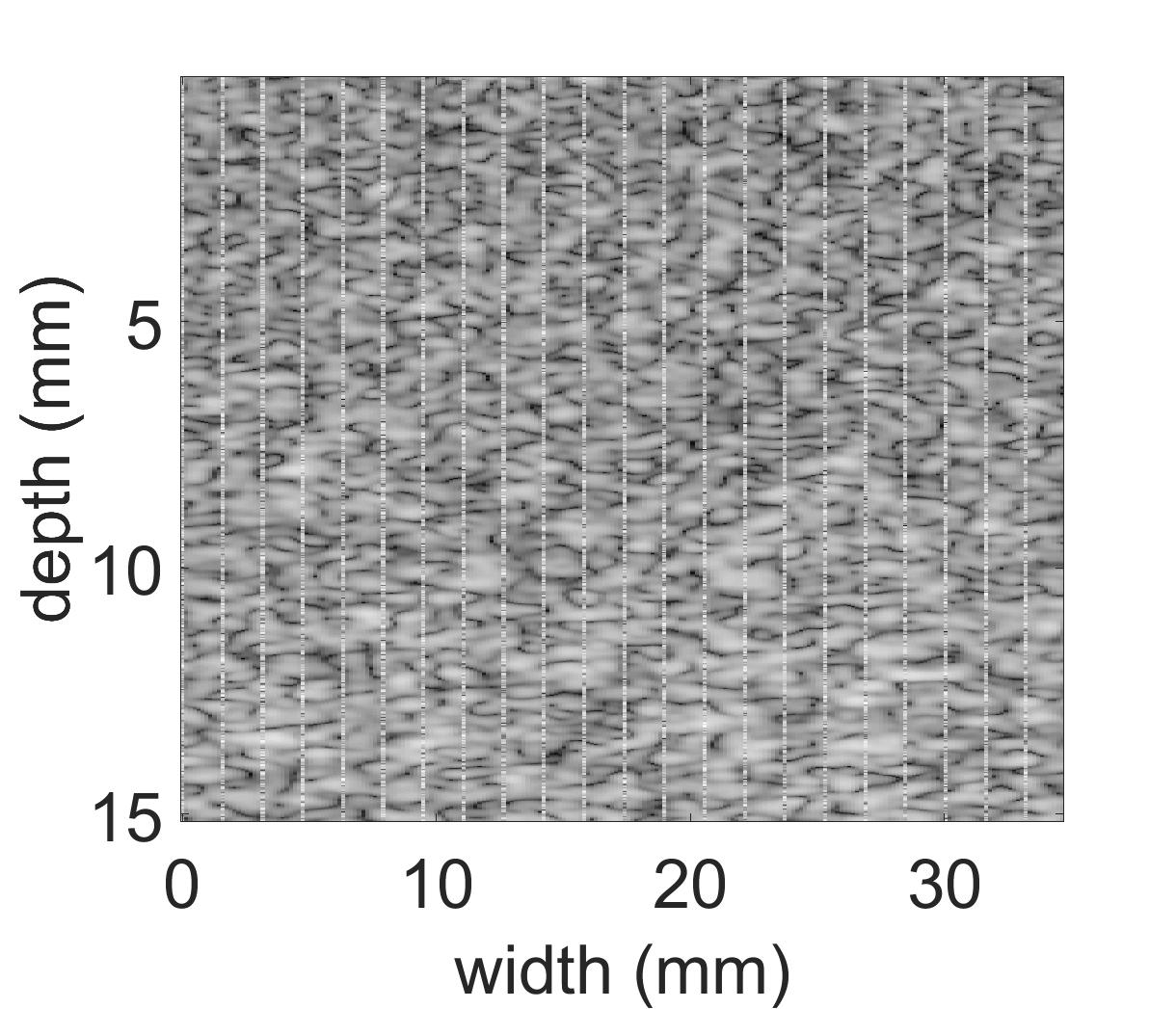}}\hspace*{-0.5em}
\subfigure[ground truth]{\includegraphics[height=3.25 cm,width=4.8 cm]{ground_truth_cropped}}\hspace*{-0.5em}
\subfigure[GLUE]{\includegraphics[height=3.25 cm,width=4.8 cm]{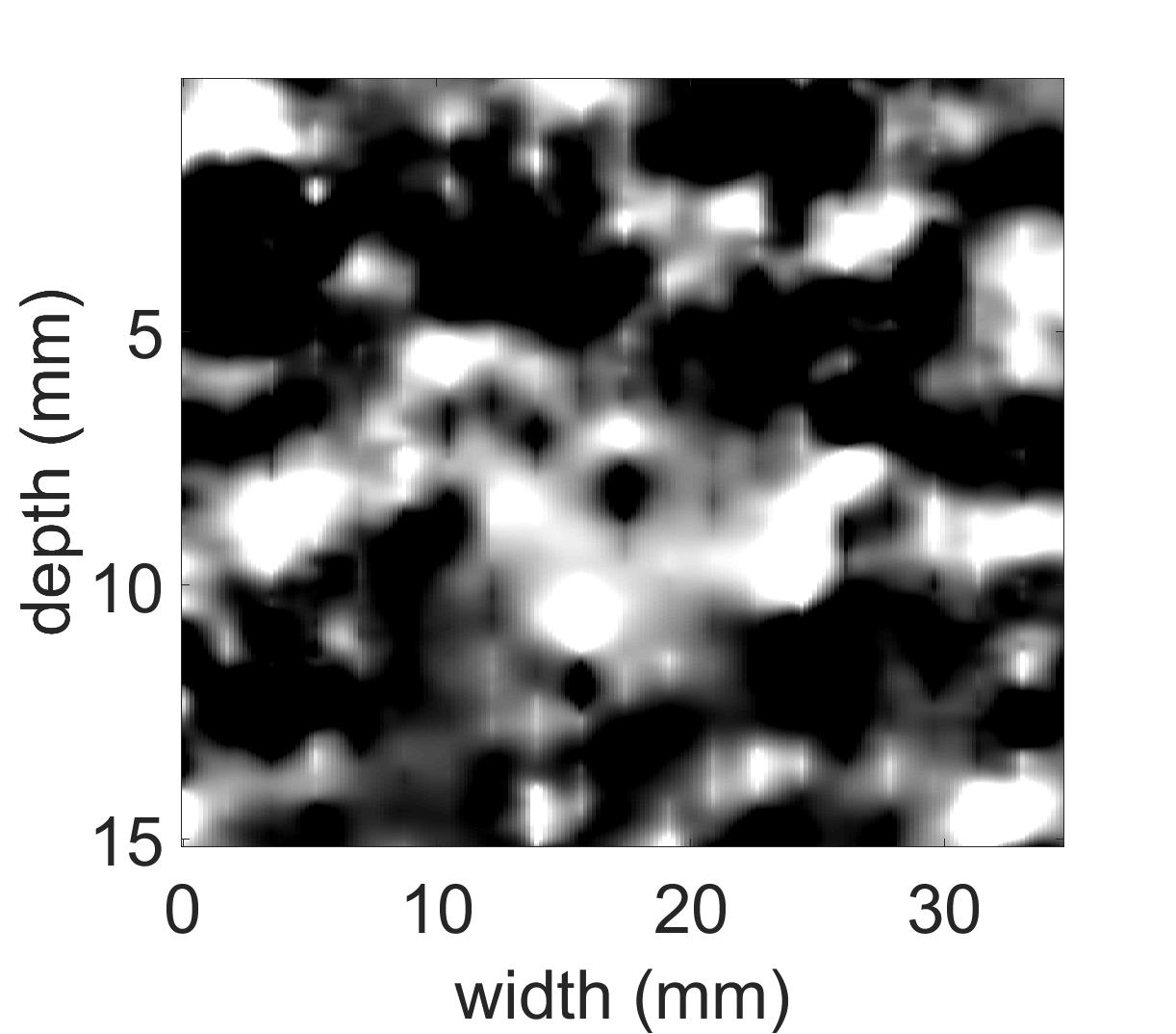}}\hspace*{-0.5em}
\subfigure[PCA-GLUE]{\includegraphics[height=3.25 cm,width=4.8 cm]{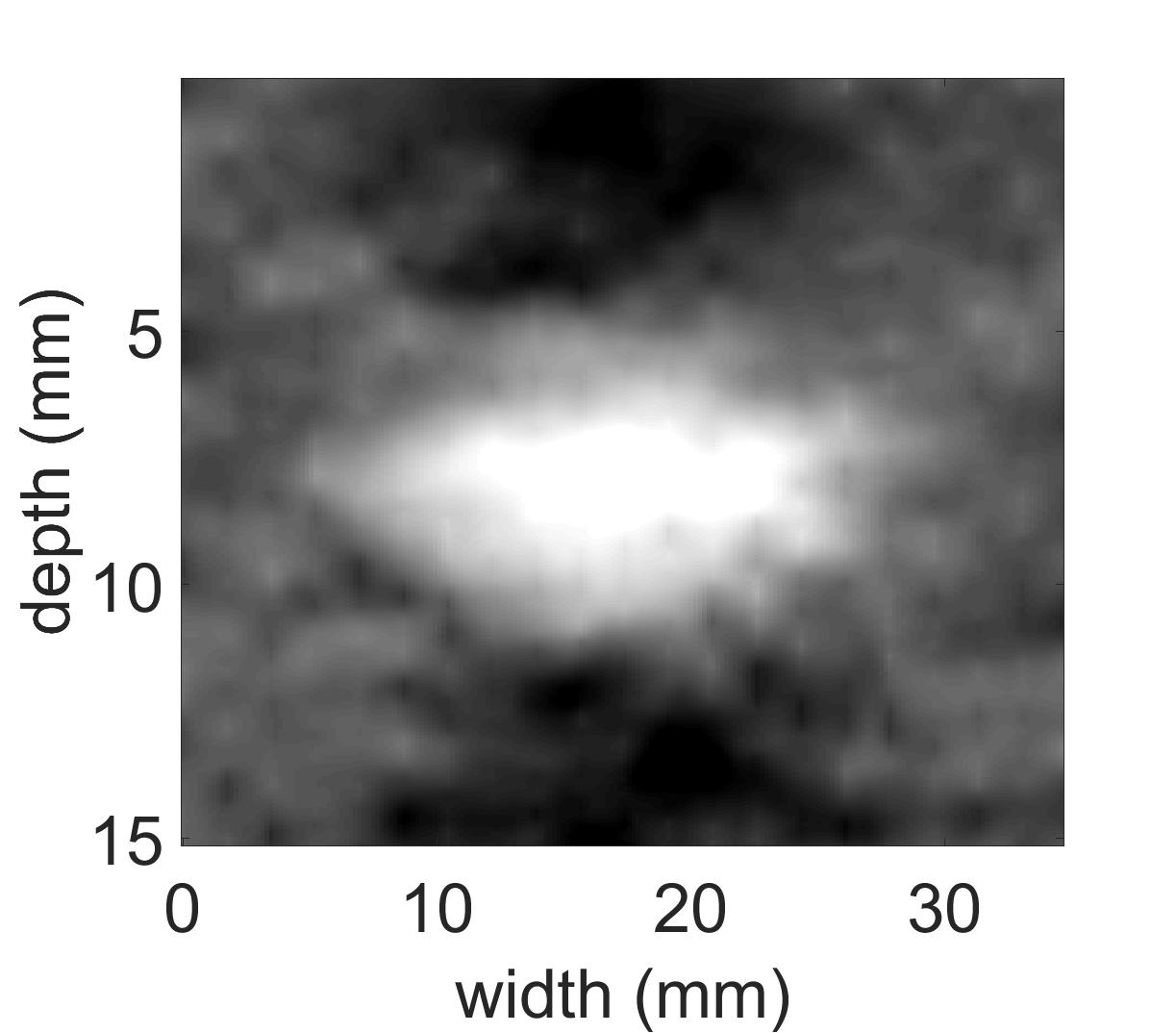}}\vspace*{-0em}
\stackunder[5pt]{\includegraphics[height=0.65 cm,width=4 cm]{p3_after_colorbar_final}}{\textcolor{black}{Strain color bar}}\vspace*{-0.5em}

\end{center}%
\caption{The B-mode ultrasound and ground truth axial strain as well as the result of both GLUE and PCA-GLUE for the simulation data before and after adding gaussian noise with $\mu=0$ and $\sigma^{2}=0.1225$ to 10\% of the RF lines.} \label{simulation_noisy}
\end{figure*}

\begin{figure*}[]
\begin{center}
\subfigure[B-mode]{\includegraphics[height=3.25 cm,width=4.8 cm]{phantom_bmode_big}}\hspace*{-0.5em}
\subfigure[Strain using 5 RF lines]{\includegraphics[height=3.25 cm,width=4.8 cm]{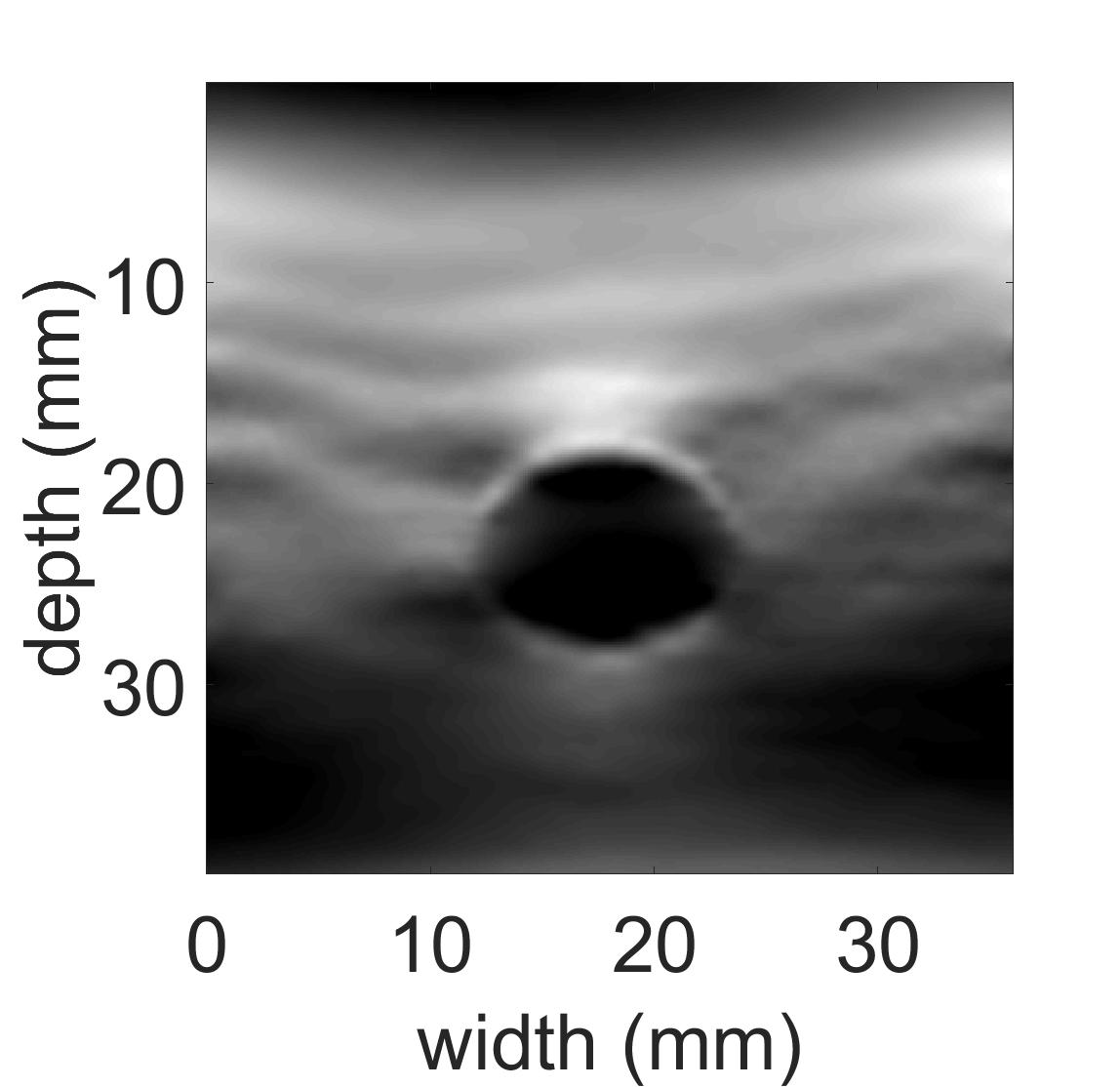}}\hspace*{-0.5em}
\subfigure[Strain using 15 RF lines]{\includegraphics[height=3.25 cm,width=4.8 cm]{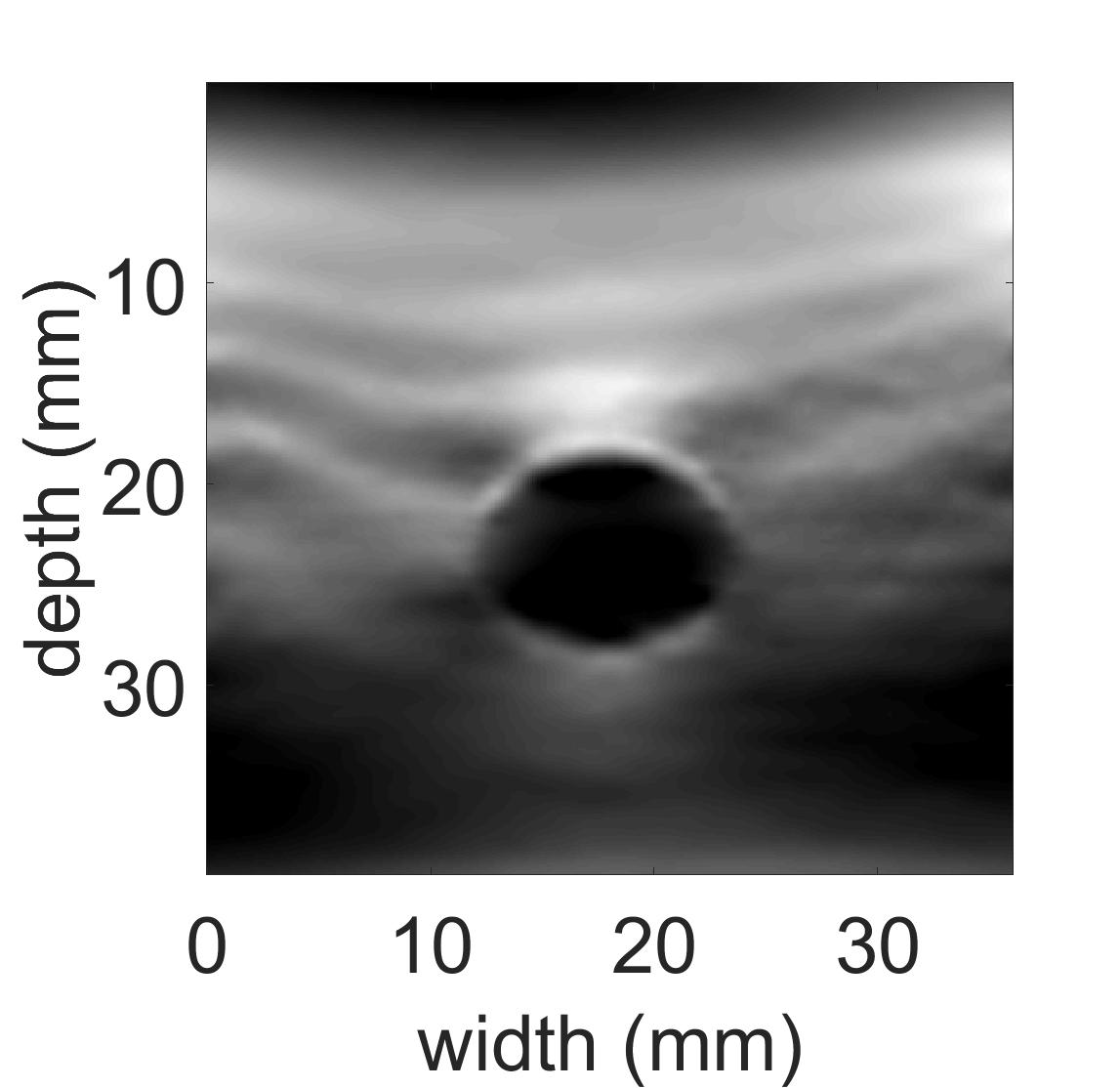}}\hspace*{-0.5em}
\subfigure[Strain using 30 RF lines]{\includegraphics[height=3.25 cm,width=4.8 cm]{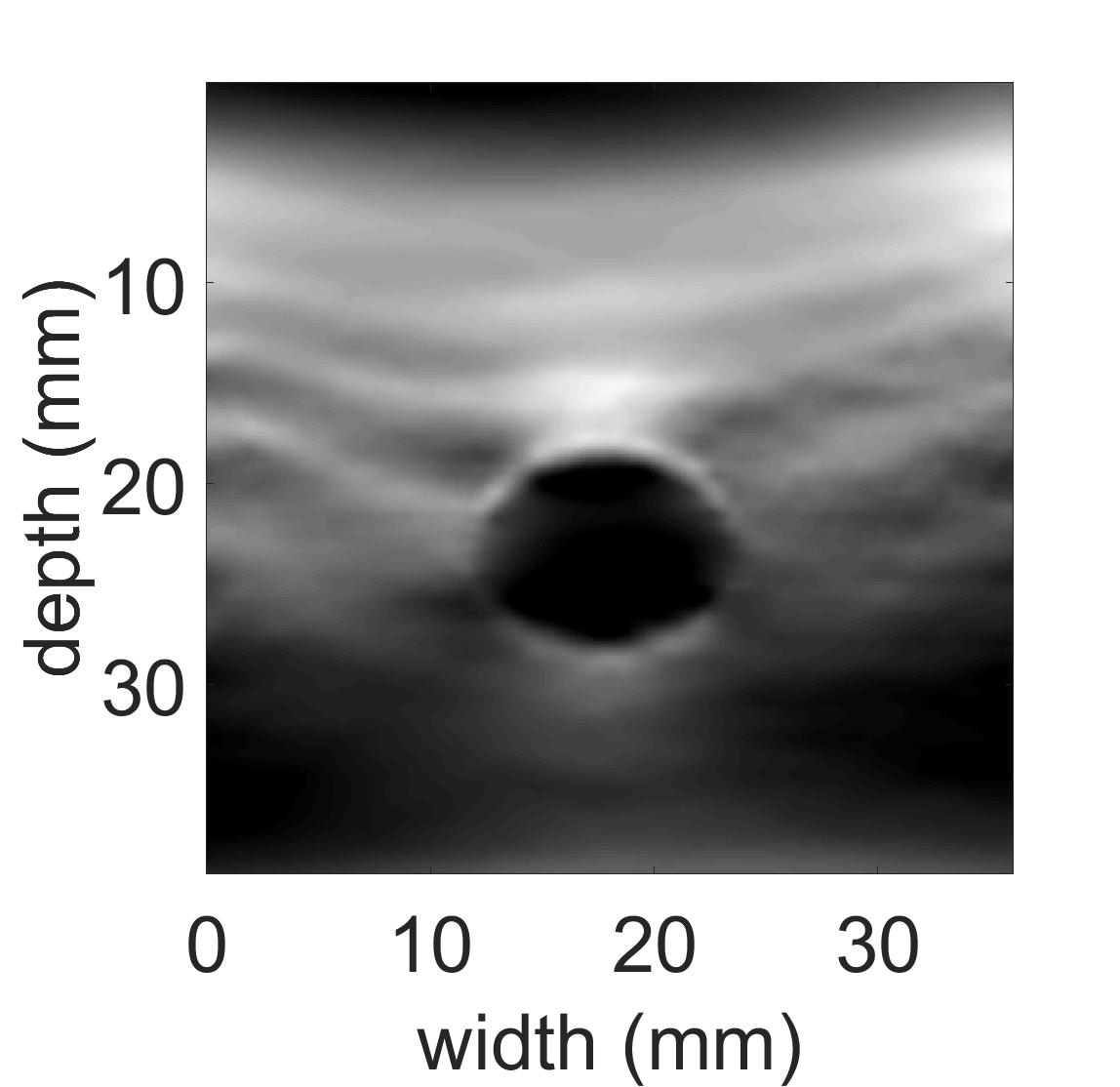}}\vspace*{-0em}
\stackunder[5pt]{\includegraphics[height=0.7 cm,width=4 cm]{phantom_colorbar2}}{\textcolor{black}{Strain color bar}}\vspace*{-0.5em}
\end{center}%
\centering
\caption{The B-mode ultrasound and axial strain image using PCA-GLUE for the real phantom experiment as we increase the number of RF lines $p$ from 5 to 30.}
\label{different_p}
\end{figure*}

\subsection{Varying the number of sparse features}
Fig.~\ref{different_p} shows the effect of running DP on more than 5 RF lines. We can conclude that the accuracy of the strain estimation does not improve any further when setting $p$ to a value more than 5. As more RF lines correspond to more features and consequently more computations, we choose the smallest value $p=5$ without sacrificing the accuracy. For more analysis and results at different values of $p$, please refer to the Supplementary Material of the paper.

\section{\textcolor{black}{Discussion}}

\textcolor{black}{We presented a novel method that can estimate a coarse displacement map from a sparse set of displacement data provided by DP. For an image of size }$\textcolor{black}{2304 \times 384}$, \textcolor{black}{DP takes 163} $\textcolor{black}{ms}$ \textcolor{black}{and estimation of coarse displacement field takes 95} $\textcolor{black}{ms}$\textcolor{black}{, for a total of 258} $\textcolor{black}{ms}$\textcolor{black}{. We also presented a novel method for frame selection that classifies a pair of RF Data as suitable or unsuitable for elastography in only 1.5} $\textcolor{black}{ms}$\textcolor{black}{. The input to our classifier is the $\textbf{w}$ vector and not the RF data or the displacement image. The reason is that inference with such low-dimensional input is very computationally efficient.}

\textcolor{black}{During training, we tried using datasets with and without inclusions to obtain the principal components. Our conclusion is that the presence or absence of inclusions in the training data does not alter the principal components and consequently the strain estimated. What is critical for learning the principal components is the presence of different types of deformation, such as axial, lateral and rotational deformation.}

\textcolor{black}{It is worth mentioning that we were only concerned with the axial displacement, as we couldn’t compute any principal components that describe the lateral displacement.} \textcolor{black}{The reason is that by following Algorithm 3 for the lateral displacement, we found that the variance is not concentrated in the first few eigenvectors, unlike the axial displacement. It is rather almost equally distributed over hundreds of eigenvectors (resembling white noise). We conclude that capturing 95\% of the variance would require us to save hundreds of principal components, which is not practical.} \textcolor{black}{Therefore, we only use the integer estimates for the }$\textcolor{black}{k=m \times p}$ \textcolor{black}{pixels computed by DP, followed by bi-linear interpolation which provides an acceptable initial lateral displacement,} \textcolor{black}{ compared to the alternative approach where we run DP on all RF lines. A comparison between the lateral displacement estimated by the two approaches is shown in the Supplementary Material. The combination of} $\textcolor{black}{N=12}$ \textcolor{black}{and} $\textcolor{black}{p=5}$ \textcolor{black}{is not a fixed choice of the hyperparameters, as different datasets would require different tuning. In our case, this choice is adequate for all the datasets used.}}

\section{Conclusion}
In this paper, we introduced a new method with two main contributions which are fast strain estimation and RF frame selection. In addition, our method is robust to incorrect initial estimates by DP. Our method is more than 10 times faster than GLUE in estimating the initial displacement image, which is the step prior to the exact displacement estimation, while giving the same or better results. \textcolor{black}{Our MLP classifier used for frame selection has been tested on 1,430 unseen pairs of RF frames from both phantom and \textit{in vivo} datasets, and the F1-measure obtained was always higher than 92\%}. This proves that our method is efficient and that it could be used commercially.
\section{Acknowledgment}
This work was by funded Richard and Edith Strauss Foundation. The authors would like to thank Drs. E. Boctor, M. Choti and G. Hager for providing us with the \textit{in vivo} patients data from Johns Hopkins Hospital. We also thank Morteza Mirzaei for his help in collecting the phantom data and for providing us with an optimized version of the NCC code, as well as the very fruitful discussions. We would like also to thank Md Ashikuzzaman for for his helpful comments and for providing us with the simulation data.


%




\ifCLASSOPTIONcaptionsoff
\newpage
\fi



%

%

\bigskip
\bigskip
\bigskip

\bstctlcite{IEEEexample:BSTcontrol}
\bibliographystyle{IEEEtran}
\bibliography{strings,ref3}
\end{document}


\title{Fast Strain Estimation and Frame Selection in Ultrasound Elastography
using Machine Learning
}

\author{Abdelrahman~Zayed, \textit{Student Member, IEEE} and Hassan~Rivaz, \textit{Senior Member}, \textit{IEEE}}

\maketitle

This supplementary material provides results that are complementary to those in the paper. We investigate the effect of changing the hyperparameter $N$, which refers to the number of principal components used, for both phantom and \textit{in vivo} data. We also simulate the effect of changing the hyperparameter $p$, which refers to the number of RF lines used for extracting the sparse features. Depending on the rate of change of the displacement image in the horizontal direction, different values of $p$ are needed. 

Using the simulation data, we also show that our method works for different compression levels. In addition, we perform an experiment where we give out-of-plane RF frame pairs (i.e. poor) to the multi-layer perceptron classifier, to see whether it can correctly classify them as bad pairs.

Finally, we perform an experiment to show that it is acceptable to only compute the lateral displacement for $p$ RF lines, followed by bi-linear interpolation, as compared to the more computationally expensive solution of calculating the displacement of all RF lines. We compare the axial strain produced by GLUE while providing it with an initial lateral displacement using both methods.

\section{Results}
\begin{figure*}[]
\captionsetup{justification=centering}
\begin{center}
\subfigure[B-mode]{\includegraphics[height=3.35 cm,width=4.8 cm]{phantom_bmode_big}}\hspace*{-0.5em}
\subfigure[Strain using $N=6$]{\includegraphics[height=3.35 cm,width=4.8 cm]{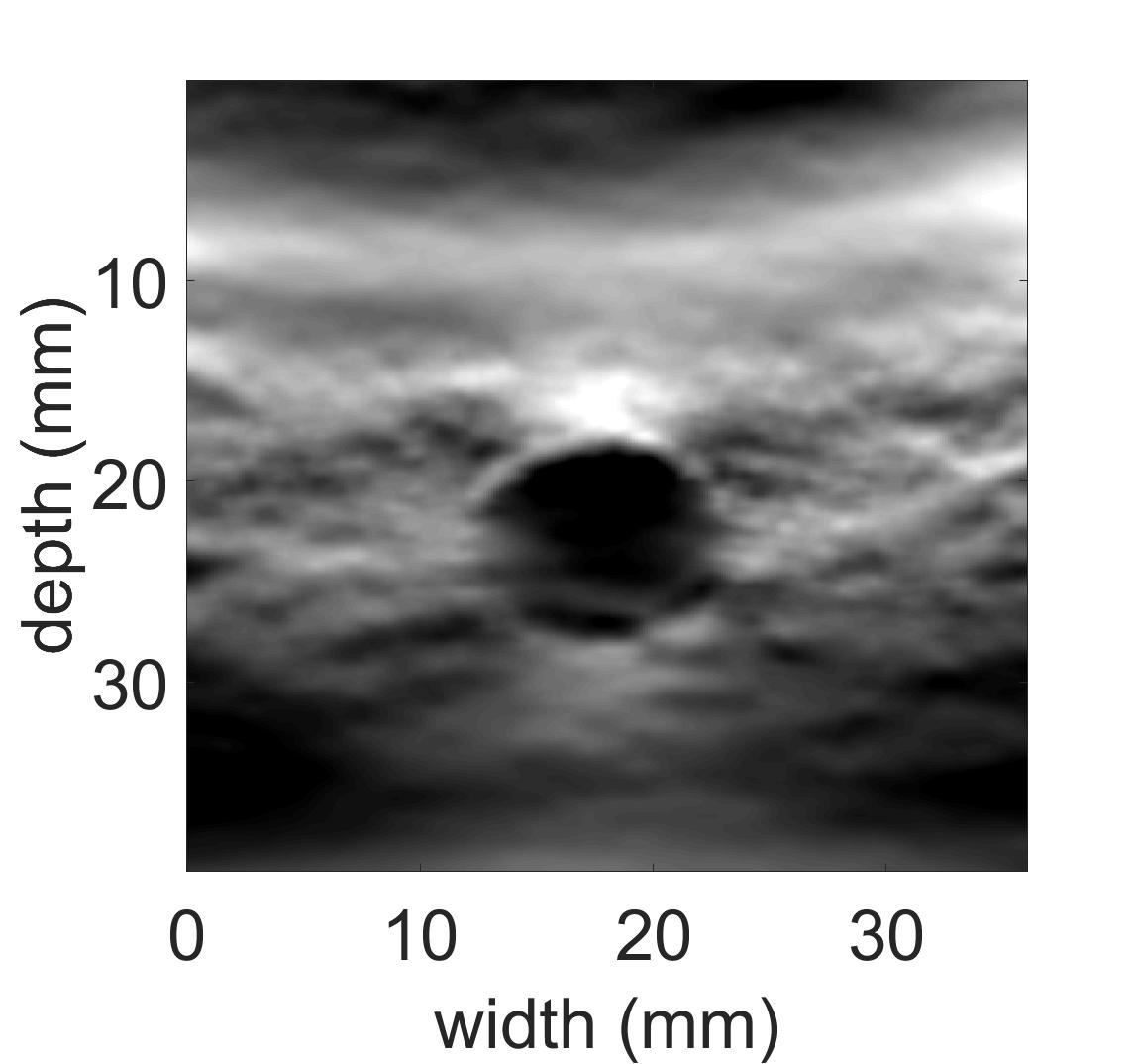}}\hspace*{-0.5em}
\subfigure[Strain using $N=12$]{\includegraphics[height=3.35 cm,width=4.8 cm]{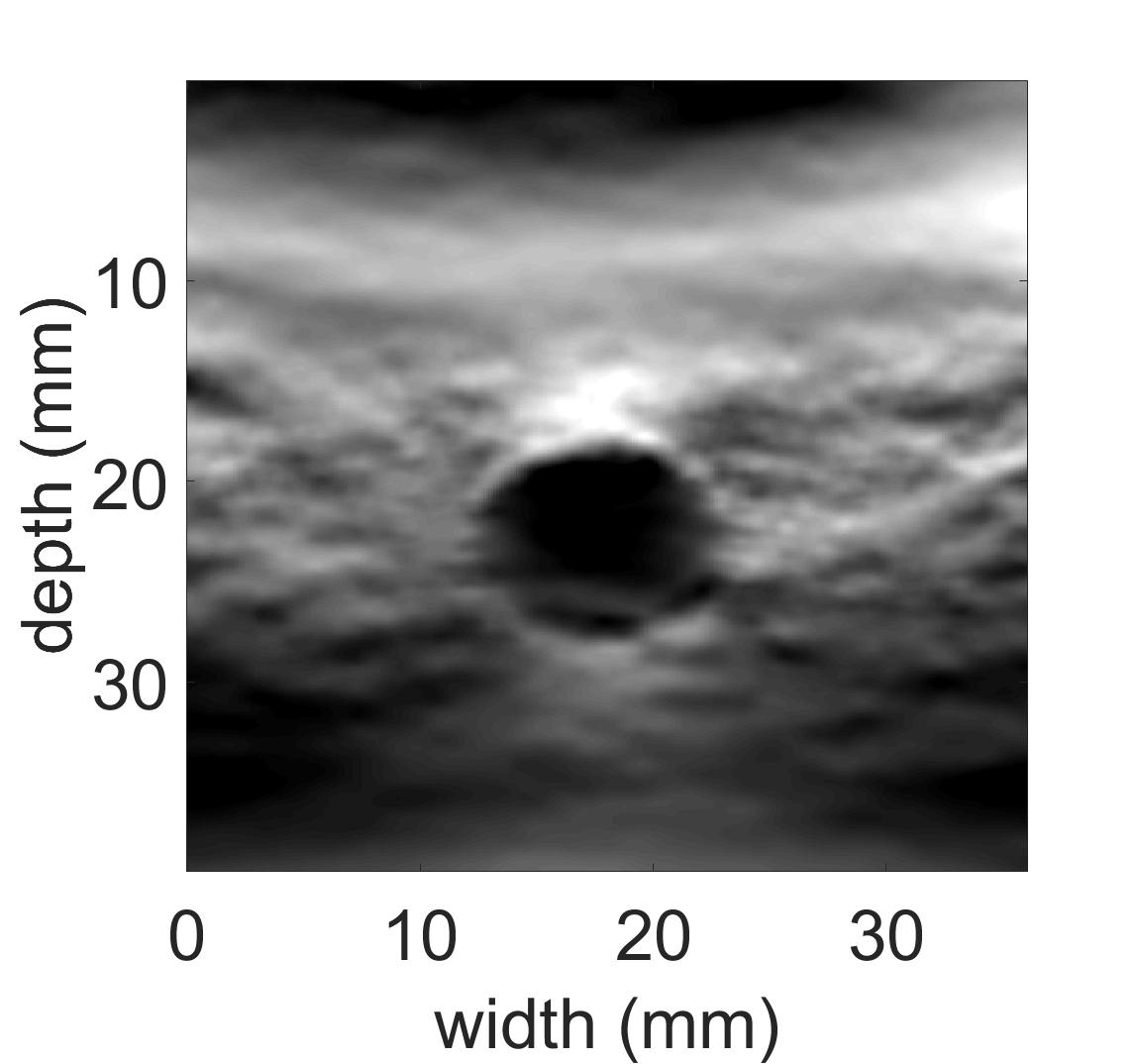}}\hspace*{-0.5em}
\subfigure[Strain using $N=24$]{\includegraphics[height=3.35 cm,width=4.8 cm]{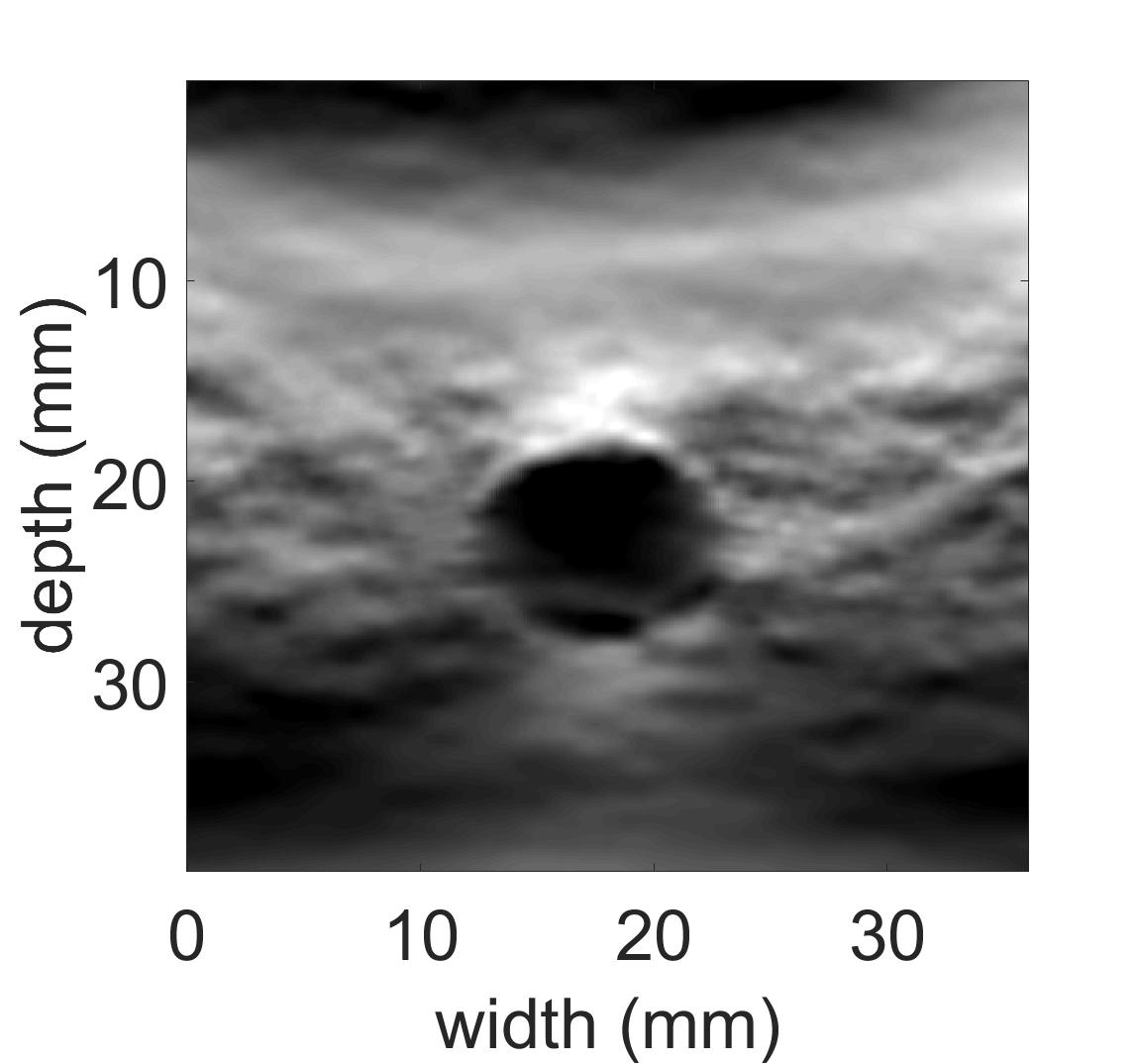}}\vspace*{-0em}
\subfigure[Color bar]{\includegraphics[height=0.8 cm,width=4 cm]{phantom_colorbar_final}}\vspace*{-1em}
\end{center}%
\centering
\caption{The B-mode ultrasound and axial strain image using PCA-GLUE for the real phantom experiment as we increase the number of principal components $N$ from 6 to 24. The color bar is for the strain images only.} 
\label{different_N_phantom}
\end{figure*}

\begin{figure*}[]
\captionsetup{justification=centering}
\begin{center}
\subfigure[B-mode]{\includegraphics[height=3.35 cm,width=4.8 cm]{p2_bmode_big}}\hspace*{-0.5em}
\subfigure[Strain using $N=6$]{\includegraphics[height=3.35 cm,width=4.8 cm]{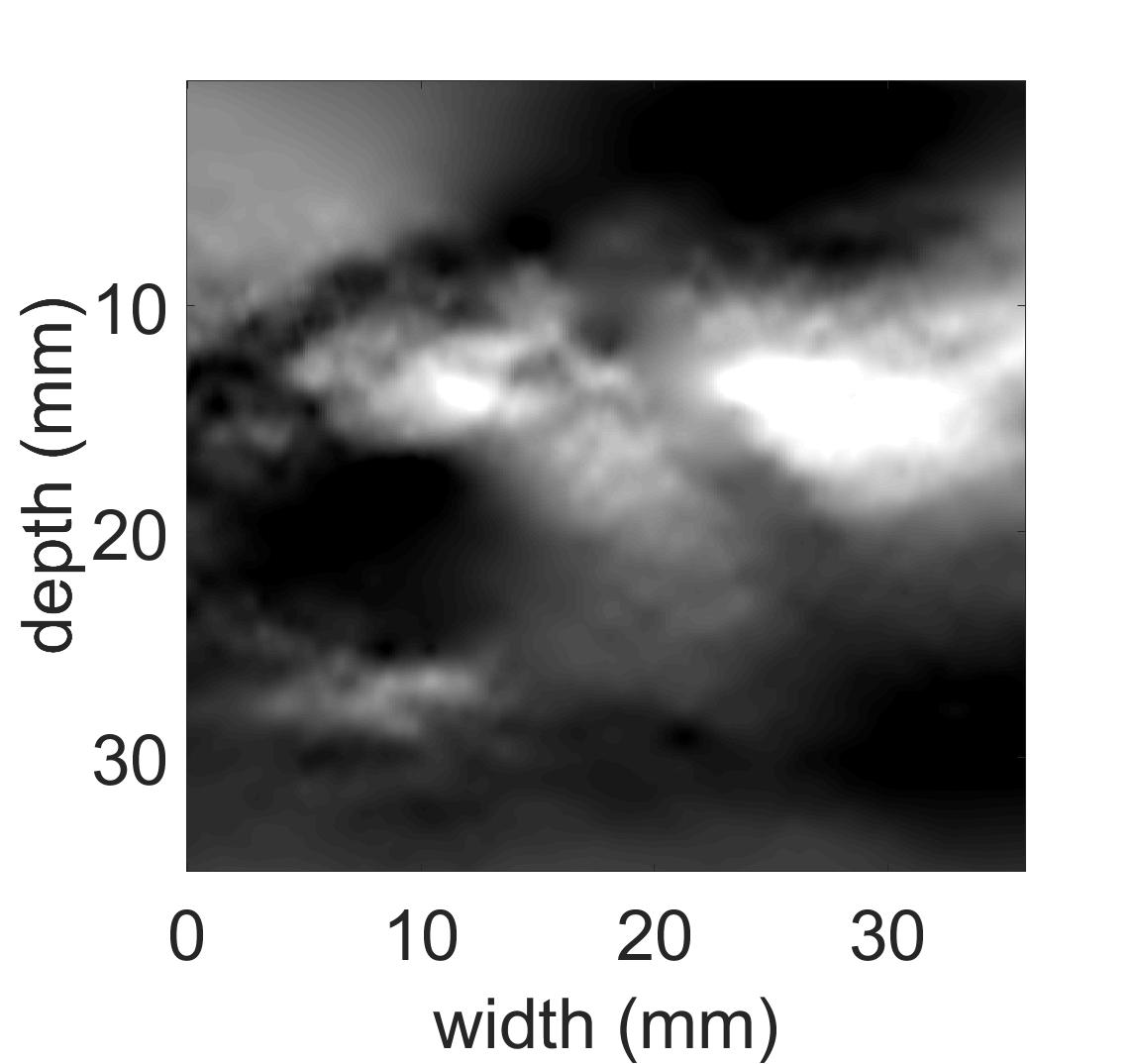}}\hspace*{-0.5em}
\subfigure[Strain using $N=12$]{\includegraphics[height=3.35 cm,width=4.8 cm]{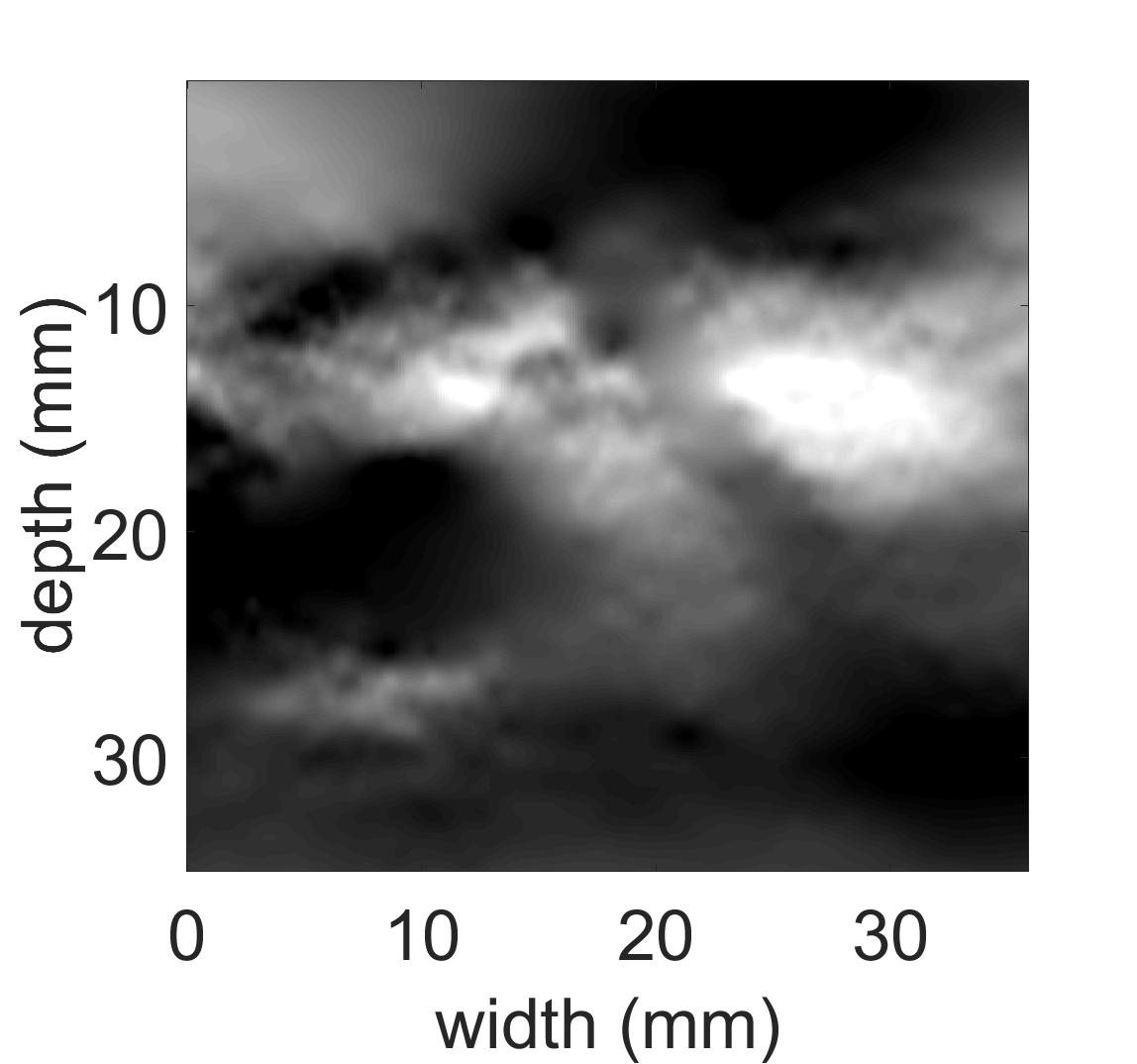}}\hspace*{-0.5em}
\subfigure[Strain using $N=24$]{\includegraphics[height=3.35 cm,width=4.8 cm]{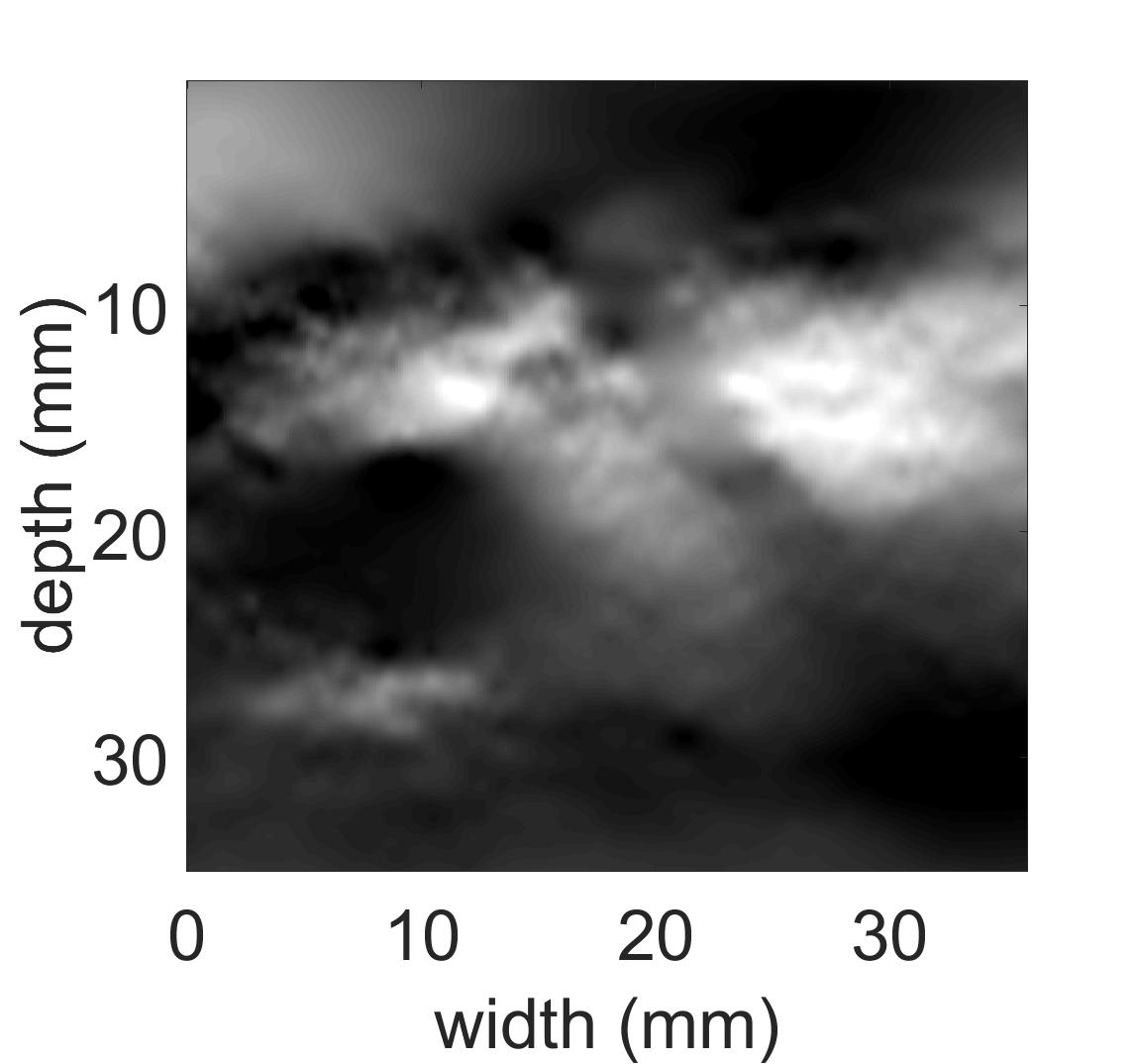}}\vspace*{-0em}
\subfigure[Color bar]{\includegraphics[height=0.9 cm,width=4 cm]{p2_colorbar_final}}\vspace*{-1em}
\end{center}%
\centering
\caption{The B-mode ultrasound and axial strain image using PCA-GLUE for the \textit{in vivo} liver data before ablation as we increase the number of principal components $N$ from 6 to 24. The color bar is for the strain images only.} 
\label{different_N_invivo_before}
\end{figure*}

\begin{figure*}[]
\captionsetup{justification=centering}
\begin{center}
\subfigure[B-mode]{\includegraphics[height=3.35 cm,width=4.8 cm]{p3_after_bmode_final}}\hspace*{-0.5em}
\subfigure[Strain using $N=6$]{\includegraphics[height=3.35 cm,width=4.8 cm]{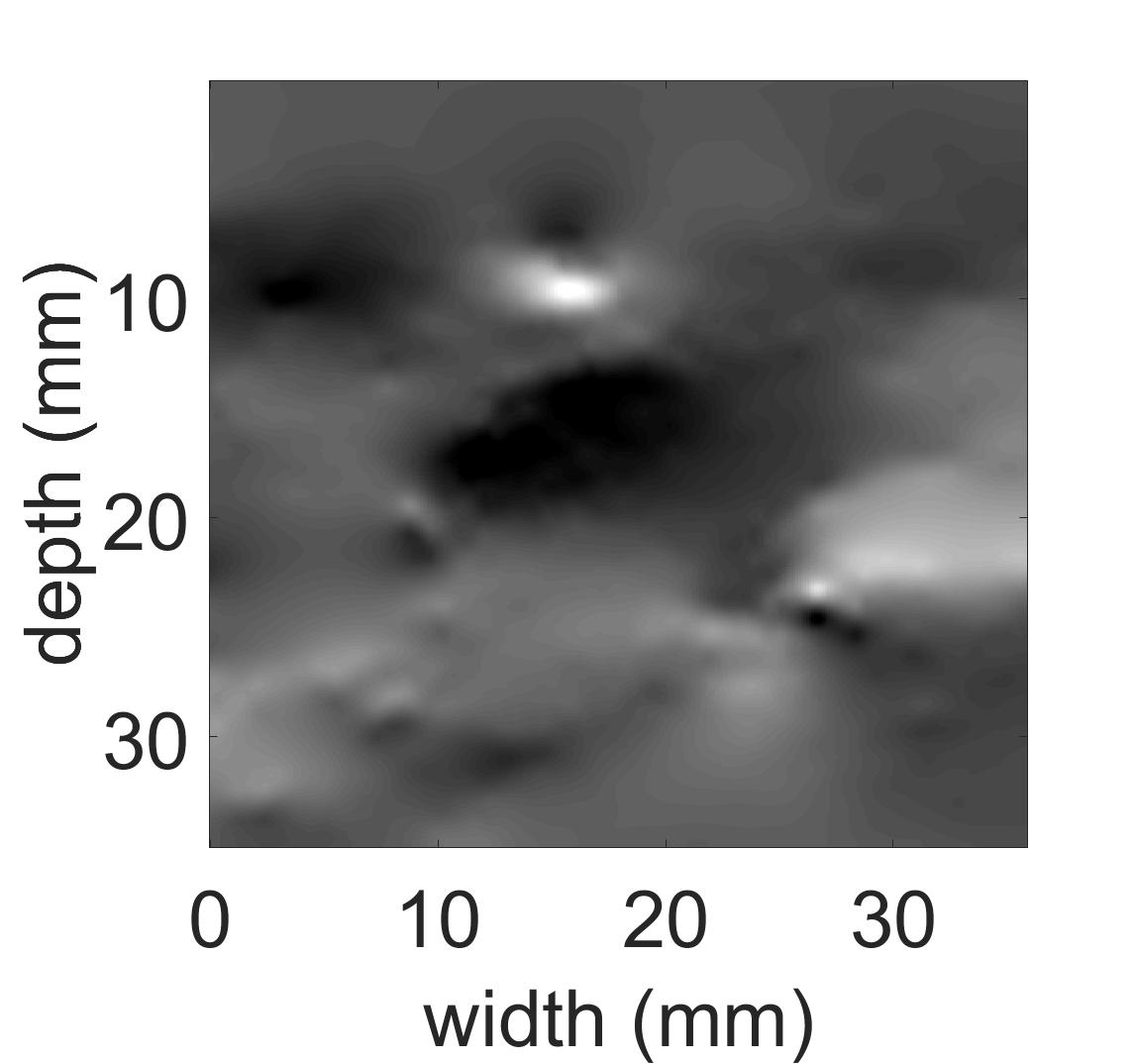}}\hspace*{-0.5em}
\subfigure[Strain using $N=12$]{\includegraphics[height=3.35 cm,width=4.8 cm]{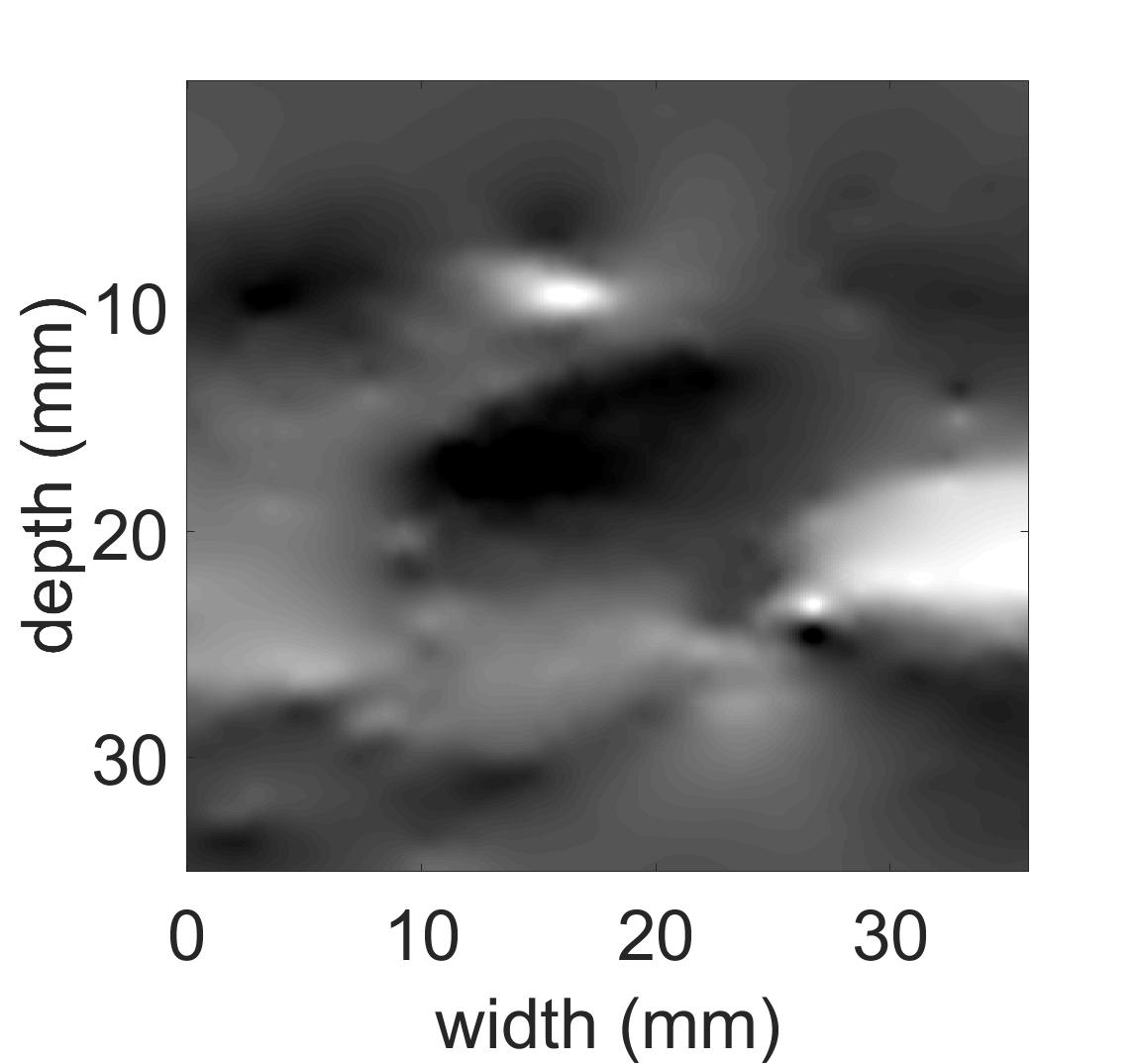}}\hspace*{-0.5em}
\subfigure[Strain using $N=24$]{\includegraphics[height=3.35 cm,width=4.8 cm]{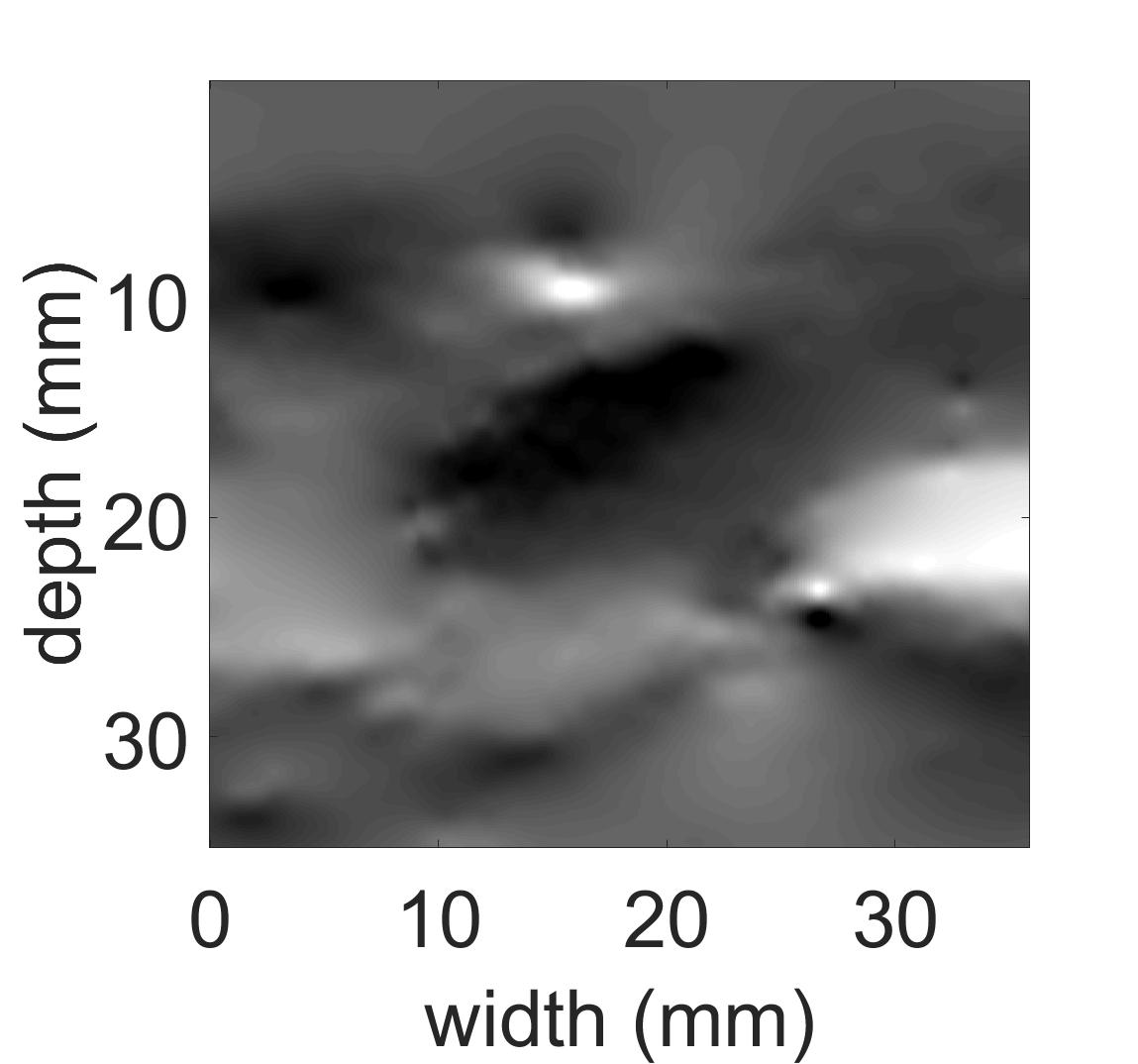}}\vspace*{-0em}
\subfigure[Color bar]{\includegraphics[height=0.7 cm,width=4 cm]{p3_after_colorbar_final}}\vspace*{-1em}
\end{center}%
\centering
\caption{The B-mode ultrasound and axial strain image using PCA-GLUE for \textit{in vivo} liver data after ablation as we increase the number of principal components $N$ from 6 to 24. The color bar is for the strain images only.} 
\label{different_N_invivo_after}
\end{figure*}

\begin{figure*}[]
\captionsetup{justification=centering}
\begin{center}
\subfigure[B-mode]{\includegraphics[height=3.35 cm,width=4.8 cm]{phantom_bmode_big}}\hspace*{-0.5em}
\subfigure[Strain using $p=2$]{\includegraphics[height=3.35 cm,width=4.8 cm]{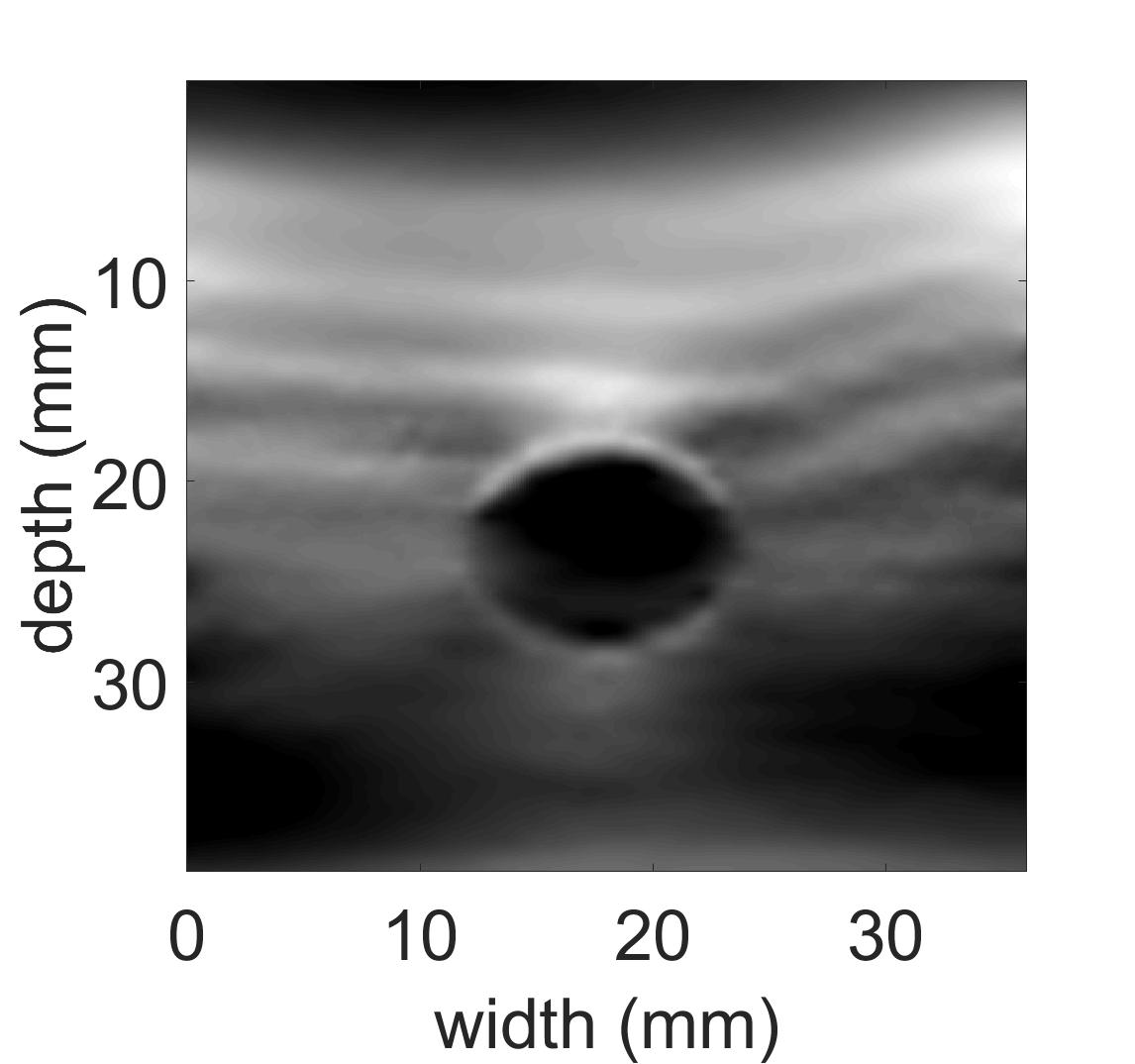}}\hspace*{-0.5em}
\subfigure[Strain using $p=5$]{\includegraphics[height=3.35 cm,width=4.8 cm]{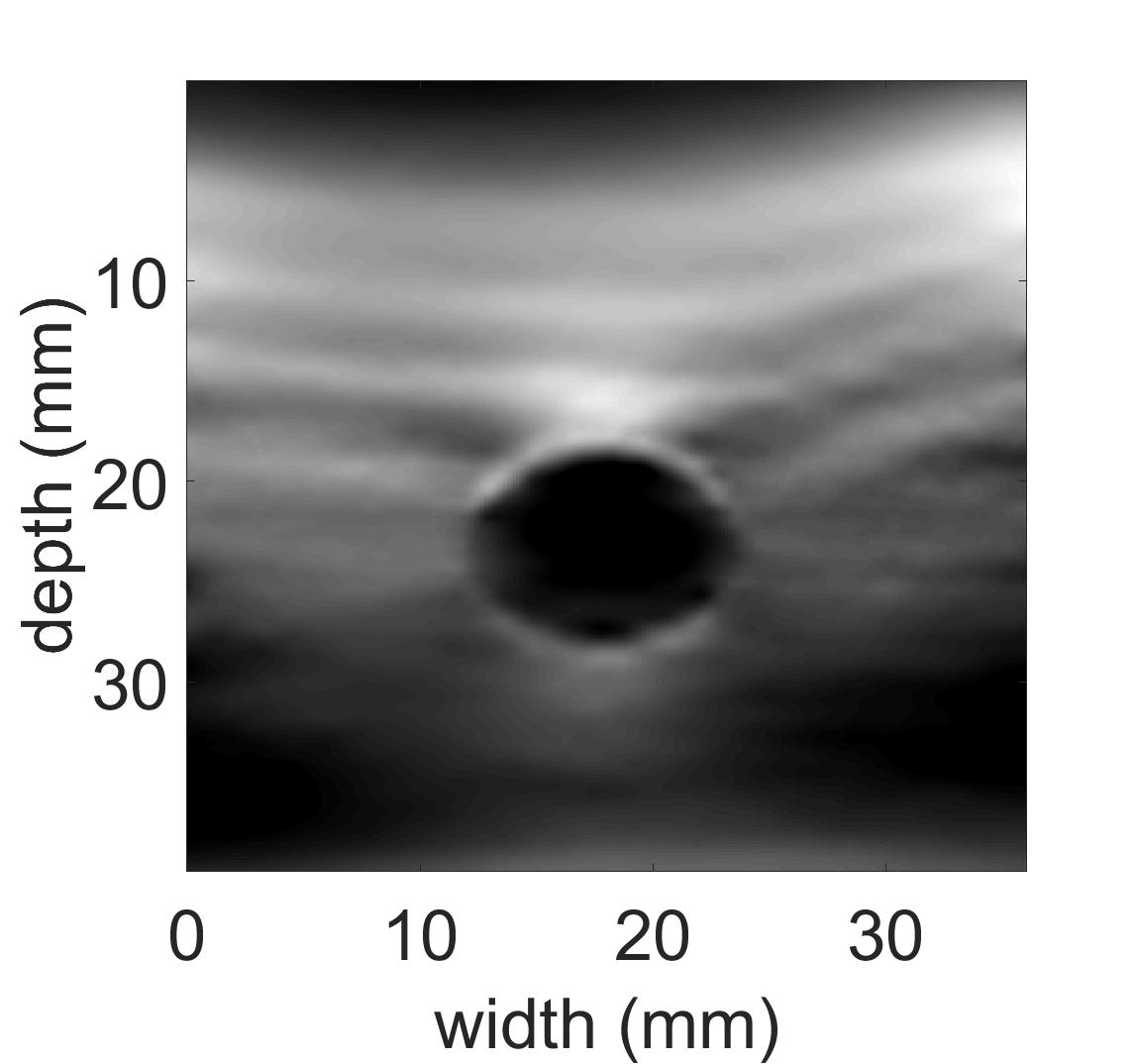}}\hspace*{-0.5em}
\subfigure[Strain using $p=10$]{\includegraphics[height=3.35 cm,width=4.8 cm]{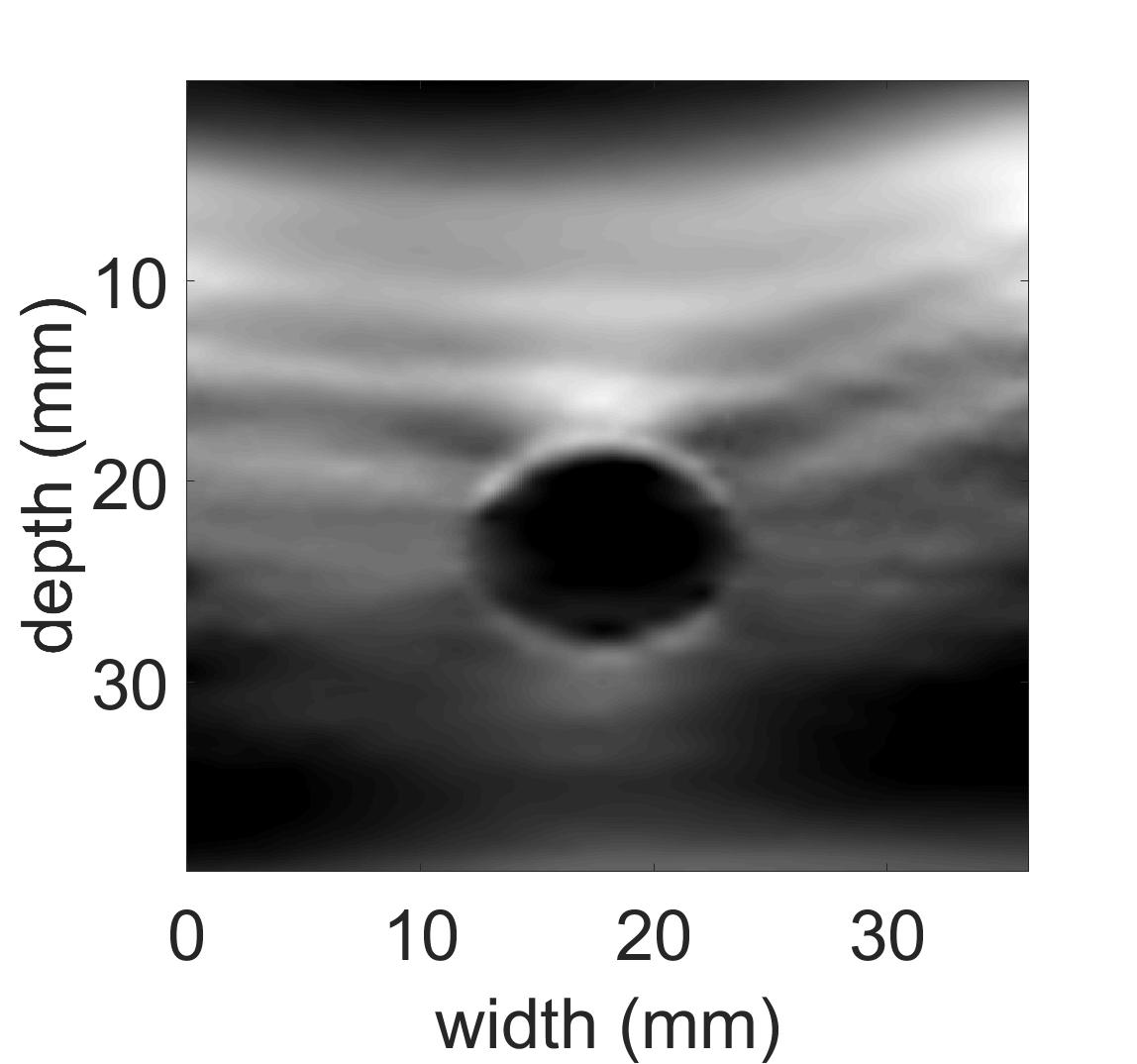}}\vspace*{-0em}
\subfigure[Color bar]{\includegraphics[height=0.7 cm,width=4 cm]{phantom_colorbar2}}\vspace*{-1em}
\end{center}%
\centering
\caption{The B-mode ultrasound and axial strain image using PCA-GLUE for the real phantom experiment as we increase the number of RF lines $p$ from 2 to 10. The color bar is for the strain images only.} 
\label{different_p_phantom}
\end{figure*}

\begin{figure*}[]
\captionsetup{justification=centering}
\begin{center}
\subfigure[B-mode]{\includegraphics[height=3.35 cm,width=4.8 cm]{p2_bmode_big}}\hspace*{-0.5em}
\subfigure[Strain using $p=2$]{\includegraphics[height=3.35 cm,width=4.8 cm]{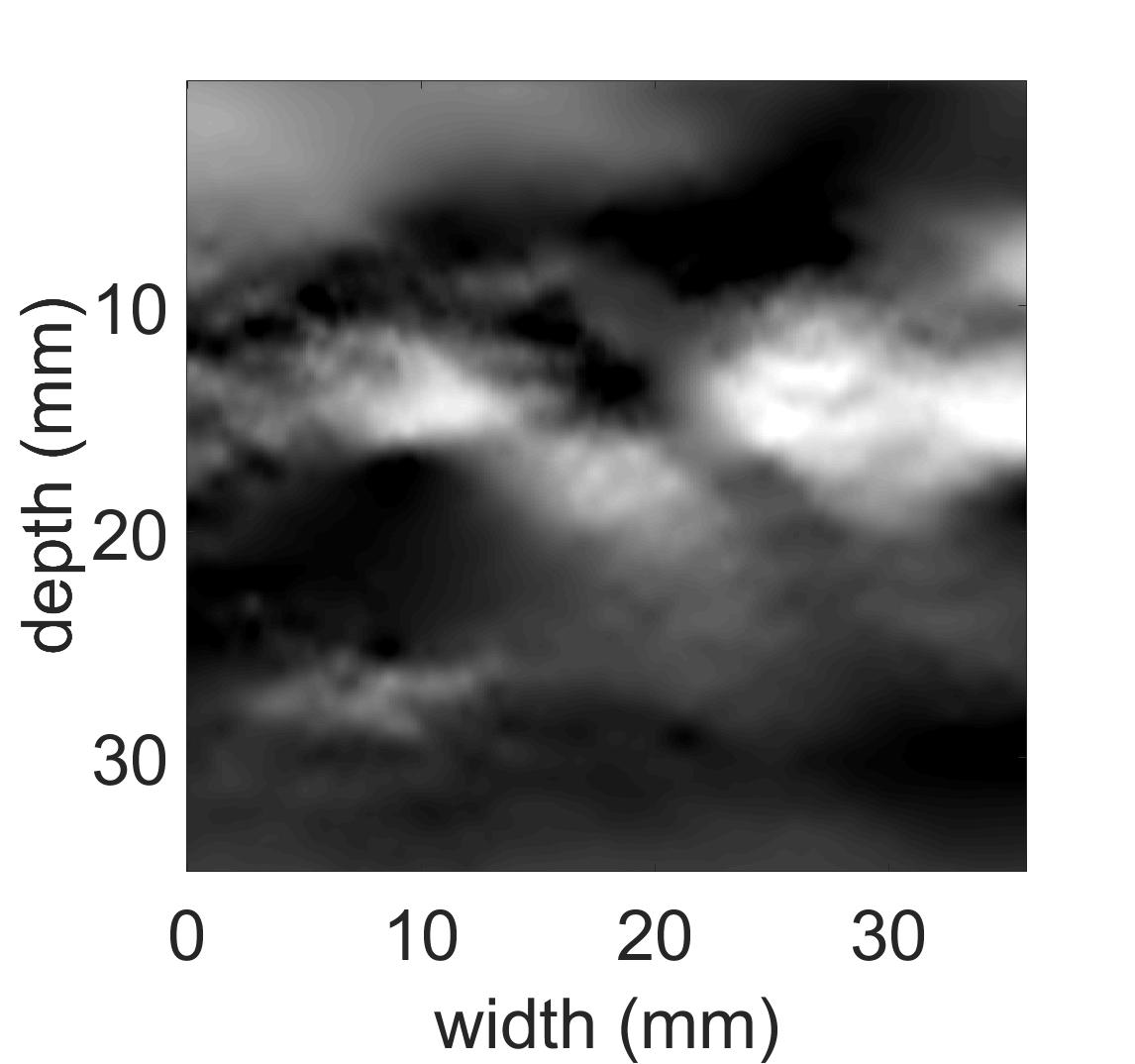}}\hspace*{-0.5em}
\subfigure[Strain using $p=5$]{\includegraphics[height=3.35 cm,width=4.8 cm]{p2_PCA-GLUE_final}}\hspace*{-0.5em}
\subfigure[Strain using $p=10$]{\includegraphics[height=3.35 cm,width=4.8 cm]{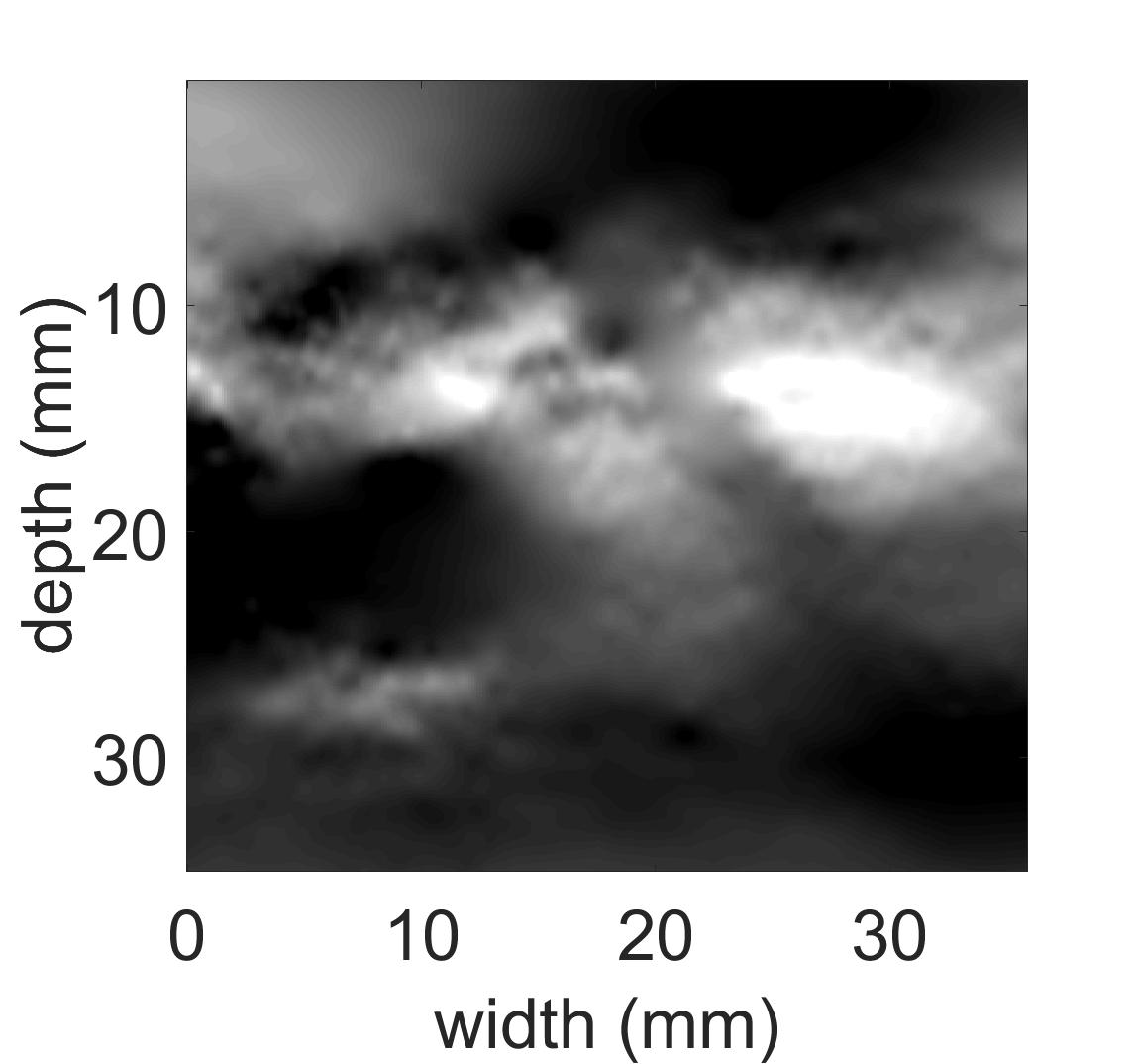}}\vspace*{-0em}
\subfigure[Color bar]{\includegraphics[height=0.9 cm,width=4 cm]{p2_colorbar_final}}\vspace*{-1em}
\end{center}%
\centering
\caption{The B-mode ultrasound and axial strain image using PCA-GLUE for the \textit{in vivo} liver data before ablation as we increase the number of RF lines $p$ from 2 to 10. The color bar is for the strain images only.} 
\label{different_p_invivo_before}
\end{figure*}

\begin{figure*}[]
\captionsetup{justification=centering}
\begin{center}
\subfigure[B-mode]{\includegraphics[height=3.35 cm,width=4.8 cm]{p3_after_bmode_final}}\hspace*{-0.5em}
\subfigure[Strain using $p=2$]{\includegraphics[height=3.35 cm,width=4.8 cm]{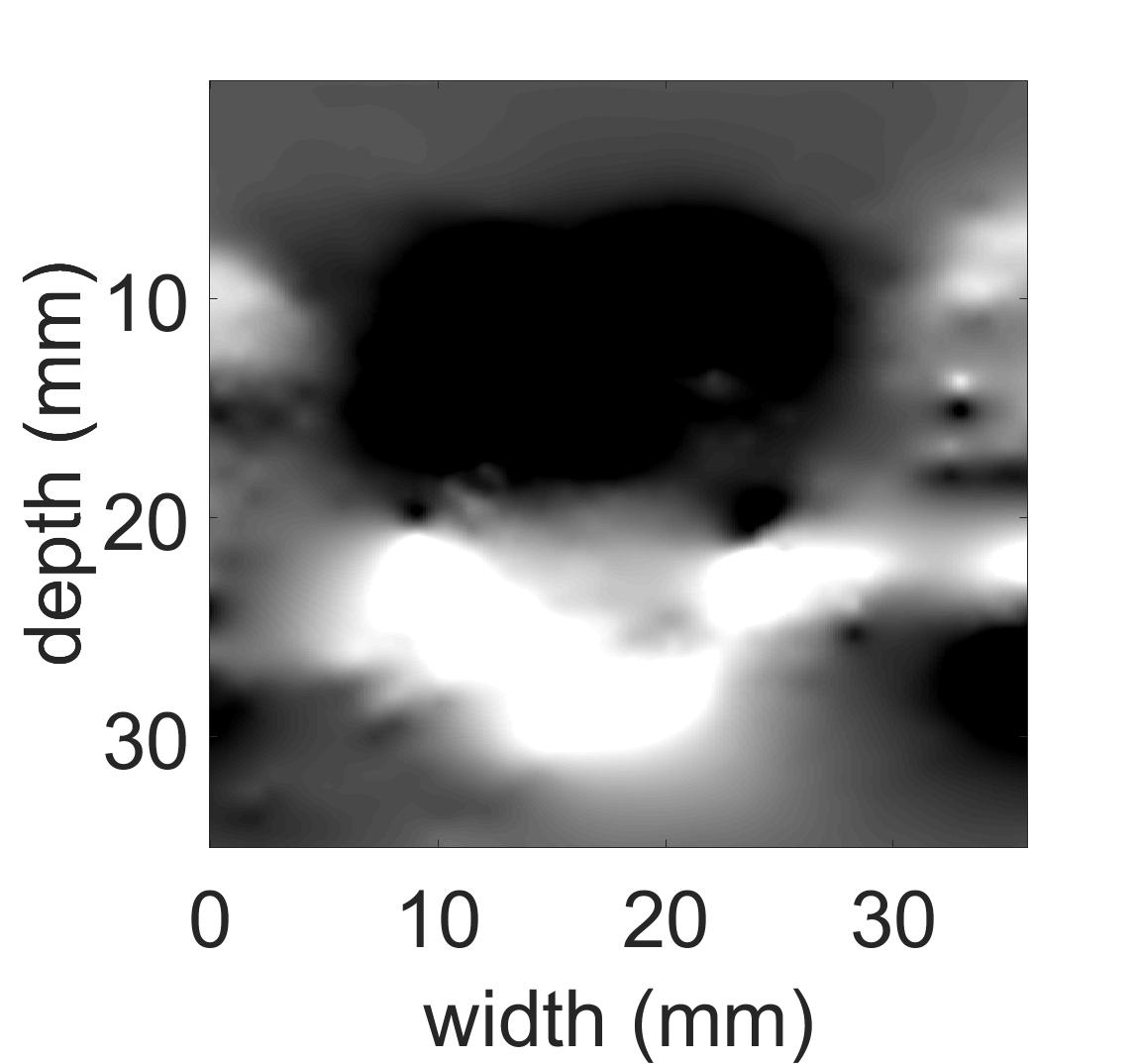}}\hspace*{-0.5em}
\subfigure[Strain using $p=5$]{\includegraphics[height=3.35 cm,width=4.8 cm]{p3_after_PCA-GLUE_final}}\hspace*{-0.5em}
\subfigure[Strain using $p=10$]{\includegraphics[height=3.35 cm,width=4.8 cm]{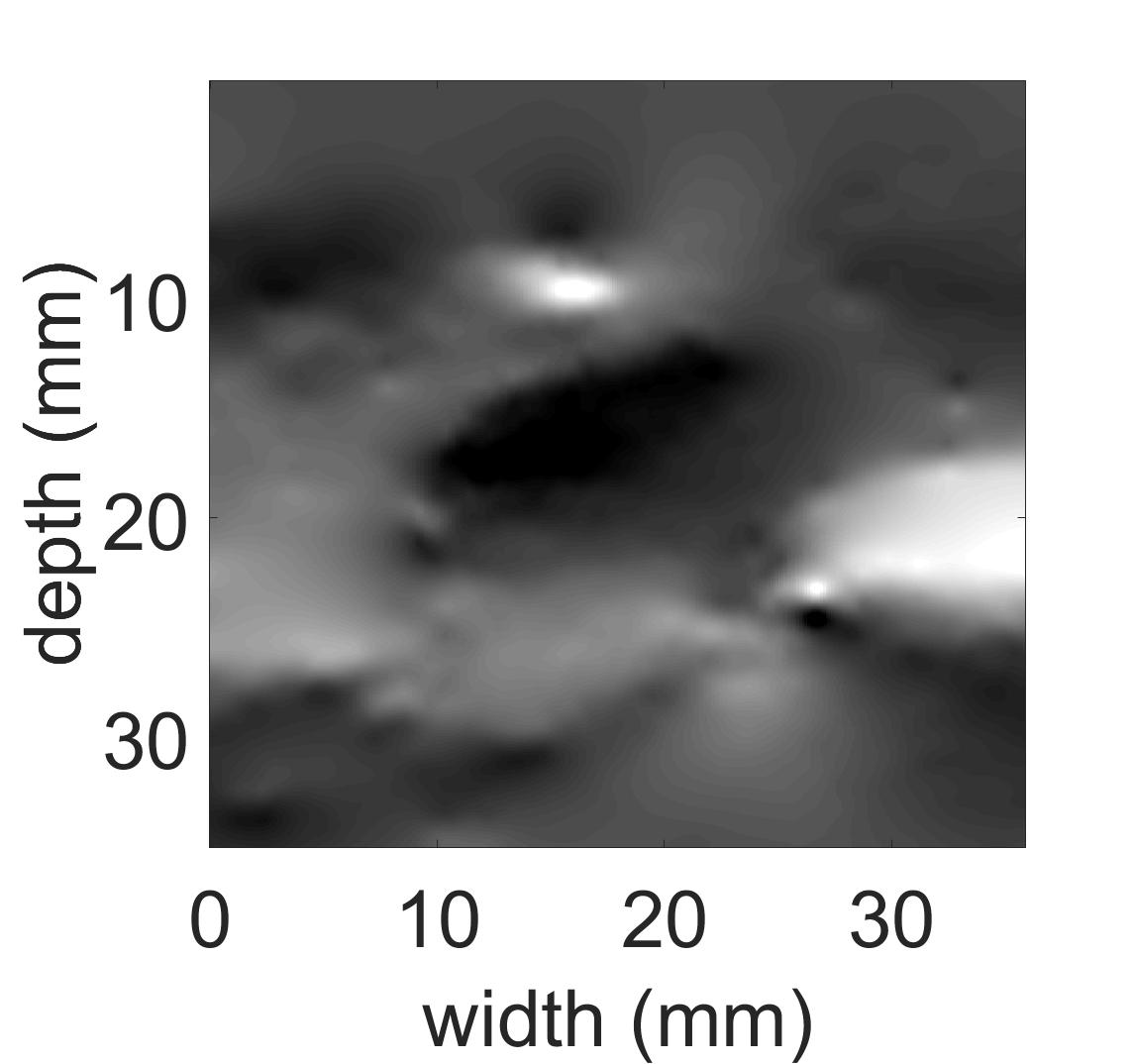}}\vspace*{-0em}
\subfigure[Color bar]{\includegraphics[height=0.7 cm,width=4 cm]{p3_after_colorbar_final}}\vspace*{-1em}
\end{center}%
\centering
\caption{The B-mode ultrasound and axial strain image using PCA-GLUE for \textit{in vivo} liver data after ablation as we increase the number of RF lines $p$ from 2 to 10. The color bar is for the strain images only.} 
\label{different_p_invivo_after}
\end{figure*}

We used different number of principal components for both phantom and \textit{in vivo} datasets to justify our choice of $N=12$. Fig. \ref{different_N_phantom}, \ref{different_N_invivo_before} and \ref{different_N_invivo_after} show that $N=6$ is a choice that works for all  the datasets except the \textit{in vivo} dataset after ablation. Therefore, we set $N=12$ for our experiments.
Fig. \ref{different_p_phantom}, \ref{different_p_invivo_before} and \ref{different_p_invivo_after} show that $p=2$ is also a choice that works for all the datasets except the \textit{in vivo} dataset after ablation. Therefore, we set $p=5$  since it doesn’t noticeably increase the computational complexity.

\begin{figure*}[]
\captionsetup{justification=centering}
\begin{center}

\subfigure[B-mode]{\includegraphics[height=3.35 cm,width=4.8 cm]{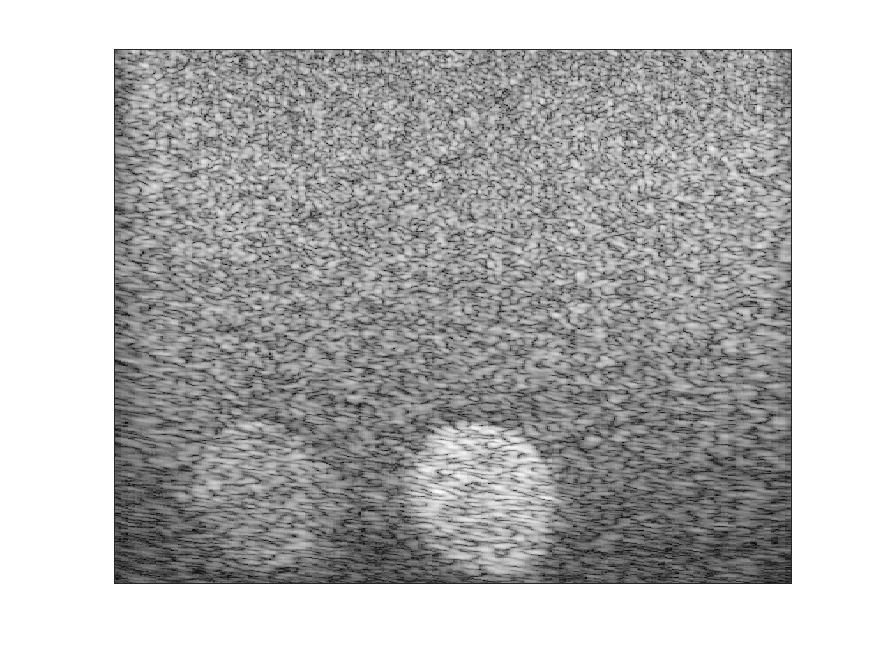}}\hspace*{-0.5em}
\subfigure[Case 1]{\includegraphics[height=3.35 cm,width=4.8 cm]{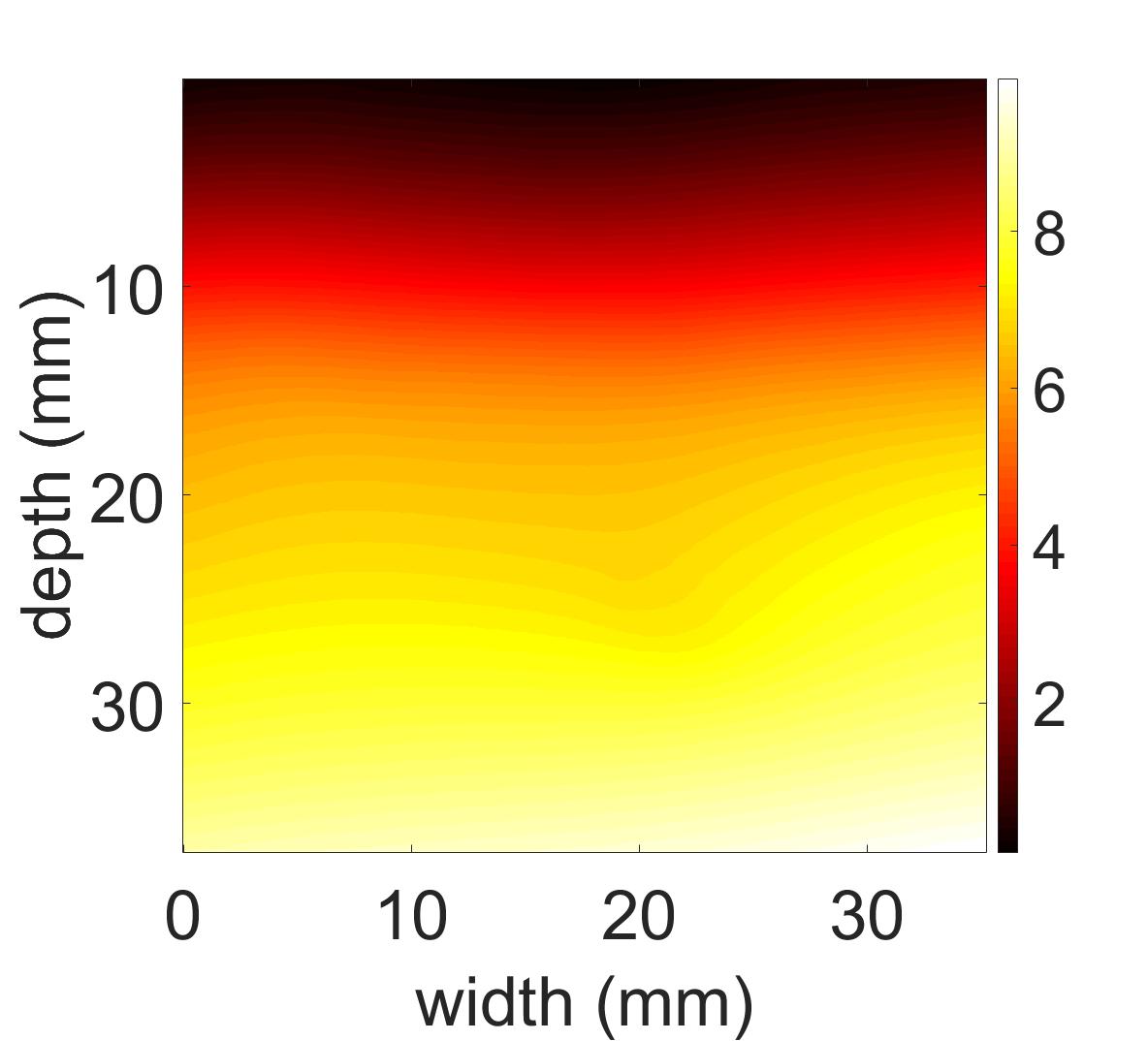}}\hspace*{-0.5em}
\subfigure[Case 2]{\includegraphics[height=3.35 cm,width=4.8 cm]{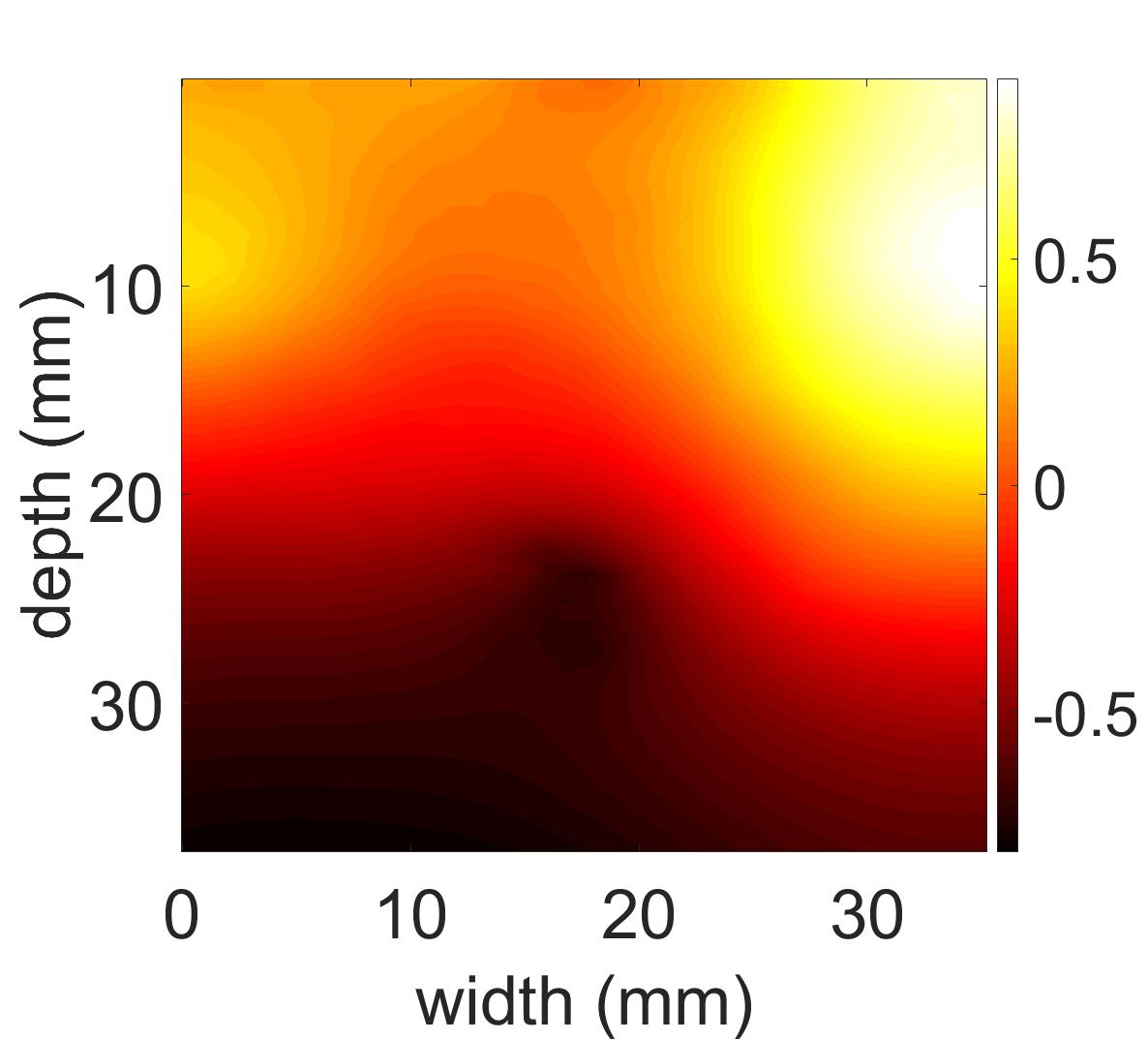}}\hspace*{-0.5em}
\subfigure[Case 3]{\includegraphics[height=3.35 cm,width=4.8 cm]{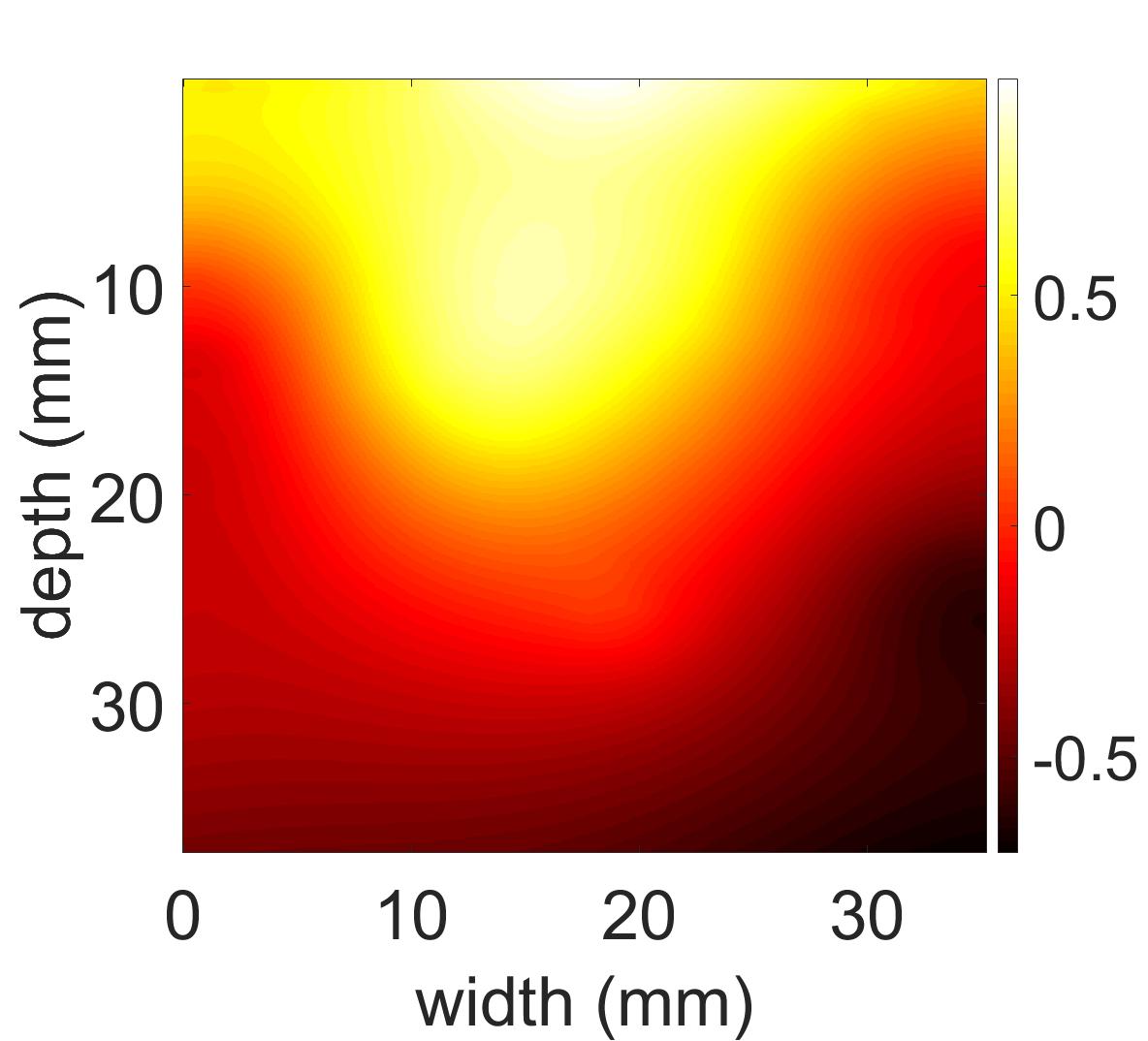}}\hspace*{1em}\\
\end{center}%
\centering
\caption{Some of the displacement images incorrectly classified as suitable for elastography.
} 
\label{bad_frames}
\end{figure*}

Tables \ref{compression_cnr} and \ref{compression_snr} show a comparison between the signal to Noise Ratio (SNR) and Contrast to Noise Ratio (CNR) of both PCA-GLUE and GLUE for different levels of compression from 1\% to 6\%. In addition, we collected new data where the probe displacement is completely out-of-plane, trying to mimic an inexperienced user. We found that out of a total of 160 out-of-plane RF frame pairs, our classifier only chose 11 pairs as suitable for elastography, achieving an accuracy of 93.125\%. This shows the ability of our classifier to distinguish the RF frames collected by an inexperienced user. Fig.~\ref{bad_frames} shows 3 of the 11 cases that were classified as suitable RF pairs. Although the compression was not in the axial direction, the displacement images are not completely out-of-plane.

\begin{figure*}[]
\captionsetup{justification=centering}
\begin{center}
\subfigure[B-mode]{\includegraphics[height=3.35 cm,width=4.8 cm]{phantom_bmode_big}}\hspace*{1em}
\subfigure[Strain using method 1]{\includegraphics[height=3.35 cm,width=4.8 cm]{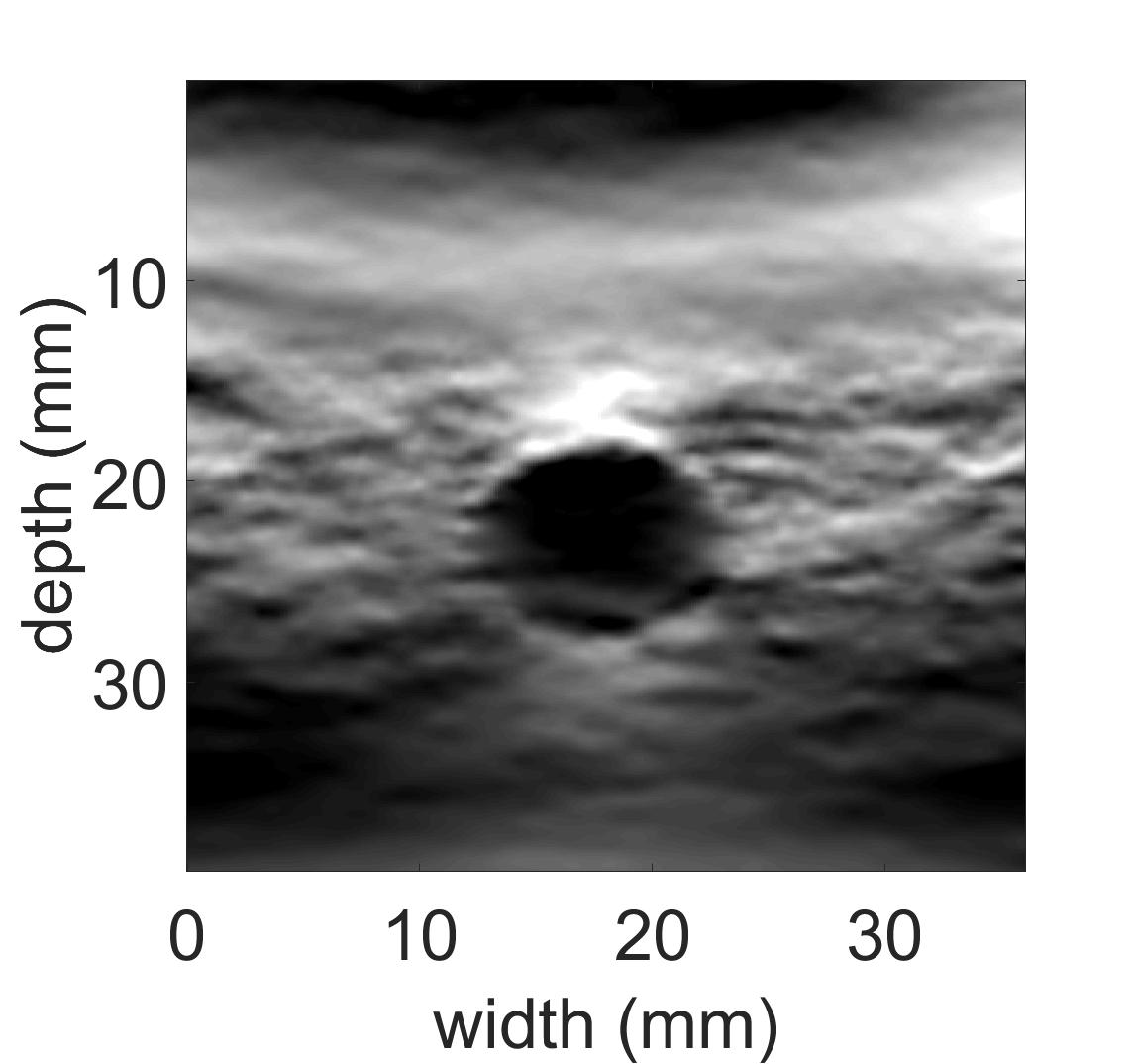}}\hspace*{1em}
\subfigure[Strain using method 2]{\includegraphics[height=3.35 cm,width=4.8 cm]{phantom_PCA-GLUE_final}}\hspace*{1em}\\
\subfigure[Color bar]{\includegraphics[height=0.8 cm,width=4 cm]{phantom_colorbar_final}}\vspace*{-1em}\\
\end{center}%
\centering
\caption{A comparison between the axial strain estimated using 2 methods for the phantom experiment. In method 1, the lateral displacement given to GLUE is obtained by passing on all RF lines. In the method 2, the lateral displacement given to GLUE is obtained by passing only on 5 RF lines, followed by bi-linear interpolation.  The color bar is for the strain images only.
} 
\label{how_acceptable_phantom}
\end{figure*}

\begin{figure*}[]
\captionsetup{justification=centering}
\begin{center}
\subfigure[B-mode]{\includegraphics[height=3.35 cm,width=4.8 cm]{p2_bmode_big}}\hspace*{1em}
\subfigure[Strain using method 1]{\includegraphics[height=3.35 cm,width=4.8 cm]{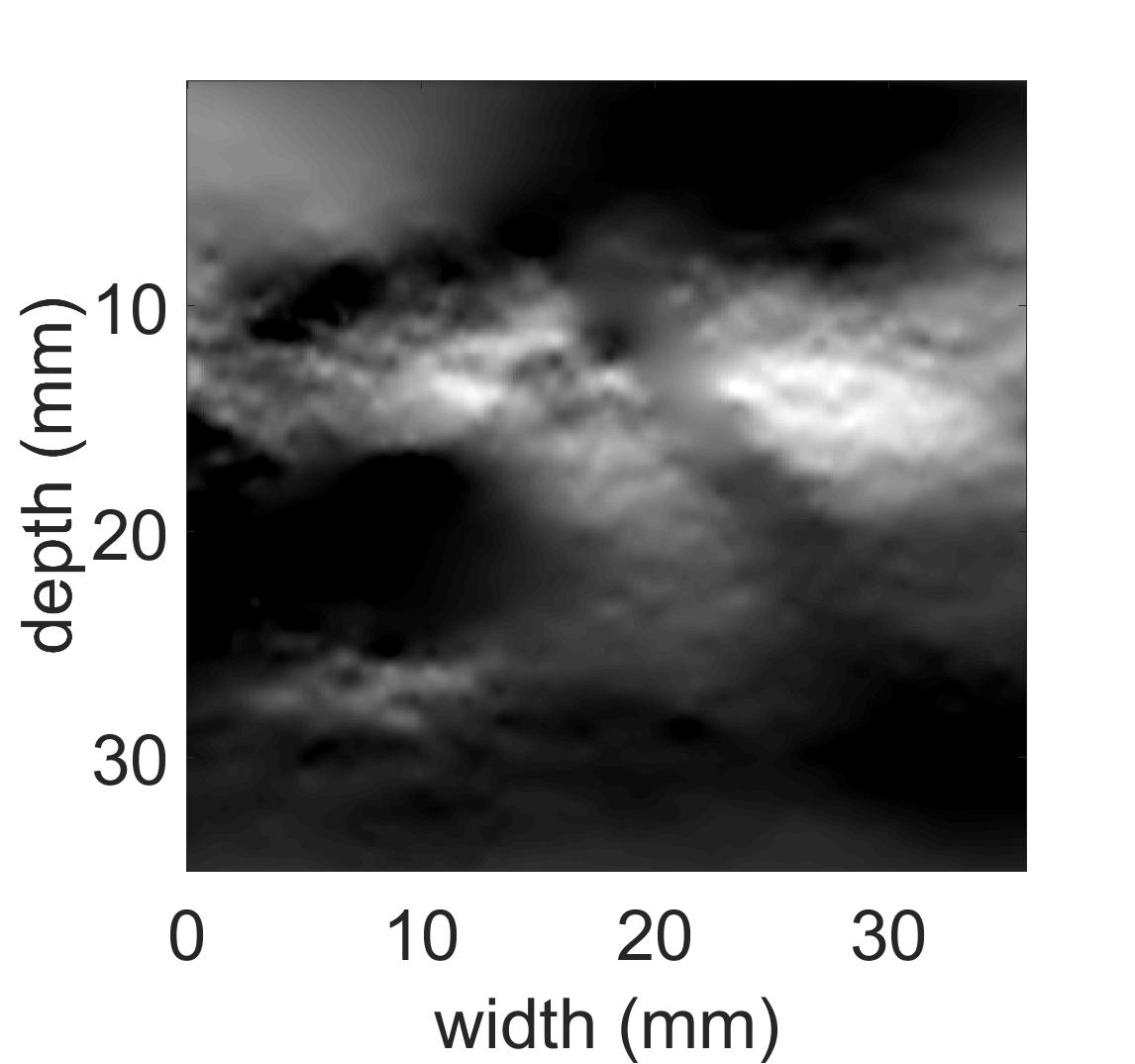}}\hspace*{1em}
\subfigure[Strain using method 2]{\includegraphics[height=3.35 cm,width=4.8 cm]{p2_PCA-GLUE_final}}\hspace*{1em}\\
\subfigure[Color bar]{\includegraphics[height=0.8 cm,width=4 cm]{p2_colorbar_final}}\vspace*{-1em}\\
\end{center}%
\centering
\caption{A comparison between the axial strain estimated using 2 methods for the \textit{in vivo} liver data before ablation. In method 1, the lateral displacement given to GLUE is obtained by passing on all RF lines. In method 2, the lateral displacement given to GLUE is obtained by passing only on 5 RF lines, followed by bi-linear interpolation. The color bar is for the strain images only.
} 
\label{how_acceptable_invivo_before}
\end{figure*}

\begin{figure*}[]
\captionsetup{justification=centering}
\begin{center}
\subfigure[B-mode]{\includegraphics[height=3.35 cm,width=4.8 cm]{p3_after_bmode_final}}\hspace*{1em}
\subfigure[Strain using method 1]{\includegraphics[height=3.35 cm,width=4.8 cm]{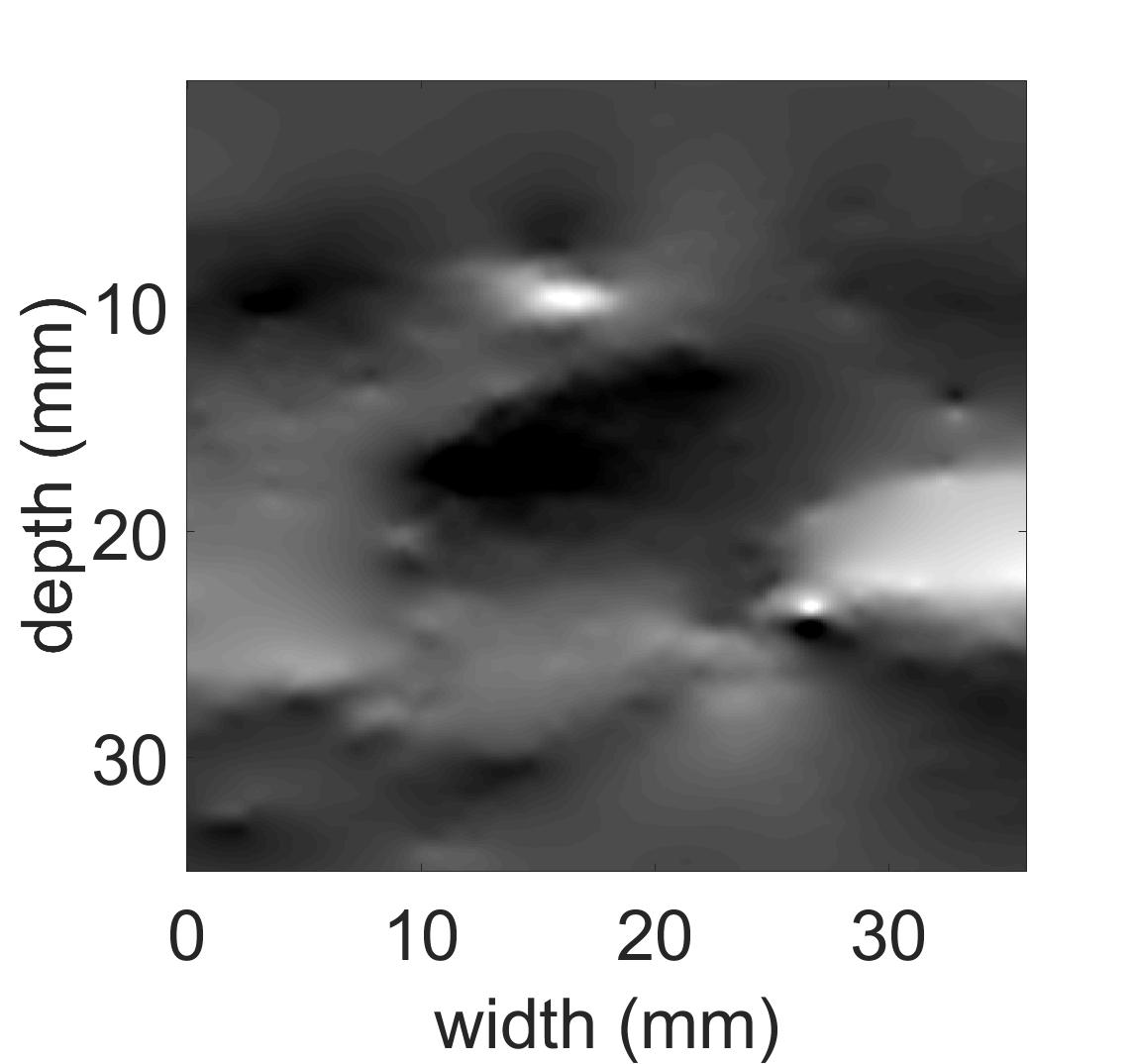}}\hspace*{1em}
\subfigure[Strain using method 2]{\includegraphics[height=3.35 cm,width=4.8 cm]{p3_after_PCA-GLUE_final}}\hspace*{1em}\\
\subfigure[Color bar]{\includegraphics[height=0.7 cm,width=4 cm]{p3_after_colorbar_final}}\vspace*{-1em}\\
\end{center}%
\centering
\caption{A comparison between the axial strain estimated using 2 methods for the \textit{in vivo} liver data after ablation. In method 1, the lateral displacement given to GLUE is obtained by passing on all RF lines. In method 2, the lateral displacement given to GLUE is obtained by passing only on 5 RF lines, followed by bi-linear interpolation. The color bar is for the strain images only.
} 
\label{how_acceptable_invivo_after}
\end{figure*}

We also compare our approach to the more computationally expensive method where the lateral displacement images given to GLUE are obtained by passing on all RF lines. Fig. \ref{how_acceptable_phantom}, \ref{how_acceptable_invivo_before} and \ref{how_acceptable_invivo_after} show that the 2 methods yield the same results on both phantom and \textit{in vivo} datasets.

\begin{table}[]
\centering
\caption{The CNR values of the axial strain images for the simulation data at different compression levels.}
\begin{tabular}{lrrrrr}
\toprule
\textbf{Compression level} & \textbf{GLUE} & \textbf{PCA-GLUE}\\
\midrule
\hspace*{3em} 1\% & \textbf{21.56} & 19.74\\
\hspace*{3em} 3\% & \textbf{21.59} & 20.23\\
\hspace*{3em} 6\% & \textbf{22.06} & 19.58 \\
\bottomrule
\end{tabular}
\label{compression_cnr}
\end{table}

\begin{table}[]
\centering
\caption{The SNR values of the axial strain images for the simulation data at different compression levels.}
\begin{tabular}{lrrrrr}
\toprule
\textbf{Compression level} & \textbf{GLUE} & \textbf{PCA-GLUE}\\
\midrule
\hspace*{3em} 1\% & 22.58 & \textbf{22.86}\\
\hspace*{3em} 3\% & \textbf{25.47} & 23.53\\
\hspace*{3em} 6\% & \textbf{28.38} & 27.98 \\
\bottomrule
\end{tabular}
\label{compression_snr}
\end{table}

%




\ifCLASSOPTIONcaptionsoff
\newpage
\fi



%

%

\bigskip
\bigskip
\bigskip

\bstctlcite{IEEEexample:BSTcontrol}
\bibliographystyle{IEEEtran}
\bibliography{strings,ref3}